\documentclass[a4paper,11pt]{article}
\pdfoutput=1 % if your are submitting a pdflatex (i.e. if you have
\usepackage{jheppub}
\usepackage{amsmath,amsfonts,amssymb,bbm,epsfig}
\usepackage{tabularx,longtable}
\usepackage{mathrsfs}
\usepackage{slashed}
\usepackage{booktabs}
\usepackage{graphicx}

\renewcommand\slash[1]{\not \! #1}

 \newcolumntype{L}[1]{>{\raggedright\arraybackslash}p{#1}}
 \newcolumntype{C}[1]{>{\centering\arraybackslash}p{#1}}
 \newcolumntype{R}[1]{>{\raggedleft\arraybackslash}p{#1}}

\def\be{\begin{equation}}
\def\ee{\end{equation}}
\def\bea{\begin{eqnarray}}
\def\eea{\end{eqnarray}}
 \newcommand{\ba}{\begin{eqnarray}}
 \newcommand{\ea}{\end{eqnarray}}

\def\nn{\nonumber}

\renewcommand\slash[1]{\not \! #1}

\newcommand{\gevsq}{\, \mathrm{GeV}^2}
\newcommand{\gev}{\, \mathrm{GeV}}
\newcommand{\mev}{\, \mathrm{MeV}}

\newcommand{\pipi}{\pi^+ \pi^-}
\newcommand{\mpipi}{m_{\pi^+\pi^-}}
\newcommand{\mPi}{m_{\pi} }
\newcommand{\Wgp}{W_{\gamma p}}

\newcommand{\fTwo}{f_{2}}
\newcommand{\pomeron}{\mathbb{P}}
\newcommand{\odderon}{\mathbb{O}}
\newcommand{\reggeon}{\mathbb{R}}
\newcommand{\fTwoR}{f_{2R}}

\newcommand{\aTwoR}{a_{2R}}
\newcommand{\omegaR}{\omega_{R}}
\newcommand{\rhoR}{\rho_{R}}

\newcommand{\sInd} {\mathfrak{s}}
\newcommand{\sIndPrim} {\mathfrak{s'}}

\newcommand{\alphaPrimPom}{\alpha'_\pomeron}
\newcommand{\alphaPom}{\alpha_\pomeron}
\newcommand{\alphaPrimOdd}{\alpha'_\odderon}
\newcommand{\alphaOdd}{\alpha_\odderon}

\newcommand{\alphaPrimRplus}{\alpha'_{\mathbb{R}_+}}
\newcommand{\alphaRplus}{\alpha_{\mathbb{R}_+}}
\newcommand{\alphaPrimRminus}{\alpha'_{\mathbb{R}_-}}
\newcommand{\alphaRminus}{\alpha_{\mathbb{R}_-}}

\title{\boldmath Photoproduction of $\pi^+ \pi^-$ pairs 
in a model with tensor-pomeron and vector-odderon exchange}

\author[a]{Arthur Bolz,}
\author[b,c]{Carlo Ewerz,}
\author[d]{Markos Maniatis,}
\author[b]{Otto Nachtmann,}
\author[a]{Michel Sauter,}
\author[a]{Andr\'e  Sch\"oning}

\affiliation[a]{Physikalisches Institut, Universit\"at Heidelberg,\\ 
Im Neuenheimer Feld 226, D-69120 Heidelberg, Germany}
\affiliation[b]{Institut f\"ur Theoretische Physik, Universit\"at Heidelberg,\\
Philosophenweg 16, D-69120 Heidelberg, Germany}
\affiliation[c]{ExtreMe Matter Institute EMMI, GSI Helmholtzzentrum f\"ur Schwerionenforschung,\\
Planckstra{\ss}e 1, D-64291 Darmstadt, Germany}
\affiliation[d]{Departamento de Ciencias B\'a{}sicas, Universidad del B\'i{}o-B\'i{}o,\\
Avda.\ Andr\'es Bello s/n, Casilla 447, Chill\'a{}n 3780000, Chile}

\emailAdd{abolz@physi.uni-heidelberg.de}
\emailAdd{C.Ewerz@thphys.uni-heidelberg.de}
\emailAdd{mmaniatis@ubiobio.cl}
\emailAdd{O.Nachtmann@thphys.uni-heidelberg.de}
\emailAdd{Michel.Sauter@desy.de}
\emailAdd{schoening@physi.uni-heidelberg.de}

\abstract{
We consider the reaction $\gamma p \rightarrow \pipi p$ at high energies. 
Our description includes dipion production via the resonances $\rho$, $\omega$, 
$\rho'$ and $f_2$, and via non-resonant mechanisms. The calculation is based 
on a model of high energy scattering with the exchanges of photon, pomeron, 
odderon and reggeons. The pomeron and the $C=+1$ reggeons are described 
as effective tensor exchanges, the odderon and the $C=-1$ reggeons as effective 
vector exchanges. 
We obtain a gauge-invariant version of the Drell-S\"oding mechanism 
which produces the skewing of the $\rho$-meson shape. 
Starting from the explicit formulae for the matrix element for dipion production 
we construct an event generator which comprises all contributions mentioned 
above and includes all interference terms. 
We give examples of total and differential cross sections and discuss asymmetries 
which are due to interference of $C=+1$ and $C=-1$ exchange 
contributions. These asymmetries can be used to search for odderon effects. 
Our model is intended to provide all necessary theoretical tools for a detailed 
experimental analysis of elastic dipion production 
for which data exist from fixed target experiments, from 
HERA, and are now being collected by LHC experiments. 
}

\begin{document} 
\maketitle
\flushbottom

\section{Introduction}
\label{sec: Introduction}
Photoproduction of $\pipi$ pairs on protons at high energies,
that is the reaction
\begin{align}
\label{eq: 1.1}
\gamma(q) + p(p) \longrightarrow \pi^+(k_1) + \pi^-(k_2) + p(p') \, ,
\end{align}
has been studied for a long time, both in theory and experiment. 
For reviews of topics relevant for this reaction see, for instance, 
\cite{Bauer:1977iq,Donnachie:2002en,Close:2007zzd}, and for model 
calculations \cite{Fiore:2003iv,Szczurek:2004xe,Forshaw:2012im}, for example. 
The reaction~\eqref{eq: 1.1} has been investigated by fixed target experiments 
\cite{Berger:1972an,Ballam:1971wq,Park:1971ts,Ballam:1971yd,Ballam:1972eq,Gladding:1974zc,Struczinski:1975ik,Egloff:1979mg,Aston:1982hr} 
and by the experiments H1 and ZEUS at HERA~\cite{Derrick:1995vq, Aid:1996bs, Breitweg:1997ed}. 
It can also be studied by the LHC experiments in ultra-peripheral $pp$ and $Ap$ collisions; 
cf.\ \cite{Baur:2001jj,Bertulani:2005ru,Baltz:2007kq,Staszewski:2011bg}. 
Indeed, first results for the 
reaction~\eqref{eq: 1.1} from such collisions have been presented in~\cite{Nystrand:2014vra}. 
The purpose of our paper is to give detailed formulae for the reaction~\eqref{eq: 1.1} 
as obtained in the recently proposed model~\cite{Ewerz:2013kda}. 
This model is constructed to describe soft high-energy reactions including 
Regge behaviour of the amplitudes. The pomeron and the charge conjugation 
$C=+1$ reggeons are described as effective tensor-exchange objects, the odderon and 
the $C=-1$ reggeons as effective vector-exchange objects. 
Due to these properties the amplitudes obtained in the model \cite{Ewerz:2013kda} 
naturally satisfy the rules of quantum field theory. In particular, they exhibit the 
correct behaviour under crossing and under charge conjugation. 

The reaction~\eqref{eq: 1.1} offers the interesting possibility to find odderon effects. 
Let us, therefore, first discuss the status of the odderon. The odderon, 
the charge conjugation $C=-1$ counterpart of the $C=+1$ pomeron, 
was introduced on theoretical grounds in~\cite{Lukaszuk:1973nt,Joynson:1975az}. 
More than forty years after its introduction the odderon is seen 
in theoretical papers, but not yet clearly in experiments. 
For a review see~\cite{Ewerz:2003xi}; for a general discussion of the 
pomeron and the odderon in high-energy reactions
and QCD see~\cite{Donnachie:2002en}. 
Various reactions have been proposed in order to look for odderon effects. 
We are concerned here with two of these proposals. 
One, suggested in 
\cite{Schafer:1992pq,Barakhovsky:1991ra,Kilian:1997ew,Berger:1999ca,Berger:2000wt}, 
is to study the exclusive photoproduction of $C=+1$ mesons 
and in particular the production of the  $f_2 \equiv f_2(1270)$ meson. 
The other is to look for certain asymmetries of the final state caused 
by the interference of $C$-even and $C$-odd exchanges. Various asymmetries 
of this kind have been described in 
\cite{Brodsky:1999mz,Ivanov:2001zc,Hagler:2002nh,Hagler:2002sg,Hagler:2002nf,Ginzburg:2002zd,Ginzburg:2002fy,Ginzburg:2005ay}. 
For a review of these and other observables suited for odderon 
searches see \cite{Ewerz:2003xi}.

\begin{figure}[thb]
\center
\includegraphics[width=0.96\textwidth]{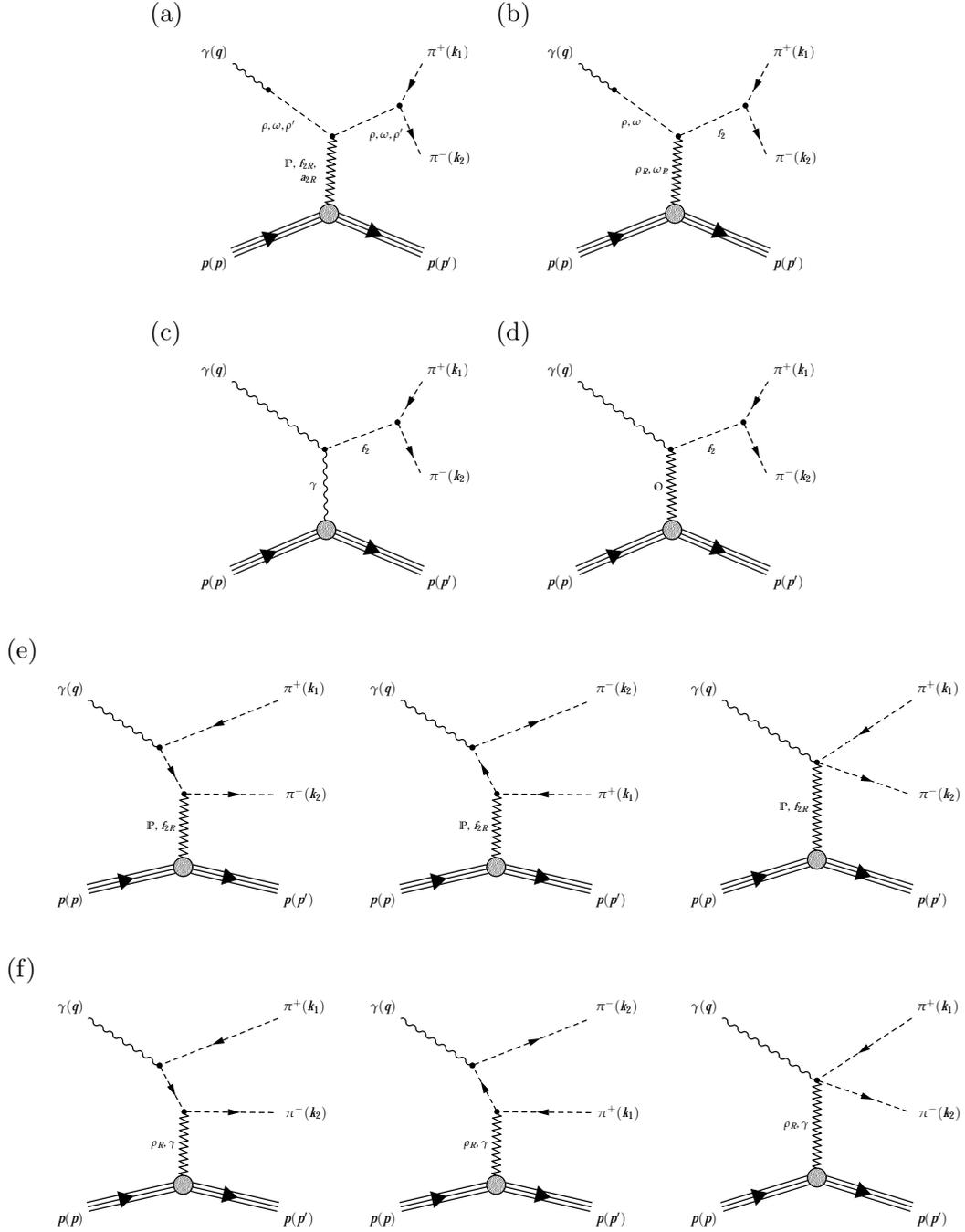}
\caption{Diagrams for $\pipi$ photoproduction: 
(a) vector-meson $\rho(770)$, $\omega(782)$, $\rho'(1450)$ production;
(b), (c), (d) $\fTwo$ production via reggeon, photon (Primakoff effect), and odderon exchanges, respectively;
(e) non-resonant $\pipi$ production via pomeron and $\fTwoR$ reggeon exchanges;
(f) non-resonant $\pipi$ production via $\rhoR$ and $\gamma$ exchanges. 
The diagrams (a) and (e) correspond to $C=+1$ exchange, the diagrams (b), (c), (d), and (f) 
to $C=-1$ exchange. 
\label{fig: 1} }
\end{figure}

We shall discuss now the relevant production mechanisms and 
diagrams for the reaction~\eqref{eq: 1.1} for high energies, $s \gg m_p^2$, and for $\pipi$
invariant mass, $\mpipi$, from threshold to $\mpipi \approx 2 \gev$. 
The kinematic quantities for this reaction are collected in appendix~\ref{app A}.
In this kinematic region we expect to see the production of the  $\rho$ meson, including 
$\rho$--$\omega$ interference effects, and the resonant production of the $f_2$
and $\rho' \equiv \rho(1450)$ mesons.
The exchanges to be considered are: the pomeron $\pomeron$,  
the reggeons $\fTwoR$, $\aTwoR$, $\omegaR$, $\rhoR$, the photon $\gamma$ and -- if it exists --
the odderon $\odderon$. 
In addition to resonance production, the $\pipi$ continuum production is considered.
The diagrams for all these subprocesses are shown in figure~\ref{fig: 1}. 
In the following these diagrams 
are evaluated in the model presented in~\cite{Ewerz:2013kda} and we shall give 
explicit formulae which can serve as basis for experimental analyses.
The photoproduction of a neutral meson by photon exchange is called the 
Primakoff effect \cite{Primakoff}. Our diagram \ref{fig: 1}(c) corresponds 
to this effect with $f_2$ as the neutral meson. 
We see from figure \ref{fig: 1} that exchanges with $C=+1$ ($\pomeron$, $\fTwoR$, $\aTwoR$) and 
$C=-1$ ($\omegaR$, $\rhoR$, $\gamma$, $\odderon$) contribute to the reaction~\eqref{eq: 1.1}.
This leads to asymmetries in the $\pipi$ distributions, as we shall  discuss at length below.
Such asymmetries have been proposed for odderon searches in  
\cite{Ivanov:2001zc,Hagler:2002nh,Hagler:2002sg,Hagler:2002nf,Ginzburg:2002zd,Ginzburg:2002fy,Ginzburg:2005ay}. 
But, as we see from figure~\ref{fig: 1}, odderon exchange is not the only $C=-1$
exchange contribution. Thus, a careful analysis of all contributing $C=+1$ and $C=-1$
exchanges is necessary in order to assess the r\^ole of the odderon. 
In this paper we present such a description as is a prerequisite for an 
experimental search for odderon effects in the reaction~\eqref{eq: 1.1}. 

\begin{figure}[thb]
\center
\includegraphics[height=6cm] {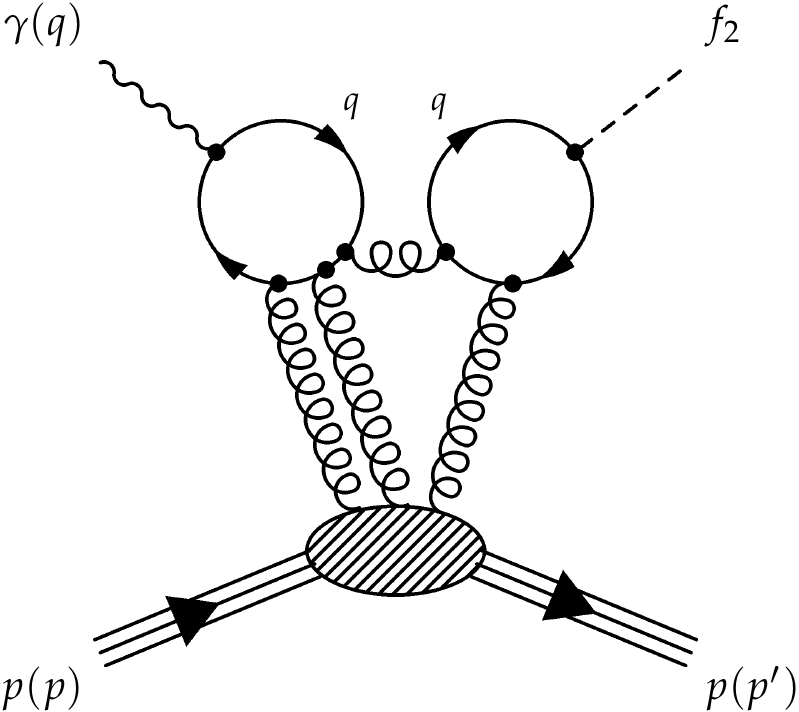}
\caption{A QCD diagram contributing to $\fTwo$ production via odderon exchange.
\label{fig: 2} 
}
\end{figure}

Before we do this we give an argument why the photoproduction of $\fTwo$ mesons
may be particularly sensitive to odderon effects.
We show in figure \ref{fig: 2} a QCD diagram which contributes and is 
specific to $f_2$ production. 
Via loops the photon $\gamma$ can couple to three gluons and the $\fTwo$ can
couple to two gluons, resulting with proper arrangement of the gluon lines in
a three-gluon exchange (the simplest perturbative representation of the odderon) with the proton. 
We note that this type of diagram does not exist for the photoproduction
of $\pi^0$ and $a_2$  mesons on which rather strict experimental limits
exist; see~\cite{Adloff:2002dw, Berndt:2002tw}. 
The non-observation of odderon exchange in these reactions was discussed
in~\cite{Donnachie:2005bu}, and 
it was shown that chiral symmetry implies a strong suppression of $\pi^0$ 
photoproduction via odderon exchange~\cite{Ewerz:2006gd}. 

Our paper is organised as follows. In section~\ref{sec: Matrix elements, cross sections, asymmetries}  
we give analytic expressions for the diagrams
of figure~\ref{fig: 1}. In section~\ref{sec: Results} we present numerical
results of total and differential cross sections and discuss the
$\pipi$ asymmetries. Section~\ref{sec: Conclusions} contains our conclusions.
Kinematical relations and a list of the effective propagators and vertices are
given in appendix~\ref{app A} and appendix~\ref{app B}, respectively. 
In appendix \ref{app C} we discuss the behaviour of the 
differential cross section for $t \to 0$. Appendix \ref{app D} deals 
with the determination of the Monte Carlo weights for our event generator. 

\section{Matrix elements, cross sections and asymmetries}
\label{sec: Matrix elements, cross sections, asymmetries}

We define the matrix element $\mathcal{M}^{\mu}_{\sIndPrim, \sInd}$ for the reaction~\eqref{eq: 1.1} as 
\begin{align}
  \label{eq: 2.1}
\mathcal{M}^{\mu}_{\sIndPrim, \sInd}(k_1, k_2, p', q, p ) \epsilon_{\mu} = 
\langle \pi^+(k_1), \pi^-(k_2), p(p', \sIndPrim)  \vert \mathcal {T} 
\vert  \gamma(q,\epsilon),p(p,\sInd) \rangle \, \text{.}
\end{align}
Here $k_1, k_2, p', q$ and $p$ are the four-vectors of the involved particles,
$\epsilon$ is
the photon's polarisation vector, and  $\sInd$  and
$\sIndPrim$ are the spins
of the incoming and outgoing proton, respectively. 
Details of the kinematics of the reaction are discussed in appendix \ref{app A}. 

The matrix element \eqref{eq: 2.1} gets contributions from all diagrams shown in figure~\ref{fig: 1}: 
\begin{align}
  \label{eq: 2.2}
\mathcal{M}^{\mu}_{\sIndPrim, \sInd}  = 
\mathcal{M}^{\rm (a)\mu}_{\sIndPrim, \sInd} +
\mathcal{M}^{\rm (b)\mu}_{\sIndPrim, \sInd} +
\mathcal{M}^{\rm (c)\mu}_{\sIndPrim, \sInd} +
\mathcal{M}^{\rm (d)\mu}_{\sIndPrim, \sInd} +
\mathcal{M}^{\rm (e)\mu}_{\sIndPrim, \sInd} +
\mathcal{M}^{\rm (f)\mu}_{\sIndPrim, \sInd}  \, \text{.}
  \end{align}
The diagrams~(a) and~(e) correspond to $C=+1$ exchange, the diagrams~(b), (c), (d) and~(f)  
to  $C=-1$ exchange. Gauge invariance requires
\begin{align}
\label{eq: 2.3}
q_{\mu} \mathcal{M}^{\mu}_{\sIndPrim, \sInd}  = 0 \, \text{.}
\end{align}
In our calculation we find that this gauge-invariance relation holds for each subclass of diagrams~(a) to~(f)
separately: 
\begin{align}
\label{eq: 2.4}
q_{\mu} \mathcal{M}^{\rm (a)\mu}_{\sIndPrim, \sInd}  =  \cdots = q_{\mu} \mathcal{M}^{\rm (f) \mu}_{\sIndPrim, \sInd}   = 0 
\, \text{.}
\end{align}

The cross section for~\eqref{eq: 1.1} assuming unpolarised particles in the 
initial state and no observation of polarisations in the final state reads
\begin{align}
\label{eq: 2.5}
  d\sigma^{\gamma p} = & 
 \frac{(2 \pi)^4}{2 (s - m_p^2)} 
 \left(-\frac{1}{4}\sum_{\sIndPrim, \sInd} \mathcal{M}_{\mu, \sIndPrim, \sInd}^* \, \mathcal{M}_{\sIndPrim, \sInd}^{\mu}\right)  
\nonumber \\
& \times \frac{1}{(2 \pi) ^ 9}
 \frac{d^3k_1}{2 k_1^0}
 \frac{d^3k_2}{2 k_2^0}
 \frac{d^3p'}{2 p'^0} \,
 \delta^{(4)} (k_1 + k_2 + p' - p - q)
  \, \text{.} 
\end{align}
We define
\begin{align}
\label{eq: 2.6}
\mathscr{R}\left(s, t, k^2, p \cdot \left(k_1 - k_2\right), q \cdot \left(k_1 - k_2\right) \right) = 
-\frac{1}{4} \sum_{\sIndPrim, \sInd}\mathcal{M}_{\mu, \sIndPrim, \sInd}^* \, \mathcal{M}_{\sIndPrim, \sInd}^{\mu}
 \, \text{.} 
\end{align}
$\mathscr{R}$ can only depend on the variables indicated, see~\eqref{eq:
  A.10}, where the Mandelstam variables $s$ and $t$ denote the squared center of mass
energy and the squared momentum transfer of the reaction, respectively.
The four-vector of the dipion system is defined by the four-vector sum of the
$\pi^+$ and $\pi^-$ as $k=k_1+k_2$.

We can split $\mathscr{R}$ from \eqref{eq: 2.6} into $\mathscr{R}=\mathscr{R}_++\mathscr{R}_-$
where the parts $\mathscr{R}_+$ and $\mathscr{R}_-$ 
are even and odd, respectively, under the simultaneous 
sign change of the last two arguments:
\begin{align}
\label{eq: 2.7}
p \cdot \left(k_1 - k_2\right) \:\longrightarrow\: -p \cdot \left(k_1 - k_2\right) \, \text{,} \nonumber \\
q \cdot \left(k_1 - k_2\right) \:\longrightarrow\: -q \cdot \left(k_1 - k_2\right) \, \text{.} 
\end{align}
$\mathscr{R}_+$ contains the squares of $C=+1$ and $C=-1$ exchange
amplitudes, $\mathscr{R}_-$ contains the interference terms of the $C=+1$ and $C=-1$ exchanges.
We get 
\begin{align}
\label{eq: 2.8}
 \mathscr{R}_+&\left(s, t, k^2, p \cdot \left(k_1 - k_2\right), q \cdot \left(k_1 - k_2\right) \right)   \nonumber \\
& = \mathscr{R}_+\left(s, t, k^2, - p \cdot \left(k_1 - k_2\right), - q \cdot \left(k_1 - k_2\right) \right)  \nonumber \\
& = -\frac{1}{4} \sum_{\sIndPrim, \sInd} \left[
\left(\mathcal{M}_{\mu, \sIndPrim, \sInd}^{\rm (a+e)} \right)^* \mathcal{M}_{\sIndPrim, \sInd}^{\rm(a+e) \mu} + 
\left(\mathcal{M}_{\mu, \sIndPrim, \sInd}^{\rm(b+c+d+f)}\right)^* \mathcal{M}_{\sIndPrim, \sInd}^{\rm(b+c+d+f) \mu} \right] 
\, \text{,} \\
\label{eq: 2.9}
 \mathscr{R}_- &\left(s, t, k^2, p \cdot \left(k_1 - k_2\right), q \cdot \left(k_1 - k_2\right) \right)   \nonumber \\
& = - \, \mathscr{R}_-\left(s, t, k^2, - p \cdot \left(k_1 - k_2\right), - q \cdot \left(k_1 - k_2\right) \right)  \nonumber \\
& = -\frac{1}{4} \sum_{\sIndPrim, \sInd} \left[
\left(\mathcal{M}_{\mu, \sIndPrim, \sInd}^{\rm(a+e)}\right)^* \mathcal{M}_{\sIndPrim, \sInd}^{\rm(b+c+d+f) \mu} + 
\left(\mathcal{M}_{\mu, \sIndPrim, \sInd}^{\rm(b+c+d+f)}\right)^* \mathcal{M}_{\sIndPrim, \sInd}^{\rm(a+e) \mu} \right] 
\,.
\end{align}
Observables which are odd under the transformation \eqref{eq: 2.7} are, therefore, particularly
suitable to measure the interference of $C=+1$ and $C=-1$
exchanges, and thus allow to study possible odderon contributions. Examples of such observables
are discussed in section~\ref{sec: Results}.

We shall now calculate the amplitudes corresponding to the diagrams of figure~\ref{fig: 1}. 
The effective propagators and vertices needed for these calculations are
mostly taken from~\cite{Ewerz:2013kda}.
To make the present paper self-contained we list the propagators and vertices
needed here in appendix~\ref{app B}. 
We consider it an asset of our approach that given these propagators and 
vertices we can use the standard rules of QFT to obtain the amplitudes. 
This guarantees, for instance, that all gauge-invariance and charge-conjugation 
properties of the amplitudes are automatically satisfied. 
We shall only present the final results in the following.
These explicit expressions are the basis for the construction of our event
generator for the reaction~\eqref{eq: 1.1}.

\subsection{Vector-meson production}
\label{sec: VM production}

Here we discuss the diagrams of figure~\ref{fig: 1}(a), that is, the production of the vector-mesons $\rho$, 
$\omega$ and $\rho'$ which then decay into $\pipi$. 
For $\rho$ and $\omega$ we consider both $\pomeron V V$ (pomeron) and $\fTwoR V
V$ (reggeon) couplings ($V = \rho, \omega$) whereas for the rather small 
$\rho'$ contribution only the $\pomeron \rho' \rho'$ coupling is considered. 
We include strong-isospin violating effects for
the $(\rho, \omega)$ propagator
and for the $\omega \rightarrow \pi^+ \pi^-$ decay but assume absence of such 
violations for the couplings of pomeron and reggeons to vector mesons. 
With the expressions for the propagators, vertices and form factors from appendix~\ref{app B} 
we then obtain 
\begin{align}
\label{eq: 2.11}
\mathcal{M}_{\mu, \sIndPrim, \sInd}^{\rm (a)} = &     
\sum_{\substack{V = \rho , \omega,\\ V'= \rho, \omega}}
\left\{
\mathcal{M}_{\mu, \sIndPrim, \sInd}^{(V', V, \pomeron)}  + 
\mathcal{M}_{\mu, \sIndPrim, \sInd}^{(V', V, \fTwoR)} 
\right\} + 
\mathcal{M}_{\mu, \sIndPrim, \sInd}^{(\aTwoR)} + 
\mathcal{M}_{\mu, \sIndPrim, \sInd}^{(\rho', \rho', \pomeron)} \, \text{.}   
\end{align}
Here we have for $V', V \in \{ \rho, \omega, \rho' \}$
\begin{align}
\label{eq: 2.12}
\mathcal{M}_{\mu, \sIndPrim, \sInd}^{(V', V, \pomeron)}  &= \frac{i}{4} \, e \,s \, F_1(t) \,F_M(t)\, 
\tilde{F}^{(V)}\!(k^2) \,g_{V' \pi \pi}
\left\{ \mathcal{K}_{\mu, \sIndPrim, \sInd}^{(0, V', V)} V_{\pomeron}^{(0,V)}
- \mathcal{K}_{\mu, \sIndPrim, \sInd}^{(2, V', V)} V_{\pomeron}^{(2,V)} \right\}
\, \text{,}   
\\
\label{eq: 2.13}
\mathcal{M}_{\mu, \sIndPrim, \sInd}^{(V', V, \fTwoR)}  &= \frac{i}{4} \, e \,s \, F_1(t)\, F_M(t) \, 
\tilde{F}^{(V)}\!(k^2) \, g_{V' \pi \pi}
\left\{ \mathcal{K}_{\mu, \sIndPrim, \sInd}^{(0, V', V)} V_{\fTwoR}^{(0,V)}
- \mathcal{K}_{\mu, \sIndPrim, \sInd}^{(2, V', V)} V_{\fTwoR}^{(2,V)} \right\}
\, \text{,}   
\end{align}
where for $i=0,2$ 
\begin{align}
\label{eq: 2.14}
\mathcal{K}_{\mu, \sIndPrim, \sInd}^{(i, V', V)}  = 
\frac{1}{s^2} (k_1 - k_2)^{\nu} \Delta_T^{(V', V)}\!(k^2) \, \Gamma_{\nu \mu \kappa \lambda}^{(i)}(k, -q) \, 
\bar{u}_{\sIndPrim}(p')  \gamma^\kappa \left( p' + p\right)^\lambda u_{\sInd}(p) 
\, \text{,} 
\end{align}
and
\begin{align}
\label{eq: 2.15}
V_\pomeron^{(0, V)} &=  \frac{1}{\gamma_V} \,6 \, \beta_{\pomeron N N} \, a_{\pomeron V V} 
(-i \, s \, \alphaPrimPom)^{\alphaPom(t) -1}
\, \text{,} \nonumber \\
V_\pomeron^{(2, V)} &=  \frac{1}{\gamma_V} \, 3  \,\beta_{\pomeron N N} \, b_{\pomeron V V} 
( -i \, s\, \alphaPrimPom)^{\alphaPom(t) -1}
\, \text{,}
\end{align}
\begin{align}
\label{eq: 2.16}
V_{\fTwoR}^{(0, V)} &=  \frac{1}{\gamma_V} \frac{2 \, g_{\fTwoR p p}}{M_0} \, a_{\fTwoR V V}
(-i \, s \, \alphaPrimRplus )^{\alphaRplus (t) -1}
\, \text{,} \nonumber \\
V_{\fTwoR}^{(2, V)} &=  \frac{1}{\gamma_V} \frac{g_{\fTwoR p p}}{M_0} \, b_{\fTwoR V V}
(-i \, s \, \alphaPrimRplus )^{\alphaRplus (t) -1}
\, . 
\end{align}
For  $\aTwoR$ reggeon exchange in figure~\ref{fig: 1}(a) we have four contributions. 
The photon can turn into a $\rho$ which, upon $\aTwoR$ exchange, turns into a $\omega$; 
and we can have the r\^oles of $\rho$ and $\omega$ exchanged. Both, the final 
$\omega$ and $\rho$, can then, by propagator mixing, go to $V' \in \{\rho,\omega\}$ 
which decays to $\pipi$. The amplitude taking into account all these processes is 
\begin{align}
\label{eq: 2.17}
\mathcal{M}_{\mu, \sIndPrim, \sInd}^{(\aTwoR)}  =&\,  \frac{i}{4} \, e \,s \, F_1(t) \,F_M(t)\, 
\tilde{F}^{(\rho)}\!(k^2) \,
\frac{g_{\aTwoR p p}}{M_0}  \, 
(-i \, s \, \alphaPrimRplus )^{\alphaRplus (t) -1} \nonumber \\
& \times \sum_{V'= \rho, \omega} g_{V' \pi \pi} \left\{
\left[
\mathcal{K}_{\mu, \sIndPrim, \sInd}^{(0, V', \omega)} \frac{1}{\gamma_{\rho}} + 
\mathcal{K}_{\mu, \sIndPrim, \sInd}^{(0, V', \rho)} \frac{1}{\gamma_{\omega}} 
\right] 2 a_{\aTwoR \omega \rho} 
\right.
 \nonumber \\
&\qquad \qquad \qquad \quad
\left. - \left[
\mathcal{K}_{\mu, \sIndPrim, \sInd}^{(2, V', \omega)} \frac{1}{\gamma_{\rho}} + 
\mathcal{K}_{\mu, \sIndPrim, \sInd}^{(2, V', \rho)} \frac{1}{\gamma_{\omega}} 
\right] b_{\aTwoR \omega \rho} 
\right\}
\, \text{.}   
\end{align}
Here and in the following $M_0 = 1 \gev$ is used in various places for
dimensional reasons. All quantities occurring here and their definitions
can be found in table~\ref{tab: B.1} of appendix~\ref{app B}.

We note that for the $\rho'$ meson, apart from the known mass
$m_{\rho'}$ and the width $\Gamma_{\rho'}$, only the combinations 
\begin{align}
\label{eq: 2.18}
g_{\rho' \pi \pi} \frac{1}{\gamma_{\rho'}} a_{\pomeron \rho' \rho'}
\qquad \text{and} \qquad
g_{\rho' \pi \pi} \frac{1}{\gamma_{\rho'}} b_{\pomeron \rho' \rho'}
\, \text{} 
\end{align}
enter in the expression~(\ref{eq: 2.11}) for $\mathcal{M}_{\mu, \sIndPrim,
  \sInd}^{\rm (a)} $.
Thus, only these two combinations can be determined by studying the $\rho'$ contribution
to the reaction~(\ref{eq: 1.1}). 

\boldmath
\subsection{Production of $f_2$ by reggeon exchange}
\unboldmath
\label{sec: f2 production by reggeon exchange}

Here the corresponding diagram is shown in figure~\ref{fig: 1}(b). 
We get with the expressions for the propagators and vertices from appendix~\ref{app B} 
\begin{align}
\label{eq: 2.19}
\mathcal{M}_{\mu, \sIndPrim, \sInd}^{\rm (b)}  = \frac{1}{4} \, e \,s \, F_1(t) \, F_M(t) 
\left[ F^{(\fTwo \pi \pi)}(k^2) \right]^2
g_{\fTwo \pi \pi}
\sum_{V=\rho, \omega}
\left\{ 
\mathcal{N}_{\mu, \sIndPrim, \sInd}^{(0)} W_V^{(0)}
- \mathcal{N}_{\mu, \sIndPrim, \sInd}^{(2)} W_V^{(2)} 
\right\}
\, \text{.}   
\end{align}
We find for $i=0,2$ 
\begin{align}
\label{eq: 2.20}
\mathcal{N}_{\mu, \sIndPrim, \sInd}^{(i)} =& \frac{2}{s} \left[ \left(k_1- k_2\right)^\kappa \left(k_1 - k_2\right)^\lambda
- \frac{1}{4} g^{\kappa \lambda} \left(k_1-k_2\right)^2 \right] \nonumber \\
& \times \Delta^{(f_2)}_{\kappa \lambda}{}^{\kappa' \lambda '}(k) \; \Gamma^{(i)}_{\mu \nu \kappa' \lambda'} (q, p-p') \,
\bar{u}_{\sIndPrim}(p') \gamma^\nu u_{\sInd}(p) 
\, \text{,} 
\end{align}
and for $V = \rho, \omega $
\begin{align}
W_{V}^{(0)} &= \frac{1}{\gamma_V} \frac{1}{M_0} \frac{1}{M_- ^2} \, g_{V_R p p}
\, 2\,  a_{V_R V f_2}
(-\,i\, s\, \alphaPrimRminus )^{\alphaRminus (t) -1}
\, \text{,}    \nonumber \\
W_{V}^{(2)} &= \frac{1}{\gamma_V} \frac{1}{M_0} \frac{1}{M_- ^2} \, g_{V_R p p}
 \, b_{V_R V f_2}
(-\,i\, s\, \alphaPrimRminus )^{\alphaRminus (t) -1}
\,.  
\end{align}
For the definition of the parameters occurring here we refer again
to table~\ref{tab: B.1} in appendix~\ref{app B}.

\boldmath
\subsection{Production of $f_2$ by photon exchange}
\unboldmath
\label{sec: f2 production by photon exchange}

Here we calculate the amplitude corresponding to the diagram of figure~\ref{fig: 1}(c). 
Using the expressions for the propagators and vertices from appendix~\ref{app B} 
we get the following:
\begin{align}
\label{eq: 2.22}
\mathcal{M}_{\mu, \sIndPrim, \sInd}^{\rm (c)} = &     
\frac{1}{4} \, e \,s  \, g_{\fTwo \pi \pi} \, F_M(t) 
\left[ F^{(\fTwo \pi \pi)} (k^2) \right]^2
\nonumber \\
&\times
\left\{
F_1(t) \left[ 
\mathcal{N}_{\mu, \sIndPrim, \sInd}^{(0)} W_{\gamma}^{(0)}
- \mathcal{N}_{\mu, \sIndPrim, \sInd}^{(2)} W_{\gamma}^{(2)} 
\right] 
+ F_2(t) \left[ 
S_{\mu, \sIndPrim, \sInd}^{(0)} W_{\gamma}^{(0)}
- S_{\mu, \sIndPrim, \sInd}^{(2)} W_{\gamma}^{(2)} 
\right] 
\right\}
\, \text{.}  
\end{align}
Here the $\mathcal{N}_{\mu, \sIndPrim, \sInd}^{(i)}$ $(i=0,2)$ are taken
from~(\ref{eq: 2.20}) and we have defined for $i=0,2$: 
\begin{align}
\label{eq: 2.23}
\mathcal{S}_{\mu, \sIndPrim, \sInd}^{(i)} =& 
\frac{2}{s} \left[ \left(k_1 - k_2 \right)^\kappa \left( k_1 - k_2\right)^\lambda 
-\frac{1}{4} g^{\kappa \lambda} \left(k_1 - k_2\right)^2 \right] \nonumber\\
& \times \Delta^{(f_2)}_{\kappa \lambda}{}^{\kappa' \lambda'} (k) \; 
\Gamma^{(i)}_{\mu \nu \kappa' \lambda'} (q, p-p') 
\,\bar{u}_{\sIndPrim} (p') \frac{i}{2 m_p}\, \sigma^{\nu \nu'} (p'-p)_{\nu'} u_{\sInd}(p) 
\, \text{,}  
\end{align}
and 
\begin{align}
\label{eq: 2.25}
W_{\gamma}^{(0)} = & -\frac{1}{M_0} \frac{1}{t} \, 2 \, a_{f_{2} \gamma \gamma}
\, \text{,}  \nonumber \\
W_{\gamma}^{(2)} = & -\frac{1}{M_0} \frac{1}{t}  \, b_{f_{2} \gamma \gamma}
\, \text{.} 
\end{align}

\boldmath
\subsection{Production of $f_2$ by odderon exchange}
\unboldmath
\label{sec: f2 production by odderon exchange}

Now we come to the diagram  of figure~\ref{fig: 1}(d) describing $\fTwo$ production via odderon exchange.
We get here, using the formulae from appendix~\ref{app B},
\begin{align}
\label{eq: 2.22B}
\mathcal{M}_{\mu, \sIndPrim, \sInd}^{\rm (d)} = &     
\frac{1}{4} \, e \,s  \, g_{\fTwo \pi \pi} \,F_1(t) \, F_M(t) 
\left[ F^{(\fTwo \pi \pi)}(k^2) \right]^2
\left[ 
\mathcal{N}_{\mu, \sIndPrim, \sInd}^{(0)} W_{\odderon}^{(0)}
- \mathcal{N}_{\mu, \sIndPrim, \sInd}^{(2)} W_{\odderon}^{(2)} 
\right] 
\, \text{.}  
\end{align}
Here again the $\mathcal{N}_{\mu, \sIndPrim, \sInd}^{(i)}$ $(i=0,2)$ are as in 
\eqref{eq: 2.20} and we have 
\begin{align}
\label{eq: 2.26}
W_{\odderon}^{(0)} = & - \frac{1}{M_0^2} \,3 \, \beta_{\odderon p p} \, \eta_\odderon \,
2\, \hat{a}_{\odderon \gamma f_2}
(-\, i\,  s\, \alphaPrimOdd )^{\alphaOdd (t) -1}  
\, \text{,} \nonumber \\
W_{\odderon}^{(2)} = & - \frac{1}{M_0^2} \,3\, \beta_{\odderon p p} \, \eta_\odderon  \,
\hat{b}_{\odderon \gamma f_2}
(-\,i \, s \, \alphaPrimOdd )^{\alphaOdd (t) -1} 
\,  \text{.} 
\end{align}

\boldmath
\subsection{Non-resonant production of $\pipi$ by pomeron and $\fTwoR$ exchange}
\unboldmath
\label{sec: non-resonant pipi production by pomeron f2R}

The diagrams of figure~\ref{fig: 1}(e) describe the non-resonant production of a $\pipi$ pair by exchange
of the pomeron $\pomeron$ and the $\fTwoR$. 
This type of process was first discussed by S\"oding~\cite{Soding:1965nh} following a 
suggestion by Drell~\cite{Drell:1960zz, Drell:1961zz}.
However, typically it is difficult to maintain gauge invariance in calculations of this Drell-S\"oding
term.

In our approach the effective vertices are derived from coupling
Lagrangians. In this formalism we include the coupling to photons by the
minimal substitution rule, i.\,e.\ derivatives are replaced by corresponding
covariant derivatives. Thereby we are guaranteed to have gauge invariance
automatically. 
For the pomeron exchange contribution to non-resonant
$\pi^+\pi^-$ production we start from the $\pomeron \pi \pi$ Lagrangian in
equation~(7.3) of~\cite{Ewerz:2013kda}.
We note that for writing down such a $\pomeron \pi\pi$ Lagrangian 
it is essential to consider the pomeron exchange as an effective tensor
exchange. 
The minimal substitution gives then the Lagrangian \eqref{eq: B.65} which fixes
the $\pomeron \gamma \pipi$ contact term in figure~\ref{fig: 1}(e). 
Using analogous arguments the $\fTwoR \gamma \pipi$ 
contact term in figure~\ref{fig: 1}(e) is fixed.
With the formulae of appendix~\ref{app B} we obtain 
\begin{align}
\label{eq: 2.27}
\mathcal{M}_{\mu, \sIndPrim, \sInd}^{\rm (e)} = & \frac{i}{2} \,\frac{1}{s} \,  e\,  F_1(t) \, F_M(t) \nonumber \\
& \times \left[ 6 \, \beta_{\pomeron N N} \, \beta_{\pomeron \pi \pi} ( -i \, s \, \alphaPrimPom )^{ \alphaPom(t) -1}  + 
\frac{1}{2 M_0^2} \, g_{\fTwoR p p} \, g_{\fTwoR \pi \pi} ( -i \, s \, \alphaPrimRplus)^{ \alphaRplus(t) -1} 
\right] \nonumber \\ 
& \times \bigg\{ 
- \frac{1}{(q-k_1)^2 - \mPi^2} \left(q-2k_1\right)_{\mu} \left(q-k_1+k_2\right)_\kappa \left(q-k_1+k_2\right)_\lambda  
\nonumber \\
&\qquad
+ \frac{1}{(q-k_2)^2 - \mPi^2} \left(q-2k_2\right)_{\mu} \left(q+k_1-k_2\right)_\kappa \left(q+k_1-k_2\right)_\lambda  
\nonumber \\
&\qquad
+2 \, g_{\mu \kappa} \left(k_2-k_1\right)_\lambda +2 \, g_{\mu \lambda} \left(k_2-k_1\right)_\kappa 
\bigg\} \nonumber \\
&\times \bar{u}_{\sIndPrim}(p') \left[ \frac{1}{2} \, \gamma^\kappa \left( p' + p\right)^\lambda + 
\frac{1}{2} \, \gamma^\lambda \left( p' + p\right)^\kappa - 
\frac{1}{4} \, g^{\kappa \lambda} \left( \slashed{p'} + \slashed{p}\right) \right] u_{\sInd}(p) 
\, \text{.}
\end{align}

Note that all three diagrams in figure~\ref{fig: 1}(e) come with the same energy dependence,  
that is with the same energy variable $s$ in the Regge factor. In Regge language this 
corresponds to saying that the propagation of the pion between its couplings to the photon 
and to the pomeron or $\fTwoR$ in the first two diagrams of figure \ref{fig: 1}(e) 
is part of the impact factor. This ensures the gauge invariance of the expression \eqref{eq: 2.27} 
and is in agreement with Regge factorisation. 
Due to its lower spin the $t$-channel propagation of the pion 
over a sizeable range in rapidity is strongly suppressed with respect to pomeron 
and $\fTwoR$ exchange. 

\boldmath
\subsection{Non-resonant production of $\pipi$ by $\rhoR$ and photon exchange}
\unboldmath
\label{sec: non-resonant pipi production by rhoR}

From the diagrams of figure~\ref{fig: 1}(f) we get the following amplitude for non-resonant 
$\pipi$ production via $\rhoR$ and photon exchange:
\begin{align}
\label{eq: 2.28}
\mathcal{M}_{\mu, \sIndPrim, \sInd}^{\rm (f)} = & -\frac{1}{2} e\,  F_1(t) \, F_M(t) 
\frac{1}{ M_-^2} \, g_{\rho_R \pi \pi}  \, g_{\rho_R p p}  ( -i \, s \, \alphaPrimRminus )^{ \alphaRminus(t) -1} 
\nonumber \\ 
& \times \left\{ 
 \frac{1}{(q-k_1)^2 - \mPi^2} \left(q-2k_1\right)_{\mu} \left(q-k_1+k_2\right)_\nu  
\right.
\nonumber \\
& \qquad
\left.
+ \frac{1}{(q-k_2)^2 - \mPi^2} \left(q-2k_2\right)_{\mu} \left(q+k_1-k_2\right)_\nu
-2 \, g_{\mu \nu} 
\right\} \nonumber \\
&\times \bar{u}_{\sIndPrim}(p')  \, \gamma^\nu  u_{\sInd}(p)  \nonumber \\
& + e^3 \, \frac{1}{t} \, \left\{ 
 \frac{1}{(q-k_1)^2 - \mPi^2} \left(q-2k_1\right)_{\mu} \left(q-k_1+k_2\right)_\nu  
\right.
\nonumber \\
&\qquad\qquad
\left.
+ \frac{1}{(q-k_2)^2 - \mPi^2} \left(q-2k_2\right)_{\mu} \left(q+k_1-k_2\right)_\nu
-2 \, g_{\mu \nu} 
\right\} \nonumber \\
& \times \bar{u}_{\sIndPrim}(p')  \, \left[ \gamma^\nu \, F_1(t) + 
\frac{i}{2 m_p} \sigma^{\nu \lambda} \left( p' - p \right)_{\lambda} \, F_2(t)   \right] \, u_{\sInd}(p)
\, \text{.}
\end{align}
As it must be, the amplitudes corresponding to $C=+1$ exchanges are odd under
the exchange of the pion momenta, $k_1 \leftrightarrow k_2$; 
see ~\eqref{eq: 2.11}-\eqref{eq: 2.17} and ~\eqref{eq: 2.27}.
The amplitudes  with $C=-1$ exchanges are even under $k_1 \leftrightarrow
k_2$; see~\eqref{eq: 2.19},~\eqref{eq: 2.20},~\eqref{eq: 2.22},~\eqref{eq: 2.23},~\eqref{eq: 2.25} 
and~\eqref{eq: 2.28}.
From these properties follow the symmetry relations for $\mathscr{R}_+$ and $\mathscr{R}_-$ in
\eqref{eq: 2.8} and~\eqref{eq: 2.9}, respectively.

In agreement with Regge factorisation the energy dependence of all three diagrams 
of figure \ref{fig: 1}(f) is the same. This is analogous to the case of pomeron and $\fTwoR$ 
exchange, see section \ref{sec: non-resonant pipi production by pomeron f2R}. 

This concludes our presentation of the amplitudes corresponding to the diagrams of 
figure~\ref{fig: 1}(a-f). 
For further analysis an event generator is built, based on the following
techniques. 
For the evaluation of the matrix elements the FeynCalc package~\cite{Mertig:1990an} (version 8.2.0) 
is used and the results are cross-checked with a direct implementation based
on the template-library ltensor~\cite{ltensor}. Based on that, a weight 
and the kinematic variables of an event are determined
using a standard Monte Carlo (MC) method with pre-sampling, see appendix~\ref{app D}.
The output of the event generator is stored in a standard format of the
ROOT-package~\cite{Brun:1997pa} and used to derive (partially integrated) differential
cross sections as a function of various variables.

\section{Results}
\label{sec: Results}

In this section we present selected results for cross sections, angular 
distributions and asymmetries. For the numerical evaluation of the 
formulae presented above we use a set of default parameters for 
the model \cite{Ewerz:2013kda}. These default parameters are 
given in table~\ref{tab: B.1} of appendix~\ref{app B}. Many of the 
parameters are well constrained while we can only guess 
the values of some others. In the present study we do not make 
any attempt to fit the parameters to data. That would be desirable 
in general but is not the scope of the present paper. Instead, we concentrate 
here on studying the qualitative features of the reaction \eqref{eq: 1.1} 
in the model. We find that some asymmetries seem particularly 
promising for detailed experimental studies as they allow to 
enhance interesting effects by choosing suitable cuts on the data. 
In particular, we point out promising ways to look for effects of 
the elusive odderon. 

\subsection{Cross sections}

\begin{figure}[thb]
\center
    \includegraphics[scale=0.72]{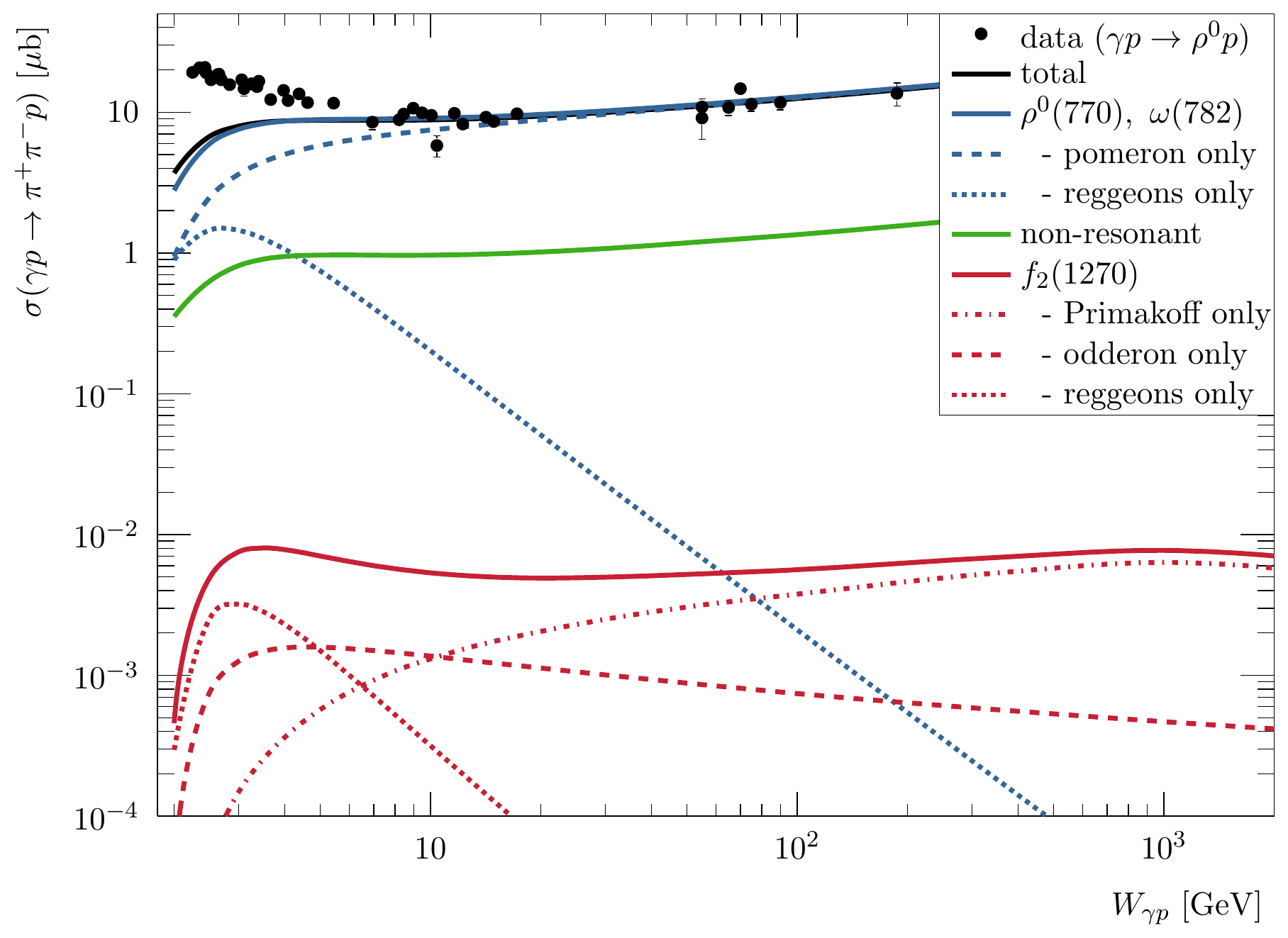}		
\caption{The total cross section $\sigma \left( \gamma p \rightarrow \pipi p
  \right)$ as a function of the center-of-mass energy $\Wgp$. The cross
  section is integrated over $2\mPi \leq \mpipi \leq 1.5 \gev$ and $-1 \gevsq \leq
  t\leq 0$. The full model and individual contributions from vector
  meson production, non-resonant processes, and $f_2$ production are
  shown. The reggeon contributions comprise $\fTwoR$ and $\aTwoR$ in case of vector
  meson, and $\rhoR$ and $\omegaR$ in case of $f_2$ production.
High energy data for $\sigma \left( \gamma p \rightarrow \rho^0 p
  \right)$ from H1 \cite{Aid:1996bs} and ZEUS \cite{Breitweg:1997ed,
    Derrick:1995vq} at HERA as well as fixed target data, referenced in \cite{Breitweg:1997ed}, are shown for illustration.\label{fig: 3}}
\end{figure}

We start with the total cross section for  the reaction~\eqref{eq: 1.1} 
which we integrate here from the dipion mass threshold to $1.5 \gev$ 
(as is done for example in the experimental study \cite{Breitweg:1997ed}): 
\begin{align}
\label{eq: 3.1}
\sigma \left( \gamma p \rightarrow \pipi p \right) \quad
\text{for } \, 2\mPi < \mpipi < 1.5 \gev \, \quad
\text{and } \, -1 \gevsq \leq t\leq 0 \,\text{.}
\end{align}
In figure~\ref{fig: 3} we show the cross section as function of $\Wgp$. 
Also the cross sections for
various subclasses of diagrams according to figure~\ref{fig: 1} are shown.
The dominant contribution to the total cross section comes from $\rho^0$
production. Total cross section data for $\gamma p \to \rho^0 p$ 
are shown for illustration. They are taken from figure~10 
of~\cite{Breitweg:1997ed} and are
defined in the same mass range as given in~\eqref{eq: 3.1}.
For $\Wgp \gtrsim 5 \gev$ our model 
describes the essential features of the data. 
At very low $\Wgp$-values an agreement with data is not expected
as further processes, not included in our high-energy model, contribute to the cross section.
We emphasise again
that our model curves are just examples and not fits.
Note that our model contains all contributions to the reaction~\eqref{eq: 1.1}
given by the diagrams of figure~\ref{fig: 1}, not just the $\rho$
contribution, which nonetheless dominates the total cross section.

\begin{figure}[thb]
\center
 \includegraphics[scale=0.72]{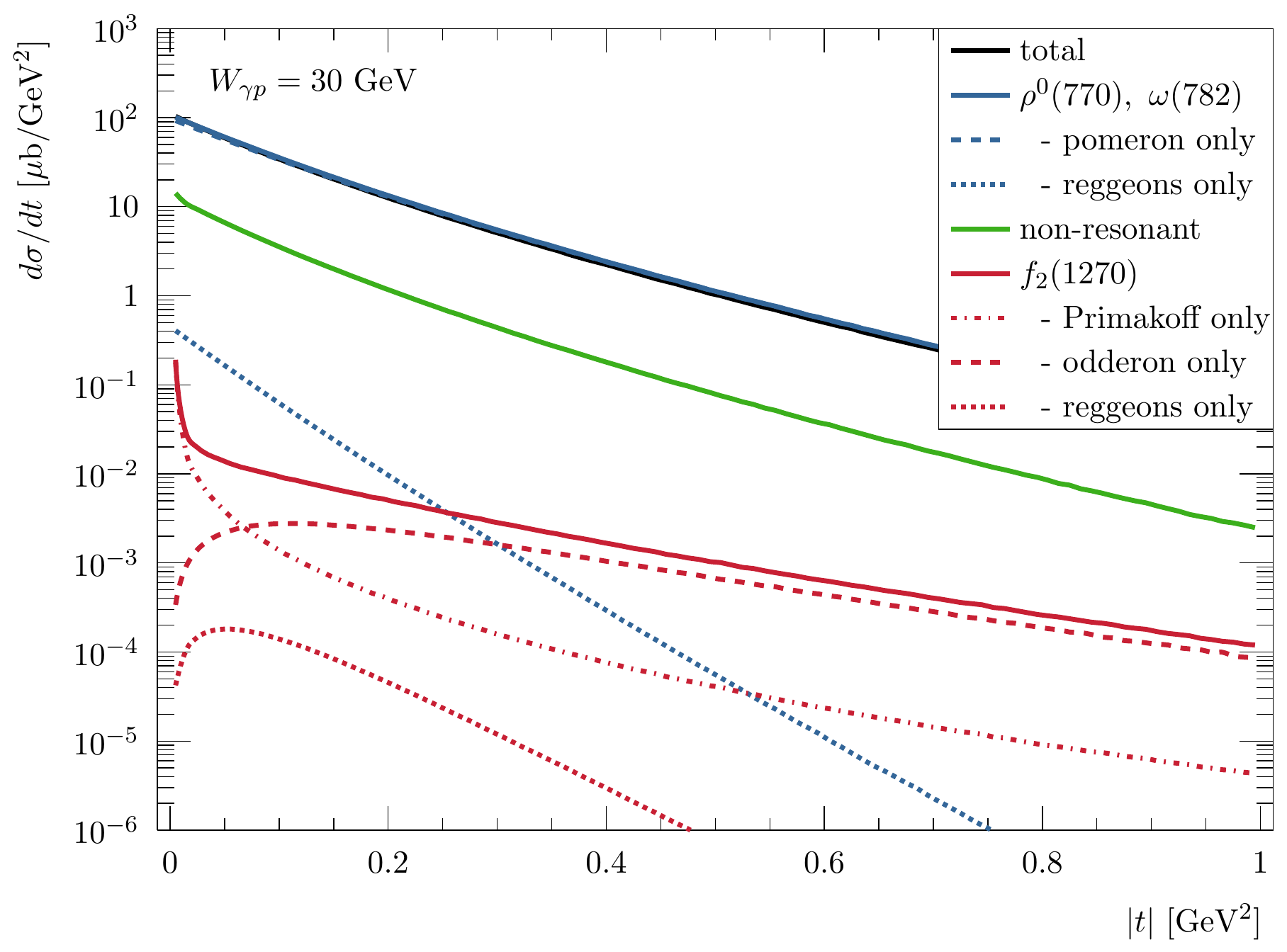}	
\caption{The differential cross section 
$ d\sigma / dt \left( \gamma p \rightarrow \pipi p \right)$ as function
of $|t|$. The cross section is integrated over the range $2\mPi \leq \mpipi
\leq 1.5 \gev$ and given for fixed $\Wgp = 30 \gev$. 
In addition to the full model results also contributions from the main
diagrams are
shown, see figure \ref{fig: 3} for explanations.
\label{fig: 4} }
\end{figure}

As a second example we show in figure~\ref{fig: 4} the differential cross section \linebreak
$d\sigma / d t \left( \gamma p \rightarrow \pipi p \right)$ as function of $|t|$. 
We integrate over $\mpipi$ as indicated in~\eqref{eq: 3.1} and choose
$\Wgp =  30 \gev$.
Again, in addition to the full differential cross section also the various contributions from
individual subprocesses are shown. 
The production of $\rho$ mesons is dominant for all values of
$t$, followed by non-resonant $\pipi$ production and the 
-- by more than three orders of magnitude -- 
smaller $\fTwo$ production. 
The cross section of all contributions falls approximately exponentially as function of
$|t|$ for $|t| \gtrsim 0.1 \gev^2$.
We note that the Primakoff contribution to $\fTwo$ production,
figure~\ref{fig: 1}(c), 
has a singularity for $|t| \rightarrow 0$. This singularity is, of course, never 
reached since for reaction \eqref{eq: 1.1} there is a minimal value for $|t|$, 
$|t_\mathrm{min}| > 0$ (see \eqref{eq:A.12a}). The singularity is driven by the photon
propagator and behaves as $1/|t|$. 
In contrast, the reggeon contributions from $\rho_R$, $\omega_R$
and odderon exchange, figures~\ref{fig: 1}(b) and~\ref{fig: 1}(d), show a dip for 
$|t| \rightarrow 0$.
An explanation for this different behaviour is given in appendix~\ref{app C}.

Also note that with the chosen default parameters odderon exchange
is the dominant contribution to $\fTwo$ production for $|t| \gtrsim 0.1 \gev^2$ which
can be exploited in an experimental search to enhance a possible odderon
signal. 
Reggeon contributions to $f_2$ production are negligible at $\Wgp=30 \gev$ or higher values.

\begin{figure}[thb]
\center
 \includegraphics[scale=0.72]{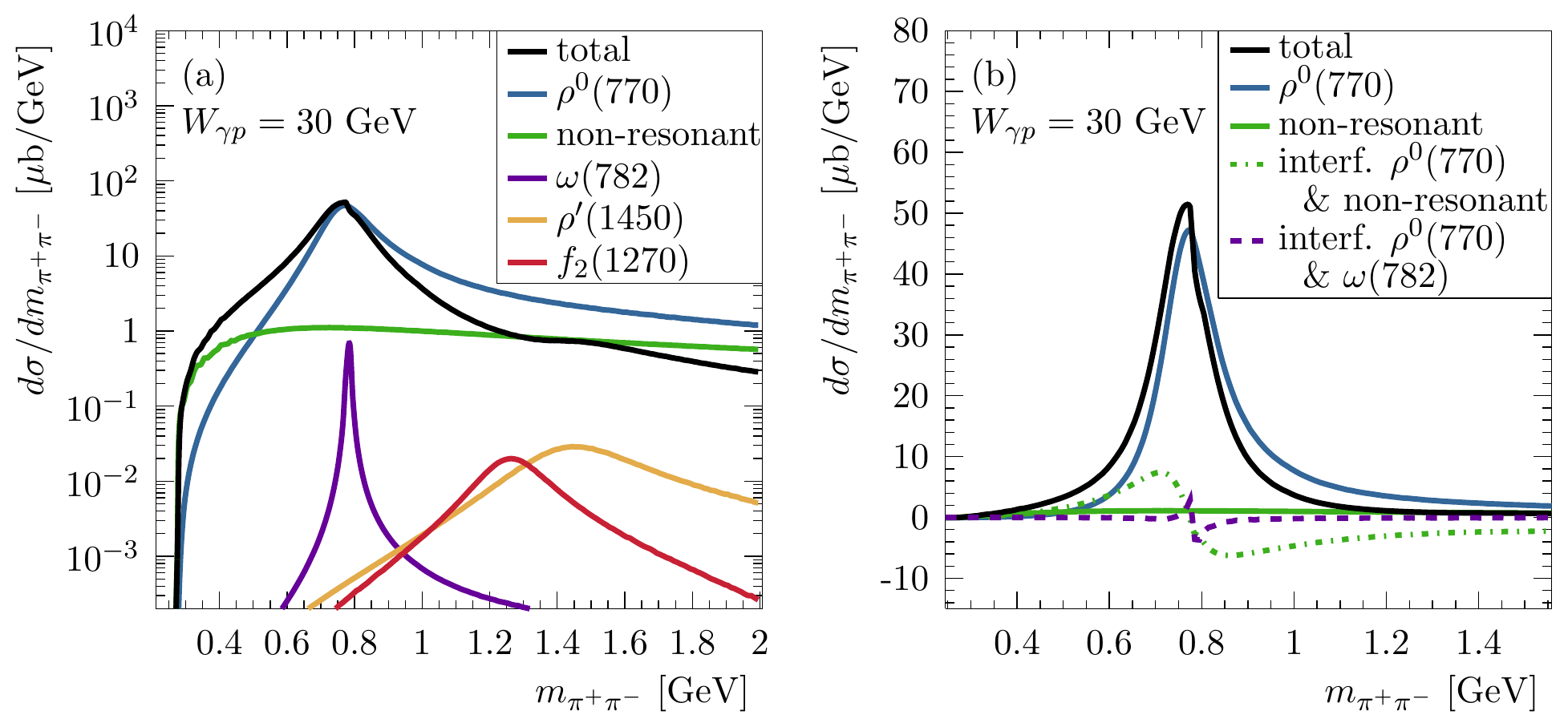}	
\caption{Differential cross sections 
$ d\sigma / d \mpipi \left( \gamma p \rightarrow \pipi p \right)$ as
  function of $\mpipi$ for fixed $\Wgp = 30 \gev$ and integrated over the
  range $-1 \gevsq \leq t \leq 0$.
(a) The full model, non-resonant contributions and the contributions from the resonances $\rho^0(770)$, $\omega(782)$, $f_2(1270)$ and $\rho'(1450)$ are shown.
(b) Dominant  contributions in the
  $\rho$ mass region 
including the leading interferences of $\rho^0(770)$ with the non-resonant
$\pipi$ production and the
$\omega(782)$ meson are shown. 
\label{fig: 5}}
\end{figure}

In figure~\ref{fig: 5}  we show the differential cross section 
$d\sigma / d \mpipi  \left( \gamma p \rightarrow \pipi p \right)$ as function of
the $\pipi$ invariant mass $\mpipi$ for the range from threshold to 
$m_{\pi^+\pi^-} = 2 \,\mbox{GeV}$, and enlarged for the $\rho$ mass region.
The differential cross sections are shown for
$\Wgp = 30 \gev$ integrated in the range $-1 \gevsq \leq t \leq 0$.
The contributions of various subclasses of diagrams are also shown as well as
the dominant contributions from interferences in the $\rho$ mass region.
The resonance structure of the $\rho$ peak 
is clearly visible. 
Note that the shape of the $\rho(770)$ peak from the diagrams
of figure~\ref{fig: 1}(a) alone is rather symmetric. The skewing of the $\rho$ shape
caused by the interference of the diagrams of figures~\ref{fig: 1}(a) and~\ref{fig: 1}(e),(f)
(non-resonant contributions)
is clearly exposed. We see here the Drell-S\"oding mechanism
\cite{Soding:1965nh,Drell:1960zz,Drell:1961zz} at work.
Note that the skewing depends crucially on the choice of the
$\rho$ form factor parameterisation, which was here 
implemented according to equation~\eqref{eq: B.83}.
Furthermore, the effect of the $\rho$--$\omega$ interference, 
the steep falloff at the top of the $\rho^0$ peak, is clearly visible.

In the mass region of the $\rho'$ meson, $1.2  \gev\lesssim
\mpipi \lesssim 1.6 \gev$, a distortion of the skewed
$\rho$ line-shape due to the interference of the $\rho'$ diagram 
in figure~\ref{fig: 1}(a), and to a lesser extent the $\fTwo$ diagrams
in figures~\ref{fig: 1}(b, c, d), with the $\rho$ meson and the 
non-resonant contribution is visible. However, 
clear resonance peaks in the $\mpipi$ distribution due to $\fTwo$ and $\rho'$
do not show up for the chosen default parameters.

\subsection{Angular distributions}

In the following we study the pion angular distribution 
in the $\pipi$ rest system. 
For illustration we choose as reference system here the proton-Jackson system, 
see figure~\ref{fig: AppendixA.1} in appendix~\ref{app A},
in which the polar angle $\theta_{k_1,p}$ of the $\pi^+$ is measured
with respect to the incoming proton direction. This and other reference systems 
are discussed in detail in appendix~\ref{app A}. 
As the decay-angle distribution is mass dependent and mainly driven by the
spin of the resonance in case of decays we study the angular distribution in
the $\rho$ and the $\fTwo$ mass regions separately.

\begin{figure}[thb]
\center
 \includegraphics[scale=0.72]{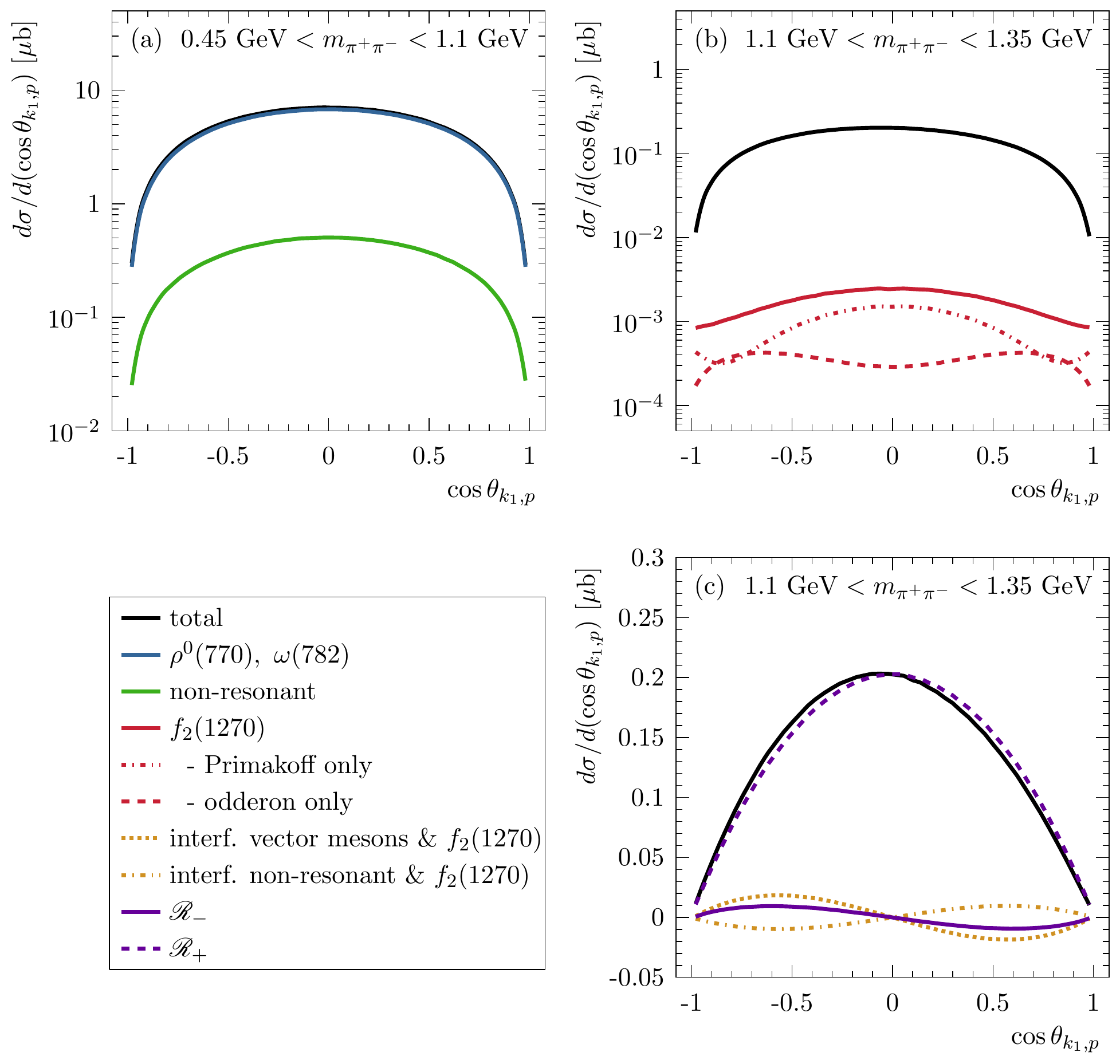}	
\hspace{0.1cm}
\caption{Differential cross section $d\sigma / d \cos \theta_{k_1,p} \left( \gamma p \rightarrow \pipi p \right)$
as function of the cosine of the polar angle $\theta_{k_1,p}$ in the proton-Jackson system for fixed $\Wgp = 30 \gev$ and integrated over the range $-1 \gevsq \leq t \leq 0$.
The full model and the dominant contributions  for the mass regions
$0.45 \gev \leq \mpipi \leq 1.1 \gev$  ($\rho$ mass region) and $1.1 \gev \leq \mpipi
\leq 1.35 \gev$ ($\fTwo$ mass region)
are shown in (a) and  (b),(c), respectively.
(c) shows in addition to the full
differential cross section the dominant interference terms on a linear scale.
All contributions are explained in the legend; for $\mathscr{R}_+$ and 
$\mathscr{R}_-$ see \eqref{eq: 2.8} and \eqref{eq: 2.9}. 
\label{fig: 6} }
\end{figure}

In figure~\ref{fig: 6}(a) the differential distribution 
$d\sigma / d \cos \theta_{k_1,p} \left( \gamma p \rightarrow \pipi p \right)$
in the $\rho$ mass region, $0.45 \gev \leq \mpipi \leq 1.1 \gev$, is shown 
as function of $\cos{\theta_{k_1,p}}$.
The distribution shows a typical $\propto \sin^2{\theta_{k_1,p}}$ behaviour as is 
expected for photo-produced vector mesons if (approximate) $s$-channel 
helicity conservation \cite{Gilman:1970vi} holds.\footnote{Strictly speaking, $s$-channel helicity 
conservation should be discussed in the helicity system (see appendix \ref{app A}) 
which for small $t$ is very close to the proton-Jackson system.} 
(For the general formalism of helicity amplitudes 
in this context see \cite{Schilling:1969um,Schilling:1973ag}.) 

In figure~\ref{fig: 6}(b) the same distribution is shown in the $\fTwo$ mass
region $1.1 \gev \leq \mpipi \leq 1.35 \gev$. 
In addition to the dominant $\rho$ contribution, which exhibits again the 
typical $\propto \sin^2{\theta_{k_1,p}}$ behaviour, the small $\fTwo$ contributions show more 
features in the angular distribution as it is expected for a $J=2$ resonance.
The interference of the $C$-even and $C$-odd exchange contributions leads to a small asymmetry of
the $\cos \theta_{k_1,p}$ distribution as shown in figure~\ref{fig: 6}(c). 
This asymmetry comes dominantly from the interference between the $\rho$ and
$\fTwo$ production diagrams. The asymmetry is partially cancelled by the interference
of the $\fTwo$ resonance with non-resonant $\pipi$ production but a net asymmetry
remains if all diagrams are included.
The resulting charge asymmetries are discussed in more detail in the following.

\subsection{Charge asymmetries}

Let us now turn  to asymmetries in the $\pipi$ rest system. 
As already noted in section~\ref{sec: Matrix elements, cross sections, asymmetries} 
the interference of diagrams with exchange of $C = +1$ and $C = -1$ objects is signalled
by an asymmetry under
\begin{align}
\label{eq: 3.9B}
\mathbf{k}_1 \rightarrow - \mathbf{k}_1 \, ;
\end{align}
see the discussion following~\eqref{eq: 2.7} and \eqref{eq: 2.28}. 

We note first that P-invariance tells us that 
the distributions of the $\pi^+$ momentum must be symmetric
under a reflection on the reaction plane, which in the
$\pipi$ system is given by
the plane spanned by the momenta of the incoming proton and 
the outgoing proton. 
We turn, therefore, to charge asymmetries which are defined 
with respect to specific directions (axes) in the reaction plane.
In the literature many different definitions of reference systems can be found, which are used
to study asymmetries. A summary of the different definitions is given 
in appendix~\ref{app A}.

We start our discussion of charge asymmetries in the proton-Jackson system,
for which the $\theta_{k_1,p}$ distributions were shown in the previous section. 
It is convenient to define a $\theta_{k_1,p}$ dependent charge asymmetry:
\begin{align}
\label{eq: asymm1}
\widehat{\cal A}(\cos \theta_{k_1,p}) = \frac{\frac{d \sigma}{d\cos \theta_{k_1,p}}\,(\cos \theta_{k_1,p})
  - \frac{d \sigma}{d\cos \theta_{k_1,p}}\,(-\cos \theta_{k_1,p})}{\frac{d \sigma}{d\cos
    \theta_{k_1,p}}\,(\cos \theta_{k_1,p}) + \frac{d \sigma}{d\cos \theta_{k_1,p}}\,(-\cos
  \theta_{k_1,p})}
\qquad \textrm{for } \cos \theta_{k_1,p}>0
\,.
\end{align}
We also define a total charge asymmetry: 
\begin{align}
\label{eq: asymm2}
{\cal A}_\textrm{tot} = \frac{\sigma_+ - \sigma_-}{\sigma_+ + \sigma_-}
\,,
\end{align}
using the definitions 
\begin{align}
\label{eq: asymm_def}
\sigma_\pm = \int_0^1 \frac{d \sigma}{d\cos \theta_{k_1,p}} \,(\pm \cos \theta_{k_1,p}) \; d\cos \theta_{k_1,p}
\,.
\end{align}

\begin{figure}[thb]
\center
\setlength{\unitlength}{1.0cm}
 \includegraphics[scale=0.72]{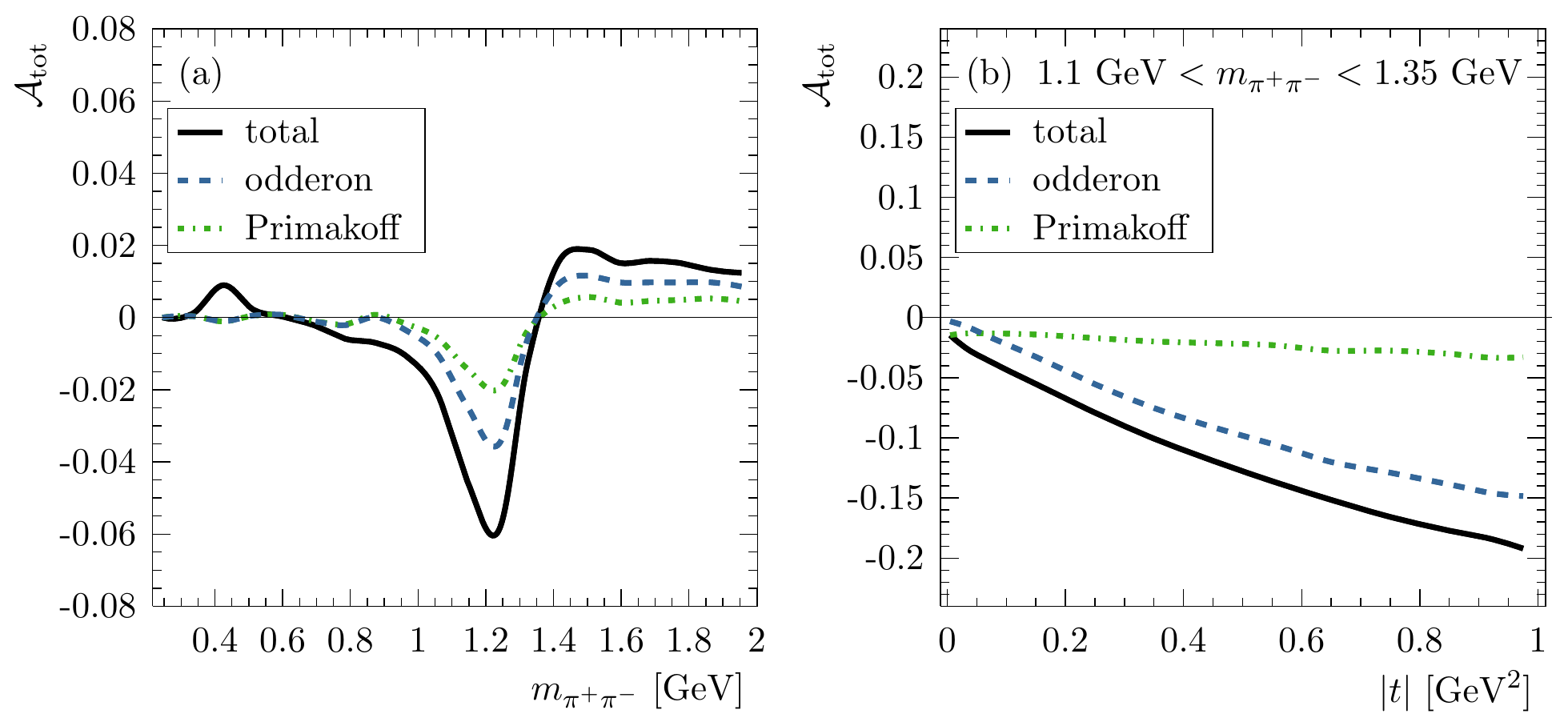}	
\caption{Total charge asymmetry ${\cal A}_\textrm{tot}$ \eqref{eq: asymm2}
in the proton-Jackson system as function of the
  (a) invariant mass of the $\pipi$ system
and (b) squared momentum transfer $t$.
The asymmetries are presented for fixed $\Wgp = 30 \gev$ and (a) integrated over the range 
$-1 \gevsq \leq t \leq 0$ and (b) integrated over the range $1.1 \gev < m_{\pipi}<1.35 \gev$.
The individual contributions to the asymmetries 
from photon (Primakoff) and odderon exchange are shown by the green
  dashed-dotted and blue dashed 
lines, respectively.
\label{fig: 7} }
\end{figure}

In figure~\ref{fig: 7}(a) the total charge asymmetry defined in the proton-Jackson system
is shown as function of the invariant mass of the $\pipi$ system.
A negative asymmetry of a few percent is visible in the $\fTwo$ resonance region
for the chosen parameter values. This asymmetry is
mainly due to the interference of the $\fTwo$ resonance with the high mass tail
of the $\rho$ resonance. For higher masses, $m_{\pipi}>1.35 \gev$, an asymmetry with
opposite sign is visible in the model, which is mainly due to the interference of the
$\fTwo$ resonance with the non-resonant $\pipi$ production with  $C=+1$ pomeron
exchange.
Another but smaller positive asymmetry is visible just above the $\pipi$
production threshold
which is mainly due to  the interference of $C=+1$ and $C=-1$
exchange diagrams of non-resonant $\pipi$ production.

We are particularly interested in the asymmetry contribution from the odderon.
In figure~\ref{fig: 7}(b) the total charge asymmetry is shown as function of 
$t$ in the $\fTwo$ mass region $1.1 \gev \leq \mpipi \leq 1.35 \gev$.
A small negative asymmetry, almost constant in $t$, is generated by the 
Primakoff contribution. In contrast, a strong increase of the absolute value of the 
negative charge asymmetry approximately linear in
$|t|$ is predicted by odderon exchange with the chosen model parameters.
As most events are located at low $|t|$, 
the experimental sensitivity to the odderon exchange diagram can be 
significantly enhanced by requiring a minimum $|t|$-cut, or better by measuring
the approximately linear increase of the absolute value of the 
charge asymmetry as function of $|t|$.

\begin{figure}[thb]
\center
 \includegraphics[scale=0.72]{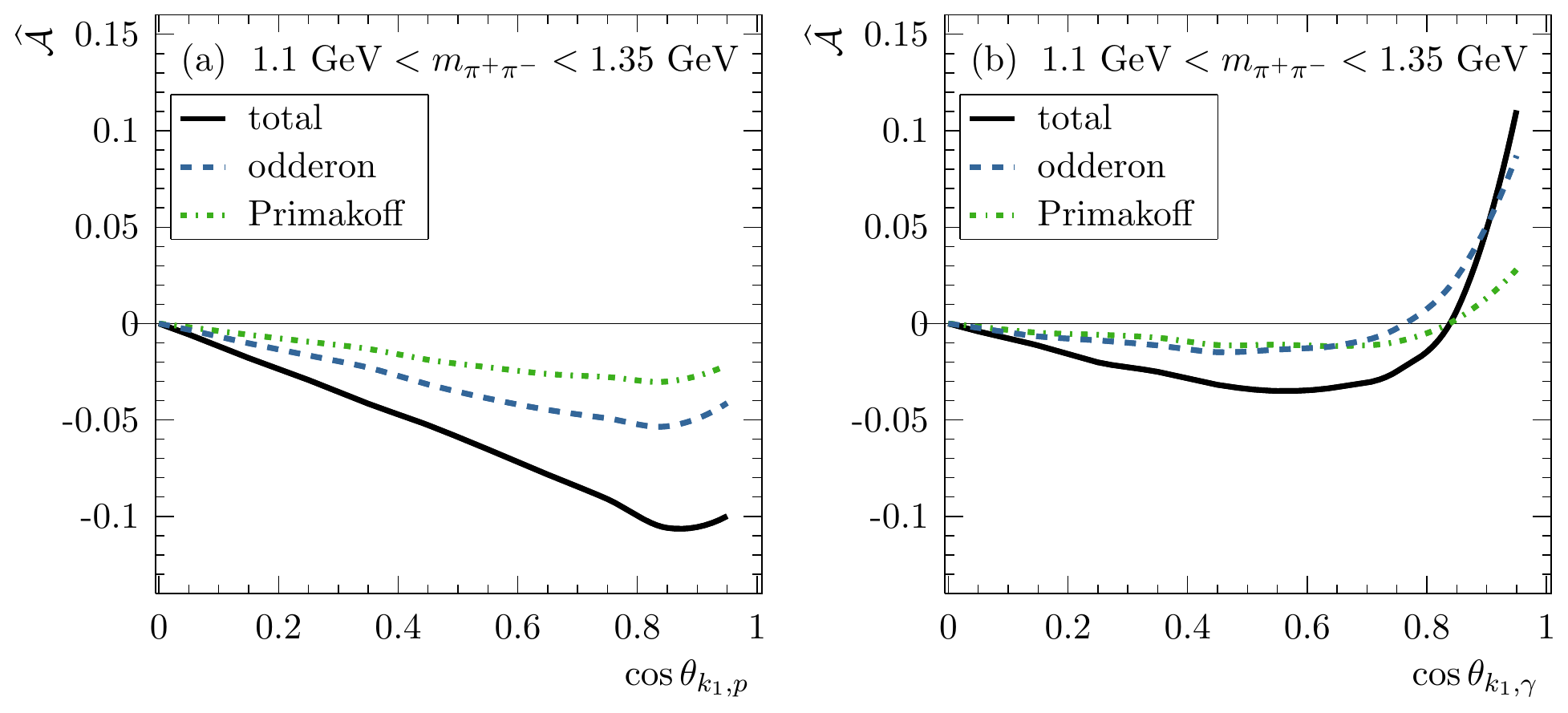}	
\caption{Charge asymmetry $\widehat{\cal A}$ \eqref{eq: asymm1} 
as function of the cosine of the polar angle of $\pi^+$ in the 
(a) proton-Jackson and (b) photon-Jackson system (see appendix~\ref{app A}).
The asymmetries are given for fixed $\Wgp = 30 \gev$ and integrated over the
range $-1 \gevsq \leq t \leq 0 $ and $1.1 \gev < m_{\pipi}<1.35 \gev$.
The individual contributions to the asymmetries 
from photon (Primakoff) and odderon exchange are shown by the green
  dashed-dotted and blue dashed lines, respectively.
\label{fig: 8} }
\end{figure}

Finally we compare the charge asymmetry in different systems. 
Figure~\ref{fig: 8}(a) shows the charge asymmetry as function of $\cos \theta_{k_1,p}$ in
the proton-Jackson system. The contributions from odderon and photon
exchange diagrams are also shown. A strong increase of the absolute value 
of the asymmetry towards the forward/backward direction occurs.

Figure~\ref{fig: 8}(b) shows the charge asymmetry as function of $\cos \theta_{k_1,\gamma}$ in
the photon-Jackson system, in which the polar angle $\theta_{k_1,\gamma}$ is defined
with respect to the incoming photon; see appendix~\ref{app A}.
For the photon-Jackson system it is interesting to note that (1) in the central
region, $\cos \theta_{k_1,\gamma} \lesssim 0.6$, the asymmetry contributions due to
Primakoff and odderon exchange
are of similar size and (2) that the asymmetry changes sign in the polar
region, $\cos \theta_{k_1,\gamma} \gtrsim 0.6$. 
At large $\cos \theta_{k_1,\gamma}$ the large positive asymmetry 
is dominated by the contribution from odderon exchange.
Therefore, the photon-Jackson system offers the
opportunity to study interference effects due to Primakoff and odderon
exchange diagrams separately by analysing different $\cos \theta_{k_1,\gamma}$ regions.
Note that due to the sign change the total asymmetry is strongly reduced 
(after integration over $\cos \theta_{k_1,\gamma}$)  in the photon-Jackson system 
and that a possible odderon signal can be  experimentally overlooked if only total 
asymmetries are studied.
In the same context it should be remarked that limited detector acceptances
in experimental searches might affect the measurement of angular distributions
and thus the sensitivity to charge asymmetries. 

The full asymmetry information of the $\pipi$ system can be exploited
by investigating the asymmetry for all solid angles.
For this study we define the angle $\alpha$, which describes in the reaction plane
the azimuthal angle of the $\pi^+$ with respect to the incoming proton beam
direction, 
and the elevation angle $\beta$; see figure \ref{fig:1} and \eqref{eq: alpha}
in appendix~\ref{app A}. 

In figure~\ref{fig: 9} the charge asymmetry 
\begin{align}
\label{eq:3.6}
{\cal A} (\alpha, \beta) = 
\frac{
\frac{d\sigma}{d\Omega} (\alpha,\beta) - \frac{d\sigma}{d\Omega} (\alpha + \pi,-\beta)}{\frac{d\sigma}{d\Omega} 
(\alpha,\beta) + \frac{d\sigma}{d\Omega} (\alpha + \pi,-\beta)}
\end{align}
of the $\pipi$ final state  is shown
as function of $\alpha$ and $\beta$
in the $\fTwo$ mass region. By construction, see \eqref{eq:3.6}, the relation 
$\cal{A}(\alpha,\beta)=-\cal{A}(\alpha+\pi,-\beta)$ holds.
From P-invariance follows $\cal{A}(\alpha,\beta)=\cal{A}(\alpha,-\beta)$. 
The asymmetry distribution exhibits two dipoles: 
one at $\alpha \approx 0~ (-\pi)$ related to the incoming proton direction and
a second one at $\alpha \approx -\pi/2~(+\pi/2)$ which is broader in both 
$\alpha$ and $\beta$. 
For illustration, lines of constant polar angle $\theta_{k_1,p}$ are shown,
along which the asymmetry was integrated to calculate the  
differential  asymmetries for
figure~\ref{fig: 8}(a).
The presence of complex structures in the asymmetry distribution, 
which are generally integrated out in one-dimensional projections,  
suggests to exploit the full 
two-dimensional information for asymmetry measurements in order to
increase sensitivity and to separate the different asymmetry sources.
\begin{figure}[thb]
\center
\setlength{\unitlength}{1.0cm}
 \includegraphics[scale=0.72]{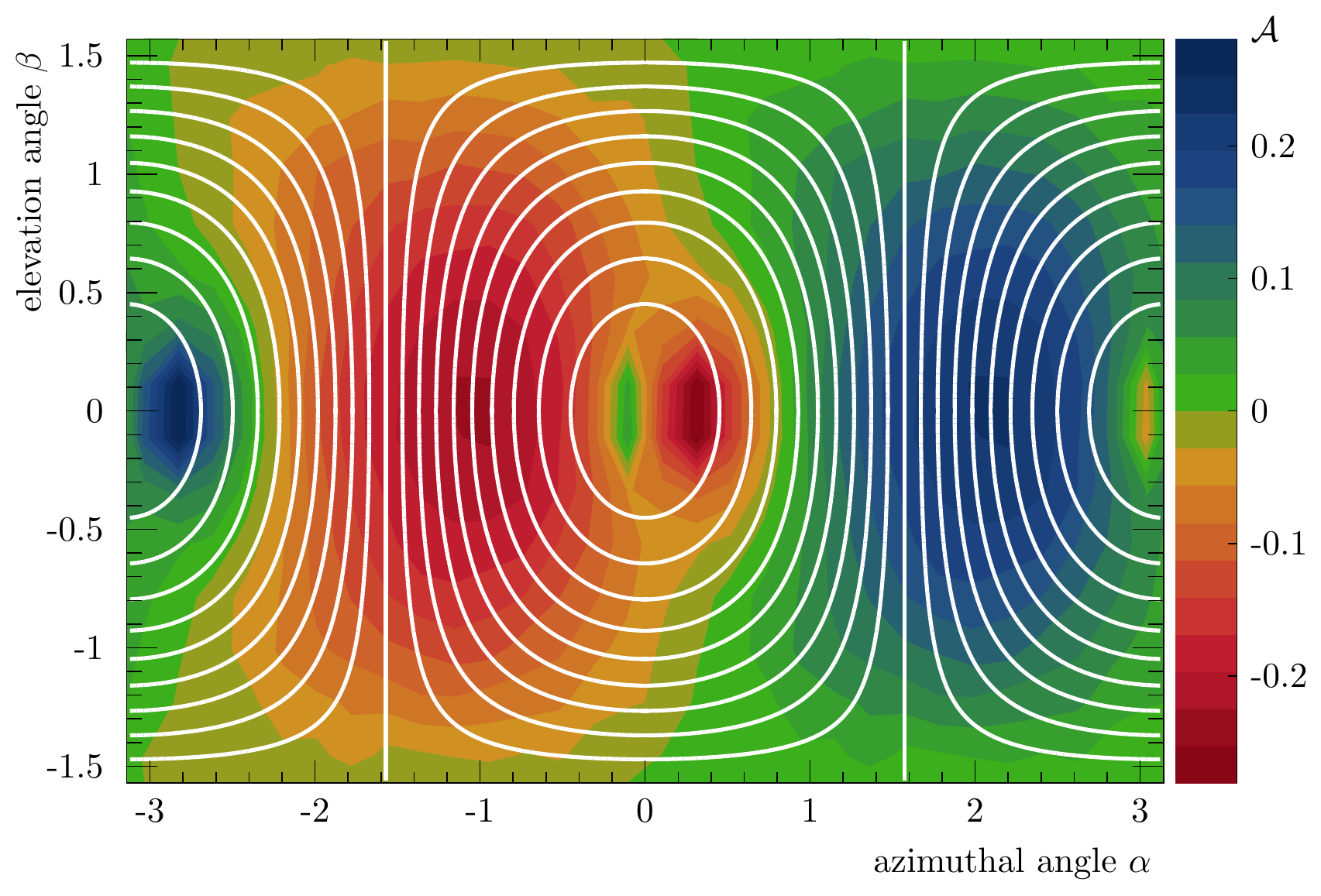}	
\caption{Charge asymmetry ${\cal A}$ \eqref{eq:3.6} as function of the azimuthal angle
$\alpha$ and the elevation angle $\beta$ defined in the proton-Jackson system.
The asymmetries are given for fixed $\Wgp = 30 \gev$ and integrated over the
range $-1 \gevsq \leq t \leq 0$ and $1.1 \gev < m_{\pipi}<1.35 \gev$.
$\alpha=0$ corresponds to the incoming proton direction.
Lines of constant polar angle $\theta_{k_1,p}$ are shown in white.
\label{fig: 9} }
\end{figure}

\section{Conclusions}
\label{sec: Conclusions}

In this article we have presented a study of exclusive photoproduction of 
$\pipi$ pairs on protons, $\gamma p \rightarrow \pipi p$, 
in the framework of a comprehensive model for soft high-energy reactions. 
We have considered $\pipi$ production via the $\rho$, $\omega$, $\rho'$ and $f_2$ 
resonances as well as production of non-resonant $\pipi$ pairs.
Taking into account photon, pomeron, odderon and reggeon exchanges we 
have obtained analytic expressions for all contributing diagrams. 
We have calculated the total and various differential cross sections, angular distributions 
and asymmetries for a set of default parameters of the model. 
We emphasise that in the present paper no attempt has been made 
to fit data. The purpose of our paper is to provide all necessary 
theoretical tools for such a comparison with data.

Our methods can easily be extended to exclusive electroproduction 
\begin{align}
\label{eq: 4.1}
e + p \longrightarrow e + \pi^+ + \pi^- + p
\end{align}
at low to moderate values of $Q^2$.
Another extension is to central production of $\pipi$ pairs in peripheral $p p$ collisions
\begin{align}
\label{eq: 4.2}
p + p \longrightarrow p + \pi^+ + \pi^- + p \, \text{.}
\end{align}
This reaction will be discussed in a forthcoming paper \cite{Leb}. Central production of scalar
and pseudoscalar mesons with techniques similar to the ones presented here 
has been discussed in~\cite{Lebiedowicz:2013ika}.

In summary, we have presented a study of the reaction $\gamma p \rightarrow \pipi p$ 
in an explicit model for soft high-energy scattering that includes the exchanges 
of pomeron, odderon, photon, and reggeons. In particular, the model
incorporates a gauge-invariant version of the Drell-S\"oding mechanism 
which is responsible for the skewing of the $\rho$ meson shape.
We have paid particular attention to the effects of the elusive odderon 
in the photoproduction of pion pairs. The odderon is expected to contribute 
to $f_2$ meson photoproduction as first suggested in \cite{Schafer:1992pq,Barakhovsky:1991ra}, 
and to asymmetries in the angular distribution of the $\pi^+\pi^-$ pairs 
as suggested in 
\cite{Ivanov:2001zc,Hagler:2002nh,Hagler:2002sg,Hagler:2002nf,Ginzburg:2002zd,Ginzburg:2002fy,Ginzburg:2005ay}. 
Our results indicate that the corresponding observables appear indeed very 
promising for an odderon search. More generally, we hope that our results 
can be used as guidance for the experimental study of interesting effects 
in the photoproduction of pion pairs. 

\acknowledgments

The authors would like to thank P.~Lebiedowicz, M.~Guzzi, C.~Royon, 
R.~Schicker, and A.~Szczurek for useful discussions. 
The work of C.\,E.\ was supported by the Alliance Program of the
Helmholtz Association (HA216/EMMI). 

\appendix

\section{Kinematics}
\label{app A}

Here we discuss kinematic relations for the reaction 
\begin{align}
\label{eq: A.1}
\gamma^{(*)}(q,\epsilon) + p(p,\sInd) \longrightarrow \pi^+(k_1) + \pi^-(k_2) + p' (p', \sIndPrim) 
\end{align}
where we consider a real or virtual photon, $\gamma$ or $\gamma^*$, 
with polarisation vector $\epsilon$. In the final state $p'$ stands for a proton or a 
diffractively excited proton, for instance the resonance $N(1520)$. The spin indices of $p$
and $p'$ are denoted by $\sInd$ and $\sIndPrim$, respectively.
We set
\begin{align}
\label{eq: A.2}
k &= k_1 + k_2 \, \text{,} \nonumber \\
s \equiv  \Wgp^2 &= (p+q)^2 = (p'+ k) ^2  \, \text{,} \nonumber \\
t &= (p' - p) ^ 2 = (q-k)^ 2\, \text{,} \nonumber \\
\mpipi^2 &=  k^2 \, \text{,} 
\end{align}
and we have, in general,
\begin{align}
\label{eq: A.3}
q^2 \leq 0 \, \text{,}  \qquad
p'^2 \geq m_p^2 \, \text{.} 
\end{align}
For the elastic  photoproduction reaction~\eqref{eq: 1.1} we have, of course,
\begin{align}
\label{eq: A.4}
q^2 =0 \, \text{,}  \qquad
p'^2 = m_p^2 \, \text{.} 
\end{align}
We denote the space-part of the four-vector $p$ by $\mathbf{p}$, etc. 

From the five momentum vectors of the particles 
in the reaction~\eqref{eq: A.1} we can form 15 scalar products and one
parity-odd (P-odd) invariant $I_P$. As the P-odd invariant we can choose
\begin{align}
\label{eq: A.9}
I_P = \epsilon_{\mu \nu \rho \sigma} \, p^{\mu}\,  q^{\nu} 
\left( k_1 - k_2 \right) ^{\rho} 
\left( k_1 + k_2 \right) ^{\sigma} 
\end{align}
where we use the convention $ \epsilon_{0 1 2 3} = +1$ 
for the totally antisymmetric 
symbol $ \epsilon_{\mu \nu \rho \sigma}$.
This parity-odd variable cannot enter the cross section calculation 
for unpolarised particles since we consider
a process where only the P-conserving strong and electromagnetic 
interactions contribute.

Not all 15 scalar products are independent.
We have energy-momentum conservation
\begin{align}
\label{eq: A.6bb}
q+p = & \, k_1 + k_2 + p'
\end{align}
and the mass-shell conditions
\begin{align}
\label{eq: A.6cc}
p^2 = m_p^2 \, \text{,} \qquad k_1^2 = k_2^2 = m_{\pi}^2 \, \text{.}
\end{align}
We take as independent variables 
\begin{align}
\label{eq: A.7}
p'^2 \, \text{,} \quad q^2 \, \text{,} \quad  s \, \text{,} \quad  t\, \text{,}  \quad
k^2 = \mpipi^2 \, \text{,} \quad  p\cdot \left( k_1 - k_2\right)\, \text{,} \quad q\cdot \left( k_1 - k_2\right) \, \text{.} 
\end{align}
All scalar products can be expressed in terms of the variables in \eqref{eq: A.7}. We get
\begin{align}
\label{eq: A.8}
p\cdot p' &= \frac{1}{2} \,(p'^2 +  m_p^2 -  t ) \,\text{,} \nonumber \\
p \cdot q &= \frac{1}{2}\, ( s - m_p^2 - q^2 ) \, \text{,} \nonumber\\
p \cdot k_1 &=  \frac{1}{4}\, (s-p'^2 -q^2 + t) + \frac{1}{2} \,p \cdot \left(k_1 - k_2\right) \, \text{,} \nonumber\\
p \cdot k_2 &= \frac{1}{4}\, (s - p'^2 -q^2 +t) - \frac{1}{2} \,p\cdot \left(k_1 - k_2\right) \, \text{,} \nonumber\\
p' \cdot q &= \frac{1}{2}\, (s - m_p^2 - k^2 +t) \, \text{,} \nonumber\\
p' \cdot k_1 &= \frac{1}{4}\, (s-p'^2 - k^2) + \frac{1}{2} \,p\cdot \left(k_1 - k_2\right) + \frac{1}{2} \,q\cdot \left(k_1 - k_2\right) \, \text{,} \nonumber\\
p' \cdot k_2 &= \frac{1}{4}\, (s-p'^2 - k^2) - \frac{1}{2}\, p\cdot \left(k_1 - k_2\right) - \frac{1}{2} \,q\cdot \left(k_1 - k_2\right) \, \text{,} \nonumber\\
q \cdot k_1 &= \frac{1}{4}\, (q^2 + k^2 -t) + \frac{1}{2} \,q\cdot \left(k_1 - k_2\right) \, \text{,} \nonumber\\
q \cdot k_2 &= \frac{1}{4}\, (q^2 + k^2 -t) - \frac{1}{2} \, q\cdot \left(k_1 - k_2\right) \, \text{,} \nonumber\\
k_1 \cdot k_2 &= \frac{1}{2}\, k^2 - m_{\pi}^2
\; \text{.} 
\end{align}

For the photoproduction reaction~\eqref{eq: 1.1} we have to insert the relations~\eqref{eq: A.4}
in~\eqref{eq: A.8}. Now we assume in~\eqref{eq: 1.1} unpolarised photon and proton in the 
initial state and no observation of the polarisation of the proton in the final state. 
Then a complete set of kinematic variables for the reaction is given by
(see~\eqref{eq: A.7}) 
\begin{align}
\label{eq: A.10}
s \,\text{,} \quad t \,\text{,} \quad k^2 \,\text{,} \quad 
p\cdot \left(k_1 - k_2 \right) \,\text{,} \quad 
q\cdot \left(k_1 - k_2 \right) \, \text{,}
\end{align}
and the sign of $I_P$. Equivalently we can chose 
\begin{align}
\label{eq: A.11}
s \,\text{,} \quad t \,\text{,} \quad k^2 \,\text{,} \quad  
\theta \,\text{,} \quad \phi\, \text{,}
\end{align}
where $\theta$, $\phi$ are the polar and azimuthal angles of $\mathbf{k}_1$ in
the $\pipi$ rest system.
We also note that $t \le t_\mathrm{min} < 0$ with 
\begin{align}
\label{eq:A.12a}
- t_\mathrm{min} &= 4 m_p^2 (k^2)^2 \left[ 2 (s-m_p^2)^2 + 2 (s-m_p^2) w(s,m_p^2,k^2) 
- 2 (s+m_p^2) k^2 \right]^{-1} \nn \\
&= \mathcal{O} (m_p^2 (k^2)^2 /s^2) 
\end{align}
where 
\begin{align}
\label{eq:A12b}
w(x,y,z) = (x^2 + y^2 + z^2 - 2xy - 2xz - 2yz)^\frac{1}{2} \,.
\end{align}
\begin{figure}[thb]
\center
\includegraphics[width=0.75\textwidth]{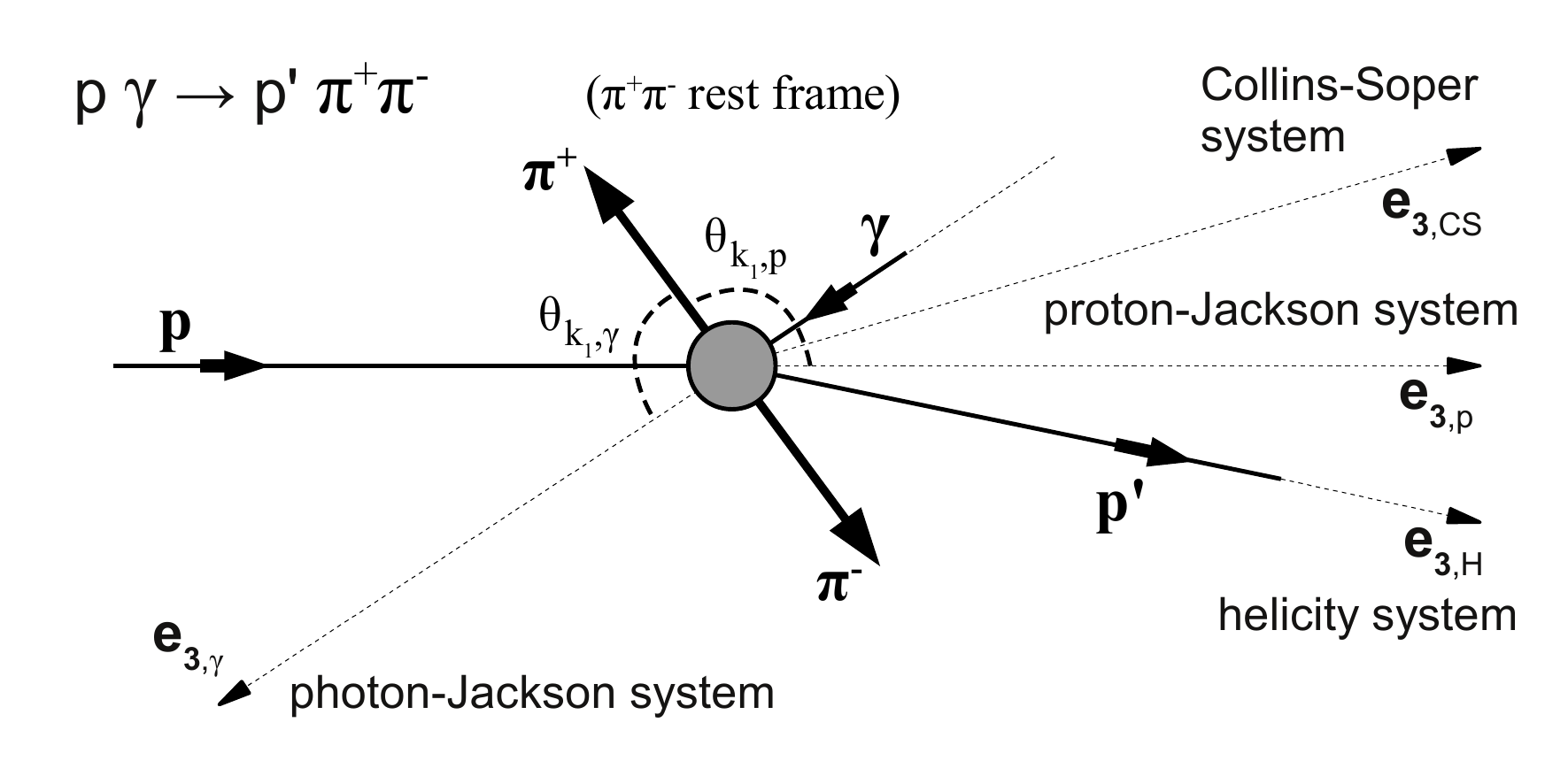}
\caption{The reaction plane in the $\pipi$ rest frame and the 
$\mathbf{e_3}$-axes of the 
proton-Jackson, photon-Jackson, 
Collins-Soper and the helicity systems.
\label{fig: AppendixA.1} }
\end{figure}

We are interested in the angular distribution of the $\pi^+$  in the centre-of-mass system
of the $\pipi$ pair. There we have 
\begin{align}
\label{eq: A.4a}
\mathbf{k} =0 \, \text{,}  \qquad
\mathbf{p} +  \mathbf{q} = \mathbf{p'} \, \text{.} 
\end{align}
Various reference systems are commonly used; 
see figure~\ref{fig: AppendixA.1} for illustration.
We list here the proton and photon-Jackson systems,
see~\cite{Gottfried:1964nx, Jackson:1964zd}, which are used in 
section~\ref{sec: Results}
to illustrate cross sections and asymmetries.
In addition we also 
mention the Collins-Soper~\cite{Collins:1977iv} and the helicity system \cite{Byckling}. 

For the proton-Jackson system we set (cf.\ p.\ 125 of \cite{Byckling}) 
\begin{align}
\label{eq: A.6dd}
\mathbf{e}_{\mathbf{1}, p} &= \frac{-\mathbf{p'} + \mathbf{\hat{p}}(\mathbf{\hat{p}}\cdot \mathbf{p'})}{ |\mathbf{\hat{p}} \times \mathbf{p'}| } \, \text{,} \nonumber \\
\mathbf{e}_{\mathbf{2}, p} &= - \frac{\mathbf{\hat{p}} \times \mathbf{p'}}{ |\mathbf{\hat{p}} \times \mathbf{p'}|} \, \text{,} \nonumber \\
\mathbf{e}_{\mathbf{3}, p} &= \mathbf{\hat{p}}  \, \text{,} 
\end{align}
with $\mathbf{\hat{p}} = \mathbf{p} / |\mathbf{p}|$, $\mathbf{\hat{q}} = \mathbf{q} /|\mathbf{q}|$ 
and $\mathbf{p}$, $\mathbf{q}$ the three-momenta of the initial proton and the $\gamma^{(*)}$ in 
the $\pipi$ rest system. 
The corresponding polar and azimuthal angles of the $\pi^+$ 
 are denoted by 
$\theta_{k_1,p}$ and $\phi_{k_1,p}$,
respectively, and we have $\cos \theta_{k_1,p} = \mathbf{\hat{k}_1} \cdot \mathbf{\hat{p}}$.

For the photon-Jackson system we set (cf.\ p.\ 125 of \cite{Byckling})
\begin{align}
\label{eq: A.6ee}
\mathbf{e}_{\mathbf{1}, \gamma} &=\frac{-\mathbf{p'} + \mathbf{\hat{q}}(\mathbf{\hat{q}}\cdot \mathbf{p'})}{|\mathbf{\hat{q}} \times \mathbf{p'}| } \, \text{,} \nonumber \\
\mathbf{e}_{\mathbf{2}, \gamma} &= - \frac{\mathbf{\hat{q}} \times \mathbf{p'}}{ |\mathbf{\hat{q}} \times \mathbf{p'}| } \, \text{,} \nonumber \\
\mathbf{e}_{\mathbf{3}, \gamma} &= \mathbf{\hat{q}}  \, \text{.} 
\end{align}
The polar and azimuthal angles of the $\pi^+$ in the photon-Jackson system are denoted by 
$\theta_{k_1,\gamma}$ and $\phi_{k_1,\gamma}$,
respectively, and we have $\cos \theta_{k_1,\gamma} = \mathbf{\hat{k}_1} \cdot \mathbf{\hat{q}}$.

We mention two more systems of reference, the Collins-Soper and the
helicity systems. The unit vectors of the CS system are chosen as
\begin{align}
\label{eq: A.5}
\mathbf{e}_{\mathbf{1}, \mathrm{CS}} &= \frac{\mathbf{\hat{p}} + \mathbf{\hat{q}}}{ |\mathbf{\hat{p}} + \mathbf{\hat{q}}| } \, \text{,} \nonumber \\
\mathbf{e}_{\mathbf{2}, \mathrm{CS}} &= \frac{\mathbf{\hat{p}} \times \mathbf{\hat{q}}}{ |\mathbf{\hat{p}} \times \mathbf{\hat{q}}| } \, \text{,} \nonumber \\
\mathbf{e}_{\mathbf{3}, \mathrm{CS}} &= \frac{\mathbf{\hat{p}} - \mathbf{\hat{q}}}{ |\mathbf{\hat{p}} - \mathbf{\hat{q}}| } \, \text{,} 
\end{align}
and the unit vectors of the helicity system are chosen as 
\begin{align}
\label{eq:A16}
\mathbf{e}_{\mathbf{1},\mathrm{H}}  &=  \frac{\mathbf{q} - 
\mathbf{\hat{p}'} (\mathbf{\hat{p}'} \cdot \mathbf{q})}
{|\mathbf{q}  \times \mathbf{\hat{p}'}|}  \, , \nn \\
\mathbf{e}_{\mathbf{2},\mathrm{H}}  &=  \frac{\mathbf{\hat{p}'} \times \mathbf{q}}{|\mathbf{\hat{p}'} \times \mathbf{q}|}
\,, \nn \\
\mathbf{e}_{\mathbf{3},\mathrm{H}}  &= \mathbf{\hat{p}'} \,,
\end{align}
with $\mathbf{\hat{p}'} = \mathbf{p}' / |\mathbf{p'}|$.

All systems discussed above have the $\mathbf{e_1}$ and $\mathbf{e_3}$ unit
vectors in the reaction plane given by the incoming and outgoing protons. 
With the above definitions we obtain the relation 
\begin{align}
\mathbf{e}_{\mathbf{2},p} =
- \mathbf{e}_{\mathbf{2},\gamma} = - \mathbf{e}_{\mathbf{2},\rm CS} =
- \mathbf{e}_{\mathbf{2},\rm H} \,.
\end{align}
The directions of the vectors $\mathbf{e}_{3}$ in the different systems
are shown in figure~\ref{fig: AppendixA.1}. 
They differ only by their orientation with respect to rotations around the  
$\mathbf{e}_{\mathbf{2}}$ axis.
\begin{figure}[thb]
        \centering
                \includegraphics[width=0.7\textwidth]{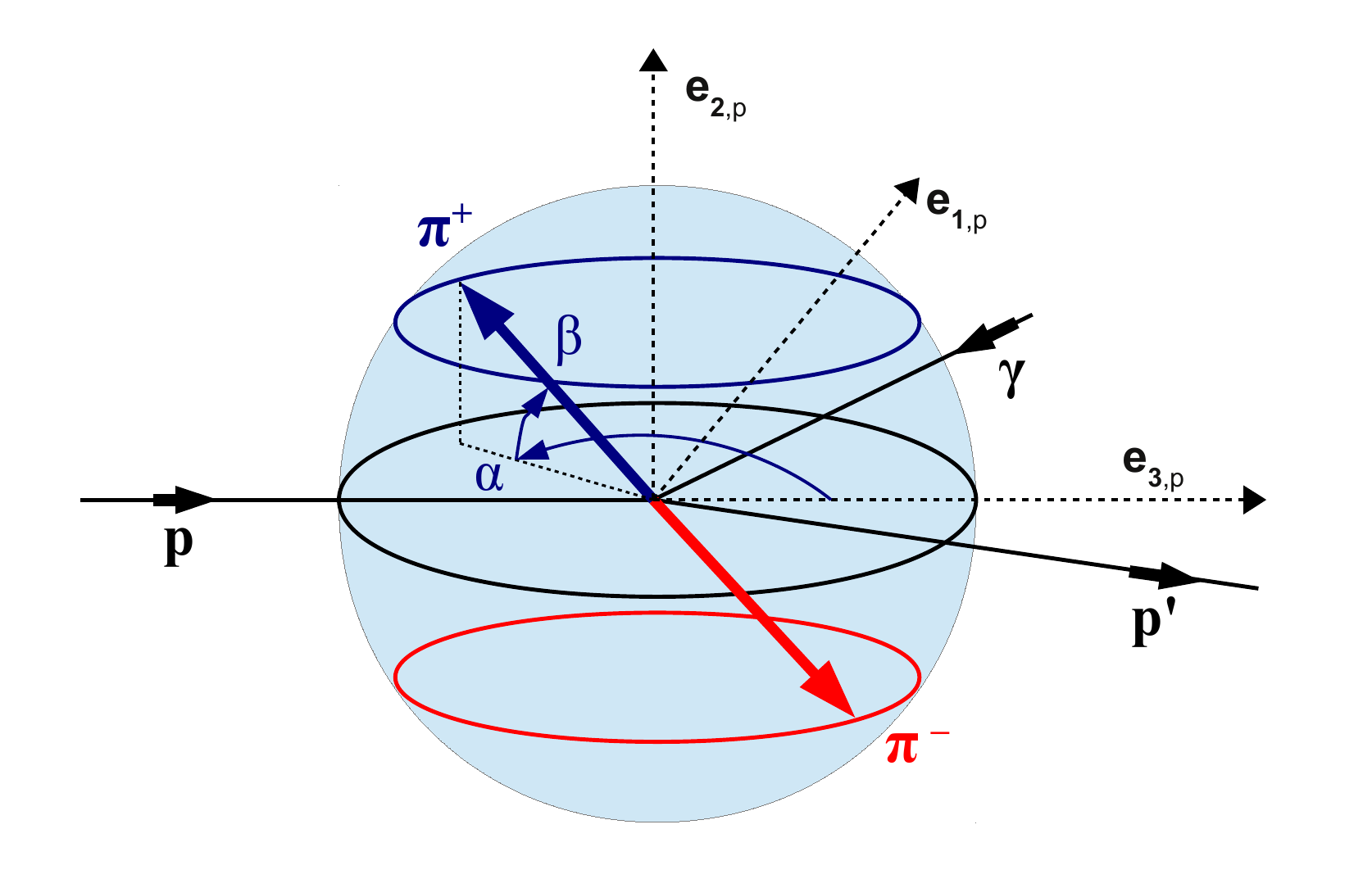}
        \caption{Definition of the angles $\alpha$ and $\beta$ in the
          $\pi^+\pi^-$ rest frame. As reference the proton-Jackson system is chosen. 
We have $0 \le \alpha < 2 \pi$ and $-\pi/2 \le \beta \le \pi/2$.
        \label{fig:1} }
\end{figure}

This property, and the fact that P-invariance with respect to reflection at
the reaction plane holds,
motivates to define two new angles $\alpha$ and $\beta$. These angles
describe the direction of the $\pi^+$ in the $\pipi$ rest frame 
and are defined as 
\begin{align}
\label{eq: alpha}
\sin{\beta}  &= \mathbf{\hat{k}_1} \cdot \mathbf{e}_{\mathbf{2},p} = 
- \frac{\mathbf{\hat{k}_1} \cdot(\mathbf{\hat{p}} \times \mathbf{p'}
  )}{|\mathbf{\hat{p}} \times \mathbf{p'}|} \,,
\nn \\
\cos \beta \,\cos \alpha &= \mathbf{\hat{k}_1} \cdot \mathbf{e}_{\mathbf{3},p} = 
\mathbf{\hat{k}_1} \cdot \mathbf{\hat{p}} \,,
\nn \\
\cos \beta \, \sin \alpha &= \mathbf{\hat{k}_1} \cdot \mathbf{e}_{\mathbf{1},p} = 
\frac{- \mathbf{\hat{k}_1} \cdot \mathbf{p'} + (\mathbf{\hat{k}_1} \cdot \mathbf{\hat{p}})(\mathbf{\hat{p}} \cdot \mathbf{p'})}{|\mathbf{\hat{p}} \times \mathbf{p'}|} \,,
\end{align}
where $\alpha=0$ is aligned to the incoming proton direction. Note 
that $\alpha$ and $\beta$ can also be interpreted as the spherical coordinates
of the $\pi^+$ in the $\pipi$ rest frame, with $\mathbf{e}_{\mathbf{2},p}$ having the r\^ole of the $z$- and
$\mathbf{e}_{\mathbf{3}, p}$ the r\^ole of the $x$-axis.

Transformation from one system $i$ to another system $j$ 
from \eqref{eq: A.6dd} to \eqref{eq:A16} can  then be represented by
rotations:
\begin{align}
\label{eq:A.22}
\alpha_{i} \rightarrow \alpha_{j} + \Delta \alpha_{ij} \,.
\end{align}
Reflections with respect to the reaction plane are described by a 
transformation of the elevation angle $\beta \rightarrow -\beta$ 
and P-invariance requires 
\begin{align}
\label{eq:A.23}
\frac{d \sigma}{d\Omega} \, (\alpha,\beta) = \frac{d \sigma}{d\Omega} \, (\alpha,-\beta) \,.
\end{align}
The following relations between the angles $\alpha$, $\beta$ and 
$\theta_{k_1,p}$, $\phi_{k_1,p}$
in the proton-Jackson system hold:
\begin{align}
\label{eq:A.24}
\tan{\alpha} \; = & \; \tan \theta_{k_1,p} \, \cos \phi_{k_1,p}\,, \\
\sin{\beta} \; = & \; \sin \theta_{k_1,p} \, \sin \phi_{k_1,p} \,.
\end{align}

\section{Propagators and vertices}
\label{app B}

In this appendix we collect the propagators and
vertices needed for the evaluation of the 
diagrams of figure~\ref{fig: 1}. Here the propagators
and vertices involving the pomeron, the odderon
and the reggeons are to be understood as effective
propagators and vertices.
Most of the relations listed in the following are
taken over from~\cite{Ewerz:2013kda}.
We reproduce them here in order to make the
present paper self-contained.

The numerical values of coupling constants and other
parameters quoted in the following are to be 
considered as default values. Adjustments of 
these parameters should come from detailed comparisons with
experiment.

All our vertices respect the standard crossing
and charge-conjugation~($C$) relations of quantum field theory.

\subsection*{Propagators}

\noindent 
$\bullet$ photon $\gamma$
\vspace*{.2cm}
\newline
\hspace*{0.5cm}\includegraphics[width=100pt]{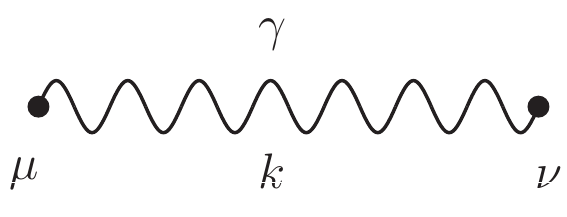} 
\begin{equation}
\label{B.1}
i \Delta_{\mu\nu}^{(\gamma)} (k) = \frac{- i g_{\mu \nu}}{k^2 + i \epsilon}\,.
\end{equation}
\newline

\noindent 
$\bullet$ pions $\pi^0$, $\pi^\pm$
\vspace*{.2cm}
\newline
\hspace*{0.5cm}\includegraphics[width=250pt]{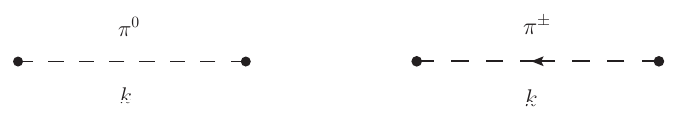} 
\begin{equation}
\label{B.2}
i \Delta^{(\pi)} (k) = \frac{i}{k^2 - m_\pi^2 + i \epsilon}\,.
\end{equation}
\newline

\noindent 
$\bullet$ vector-mesons $V=\rho^0,\omega$
\newline
Since we include in our calculations $\rho^0$--$\omega$
interference effects we rely here on the analysis
of the $\gamma$--$\rho$--$\omega$ propagator matrix as 
given in~\cite{Melikhov:2003hs}.
But here we are only interested in the
$\rho$--$\omega$, the strong-interaction, part of
the $3\times3$~propagator matrix studied there.
We get the following from appendix~B of~\cite{Melikhov:2003hs},
setting $e=0$ in the relations there, for $V', V \in \{\rho,\omega\}$:
\vspace*{.2cm}
\newline
\hspace*{0.5cm}\includegraphics[width=100pt]{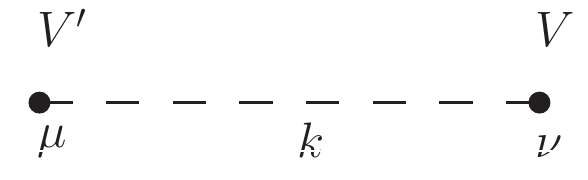} 
\begin{equation}
\label{B.3}
i \Delta_{\mu \nu}^{(V',V)}(k) = 
i \left( -g_{\mu \nu} + \frac{k_\mu k_\nu}{k^2 + i \epsilon}\right)
\Delta_T^{(V',V)} (k^2) 
- i \frac{k_\mu k_\nu}{k^2 + i \epsilon}\, \Delta_L^{(V',V)} (k^2) \,.
\end{equation}
Here and in the following $\rho$ is understood as $\rho^0$.
The longitudinal parts $\Delta_L^{(V',V)} (k^2)$
never enter in our calculations and, thus, need not 
be discussed further. For the transverse parts
we use matrix notation. 
With $k^2=s$ we set
\begin{equation}
\label{eq: B.4}
\underline{\Delta}_T(s) =
\begin{pmatrix}
\Delta_T^{(\rho,\rho)}(s) & \Delta_T^{(\rho,\omega)}(s) \\
\Delta_T^{(\omega,\rho)}(s) & \Delta_T^{(\omega,\omega)}(s)
\end{pmatrix}
\,.
\end{equation}
We have from~\cite{Melikhov:2003hs} where all these relations
are derived and discussed at length:
\begin{align}
\label{eq: B.5}
\Delta_T^{(\omega,\rho)}(s) &= \Delta_T^{(\rho,\omega)}(s)\,,\\
\label{eq: B.6}
\left(\underline{\Delta}_T(s) \right)^{-1} &=
\begin{pmatrix}
-m_\rho^2+s+B_{\rho\rho}(s) & s b_{\rho\omega}+B_{\rho\omega}(s)\\
s b_{\rho\omega}+B_{\rho\omega}(s) & -m_\omega^2+s+B_{\omega\omega}(s)
\end{pmatrix},
\end{align}
\begin{align}
\nonumber
\label{eq: B.7}
B_{V'V}(s=0) &=0\,,\\
\mathrm{Re} \,B_{\rho\rho}(m_\rho^2) &= \mathrm{Re} \,B_{\rho\omega}(m_\rho^2) =0\,,\\
\nonumber
\mathrm{Re} \,B_{\omega\omega}(m_\omega^2) &=0\,,
\end{align}
\begin{equation}
\label{eq: B.8}
\underline{\Delta}_T(s=0) =
\begin{pmatrix}
-\frac{1}{m_\rho^2} & 0 \\
0 & -\frac{1}{m_\omega^2} 
\end{pmatrix}\,,
\end{equation}
\begin{align}
\label{eq: B.9}
B_{\rho\rho}(s)=&{}\,
g_{\rho\pi\pi}^2 s \left[
R(s,m_\pi^2)-R(m_\rho^2,m_\pi^2) + \frac{1}{2}
\left( R(s,m_K^2) - R(m_\rho^2,m_K^2) \right) \right]\\
\nonumber
&+ i g_{\rho\pi\pi}^2  \left[
I(s,m_\pi^2)+\frac{1}{2}I(s,m_K^2)\right]\,,
\end{align}
\begin{align}
\label{eq: B.10}
I(s,m^2)&=\frac{1}{192 \pi} s
\left(1-\frac{4m^2}{s}\right)^{3/2}
\theta(s-4m^2)\,,\\
\label{eq: B.11}
R(s,m^2)&= \frac{s}{192 \pi^2}\, 
\mbox{V.\,P.}\int_{4m^2}^\infty 
\frac{ds'}{s' (s'-s)} \left(1-\frac{4m^2}{s'} \right)^\frac{3}{2} \,.
\end{align}
Here $\mbox{V.\,P.}$ means the principal value prescription. 
Explicitly we get the following.
For~$s>4m^2$
\begin{equation}
\label{eq: B.12}
R(s,m^2) =
\frac{1}{96\pi^2}
\left[\frac{1}{3}+\xi^2+\frac{1}{2}\xi^3 \log \left( \frac{1-\xi}{1+\xi} \right) \right] 
\quad
\text{with} \quad \xi=\left(1-\frac{4m^2}{s}\right)^{1/2}\,,
\end{equation}
for~$0<s<4m^2$
\begin{equation}
\label{eq: B.13}
R(s,m^2) =
\frac{1}{96\pi^2}
\left[
\frac{1}{3}-\xi^2+\xi^3\arctan\left(\frac{1}{\xi}\right) \right] \quad
\text{with}\quad \xi=\left(\frac{4m^2}{s}-1\right)^{1/2}\,,
\end{equation}
and for~$s<0$
\begin{equation}
\label{eq: B.14}
R(s,m^2) =
\frac{1}{96\pi^2}
\left[
\frac{1}{3}+\xi^2+\frac{1}{2}\xi^3 \log\left(\frac{\xi-1}{\xi+1}\right) \right] \quad
\text{with}\quad \xi=\left(1-\frac{4m^2}{s}\right)^{1/2}\,.
\end{equation}
Furthermore, we have from~\cite{Melikhov:2003hs} 
\begin{equation}
\label{eq: B.15}
B_{\omega\omega}(s)=
g_{\omega K K}^2 s \left[
R(s,m_K^2)-R(m_\omega^2,m_K^2) \right]
+ i \, \frac{s}{m_\omega} \Gamma_\omega\,,
\qquad g_{\omega K K} = \frac{1}{2} g_{\rho \pi \pi}\,,
\end{equation}
\begin{equation}
\label{eq: B.16}
B_{\rho\omega}(s) = 
g_{\rho \pi \pi} g_{\omega \pi \pi} 
\left[ s \left( 
R(s,m_\pi^2)-R(m_\rho^2,m_\pi^2) \right) 
+ i I(s,m_\pi^2) \right].
\end{equation}
In~\cite{Melikhov:2003hs} fits to the pion electromagnetic, weak, and 
$\pi\gamma$ transition form factors were made. From the fit~III
in table~4 of~\cite{Melikhov:2003hs}  the following values for the
parameters of $\underline{\Delta}_T(s)$~\eqref{eq: B.4}
were determined:
\begin{equation}
\label{eq: B.17}
\begin{split}
m_\rho &= 773.7\pm 0.4 \mev \, \text{,}\\
m_\omega&=782.43 \pm 0.05 \mev \, \text{,}\\
\Gamma_\omega &= 8.49 \pm 0.08 \mev \, \text{,}
\quad\text{(input from~\cite{Beringer:1900zz})}
\\
g_{\rho \pi \pi} &= 11.51 \pm 0.07\,,\\
g_{\omega\pi\pi} &= -0.35 \pm 0.10\,,\\
b_{\rho\omega} &= (3.5 \pm 0.5) \times 10^{-3}\,.
\end{split}
\end{equation}
For comparison we quote the $\rho$ and $\omega$ masses as listed 
in~\cite{Beringer:1900zz}:
\begin{equation}
\label{eq: B.18}
\begin{split}
m_\rho &= 775.26 \pm 0.25 \mev \, \text{,}\\
m_\omega &= 782.65 \pm 0.12 \mev \, \text{.}
\end{split}
\end{equation}
These values are quite consistent with the
corresponding ones from~\eqref{eq: B.17}. We recall
that we are only quoting default values 
for our calculations here and for this purpose
it makes no difference if we take those from~\eqref{eq: B.17}
or from~\eqref{eq: B.18}.
The definitions of the coupling constants 
$g_{\rho \pi \pi}$ and $g_{\omega\pi\pi}$ through
the corresponding vertices are given in~\eqref{eq: B.54}
below.

The $\rho$--$\omega$ propagator matrix as defined above should
be a good model for $|s| \lesssim \nobreak 15\gevsq$;
see section~4 of~\cite{Ewerz:2013kda}.
\newline

\noindent 
$\bullet$ vector-meson $\rho'$ 
\vspace*{.2cm}
\newline
\hspace*{0.5cm}\includegraphics[width=100pt]{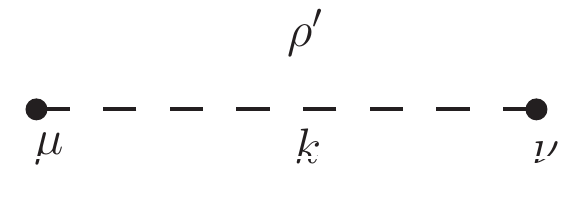} 
\begin{equation}
\label{eq: B.19}
i \Delta_{\mu \nu}^{(\rho',\rho')}(k) = 
i \left( -g_{\mu \nu} + \frac{k_\mu k_\nu}{k^2 + i \epsilon}\right)
\Delta_T^{(\rho',\rho')} (k^2) 
- i \frac{k_\mu k_\nu}{k^2 + i \epsilon} \, \Delta_L^{(\rho',\rho')} (k^2) \,.
\end{equation}
Again, $\Delta_L^{(\rho',\rho')} (k^2)$ need not be discussed.
For $\Delta_T^{(\rho',\rho')} (k^2)$ we make a simple ansatz as
suggested in~\cite{Kuhn:1990ad}
\begin{equation}
\label{eq: B.19a}
\Delta_T^{(\rho',\rho')} (s) =
\frac{1}{s-m_{\rho'}^2 + i \sqrt{s} \,\Gamma_{\rho'}(s)}\,,
\end{equation}
\begin{equation}
\label{eq: B.20}
\Gamma_{\rho'}(s) =
\Gamma_{\rho'}
\left( \frac{s- 4 m^2_\pi}{m^2_{\rho'} - 4 m^2_\pi} \right)^{3/2}
\frac{m^2_{\rho'}}{s}\,
\theta(s-4 m^2_\pi)\,.
\end{equation}
The values for~$m_{\rho'}$ and $\Gamma_{\rho'}$ listed
in~\cite{Beringer:1900zz} are
\begin{equation}
\label{eq: B.21}
\begin{split}
m_{\rho'} &= 1465 \pm 25 \mev  \,,\\
\Gamma_{\rho'} &= 400 \pm 60 \mev  \,.
\end{split}
\end{equation}

We note that in~\cite{Abramowicz:2011pk} --
contrary to what is written there --
a different form for the $\rho'$ propagator is
used, {\em not} the one from~\cite{Kuhn:1990ad}
which we reproduce here in \eqref{eq: B.19a}, \eqref{eq: B.20}.
The Breit-Wigner ansatz made in Eqs.~(4), (5) of
\cite{Abramowicz:2011pk} corresponds to replacing 
in~\eqref{eq: B.19a} $\sqrt{s}\, \Gamma_{\rho'}(s)$
by $m_{\rho'} \Gamma_{\rho'}(s)$. We shall not use such a form
for the~$\rho'$ propagator.
\newline

\noindent 
$\bullet$ tensor meson $f_2\equiv f_2 (1270)$ (see~(3.6)-(3.9)) and section~5 of
\cite{Ewerz:2013kda}) 
\vspace*{.2cm}
\newline
\hspace*{0.5cm}\includegraphics[width=100pt]{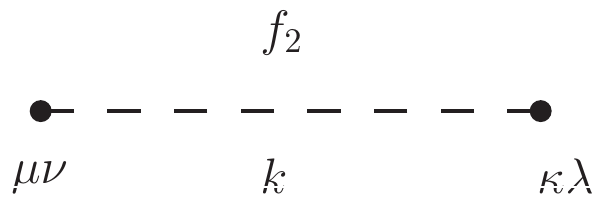}
\begin{equation}
\label{eq: B.22}
\begin{split}
i \Delta^{(f_2)}_{\mu\nu,\kappa\lambda} (k) = \, 
i \,\bigg\{ & \frac{1}{2} \left( - g_{\mu\kappa} + \frac{k_\mu k_\kappa}{k^2 + i \epsilon} \right) 
\left( - g_{\nu\lambda} + \frac{k_\nu k_\lambda}{k^2 + i \epsilon} \right)
\\
& + \frac{1}{2} \left( - g_{\mu\lambda} + \frac{k_\mu k_\lambda}{k^2 + i \epsilon} \right) 
\left( - g_{\nu\kappa} + \frac{k_\nu k_\kappa}{k^2 + i \epsilon} \right) 
\\
& - \frac{1}{3} \left( - g_{\mu\nu} + \frac{k_\mu k_\nu}{k^2 + i \epsilon} \right) 
\left( - g_{\kappa\lambda} + \frac{k_\kappa k_\lambda}{k^2 + i \epsilon} \right) 
\bigg\} \,\Delta^{(2)}(k^2)\,.
\end{split}
\end{equation}
Here we suppose~$0<k^2<10 \gev^2$. 
The invariant function $\Delta^{(2)}(k^2)$ is given
as follows, setting $k^2=s$, 
\begin{equation}
\label{eq: B.23}
\left[ \Delta^{(2)} (s) \right]^{-1} = 
-m_{f_2}^2 + s + s \left[ R_{f_2}(s) - R_{f_2}(m_{f_2}^2) \right] 
+ i \, \mathrm{Im}\, B_{f_2}(s) \,,
\end{equation}
with 
\begin{align}
\label{eq: B.24}
\mathrm{Im}\,B_{f_2}(s) &= 
\frac{\Gamma_{f_2}}{\Gamma(f_2\to \pi\pi)} \frac{1}{320 \pi} 
\left|\frac{g_{f_2\pi\pi} F^{(f_2\pi\pi)}(s)}{M_0} \right|^2
s^2 \left(1-\frac{4m_\pi^2}{s}\right)^\frac{5}{2} \theta(s-4m_\pi^2) \,,
\\
\label{eq: B.25}
R_{f_2}(s) &= \frac{s}{\pi} \,\mbox{V.\,P.}
\int_{4m_\pi^2}^\infty ds' \,
\frac{\mathrm{Im}\, B_{f_2}(s')}{s'^2 (s'-s)} \,,
\end{align}
where V.\,P.\ means  the principal value prescription
and $M_0 = 1 \gev$; see (5.19)-(5.22) of~\cite{Ewerz:2013kda}.
The coupling constant $g_{f_2\pi\pi}$ and the form factor
$F^{(f_2\pi\pi)}(s)$ are defined in~(3.37) and (5.4)
of~\cite{Ewerz:2013kda}; see~\eqref{eq: B.57}-\eqref{eq: B.59} below.

The following relations hold:
\begin{equation}
\label{eq: B.26}
\Delta^{(f_2)}_{\mu \nu, \kappa \lambda} (k) = 
\Delta^{(f_2)}_{\nu \mu, \kappa \lambda} (k) = 
\Delta^{(f_2)}_{\mu \nu, \lambda \kappa} (k) = 
\Delta^{(f_2)}_{\kappa \lambda,\mu \nu} (k) \,,
\end{equation}
\begin{equation}
\label{eq: B.27}
\begin{split}
g^{\mu\nu} \Delta^{(f_2)}_{\mu \nu, \kappa \lambda} (k) &= 0 \,,
\\
g^{\kappa\lambda} \Delta^{(f_2)}_{\mu \nu, \kappa \lambda} (k) &= 0 \,.
\end{split}
\end{equation}
From \cite{Beringer:1900zz} we get the mass and width of the $f_2$ meson as
\begin{equation}
\label{eq: B.28}
m_{f_2} = 1275.1 \pm 1.2 \mev \, \text{,} 
\qquad
\Gamma_{f_2} =185.1\,{}^{+2.9}_{-2.4} \mev \, \text{,}
\end{equation}
and the $f_2 \rightarrow \pi\pi$ branching fraction as
\begin{equation}
\label{eq: B.29}
\frac{\Gamma(f_2\to \pi\pi)}{\Gamma_{f_2}} = (84.8_{-1.2}^{+2.4} ) \,.
\end{equation}
\newline

\noindent
$\bullet$ pomeron $\pomeron$ (see~(3.10), (3.11) and section 6.1
of~\cite{Ewerz:2013kda})
\vspace*{.2cm}
\newline
\hspace*{0.5cm}\includegraphics[width=125pt]{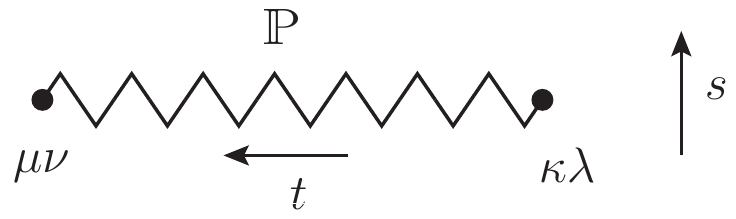} 
\begin{equation}
\label{eq: B.30}
i\Delta^{(\pomeron)}_{\mu\nu,\kappa\lambda} (s,t) 
= \frac{1}{4s} \left(g_{\mu\kappa} g_{\nu\lambda} + g_{\mu\lambda} g_{\nu\kappa} 
- \frac{1}{2} g_{\mu\nu} g_{\kappa\lambda} \right) 
\, (-i s \alpha'_\pomeron)^{\alpha_\pomeron(t)-1} \,,
\end{equation}
\begin{equation}
\label{eq: B.31}
\begin{split}
\alpha_\pomeron(t) &= 1 + \epsilon_\pomeron + \alpha'_\pomeron t \,,
\\
\epsilon_\pomeron &= 0.0808 \,,
\\
\alpha'_\pomeron &= 0.25 \gev^{-2} \,.
\end{split}
\end{equation}
\newline

\noindent 
$\bullet$ reggeons ${\reggeon}_+=f_{2R},a_{2R}$ 
(see~(3.12), (3.13)
and section~6.3 of~\cite{Ewerz:2013kda}) 
\vspace*{.2cm}
\newline
\hspace*{0.5cm}\includegraphics[width=125pt]{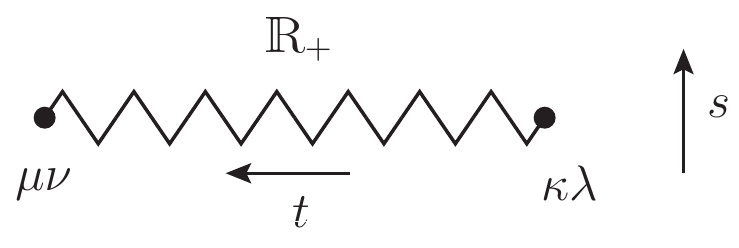} 
\begin{equation}
\label{eq: B.32}
i\Delta^{(\reggeon_+)}_{\mu\nu,\kappa\lambda} (s,t) 
= \frac{1}{4s} \left(g_{\mu\kappa} g_{\nu\lambda} + g_{\mu\lambda} g_{\nu\kappa} 
- \frac{1}{2} g_{\mu\nu} g_{\kappa\lambda} \right) 
(-i s \alpha'_{\reggeon_+})^{\alpha_{\reggeon_+}(t)-1} \,,
\end{equation}
\begin{equation}
\label{eq: B.33}
\begin{split}
\alpha_{\reggeon_+}(t) &= \alpha_{\reggeon_+} (0)+ \alpha'_{\reggeon_+} t \,,
\\
\alpha_{\reggeon_+} (0) &= 0.5475 \,,
\\
\alpha'_{\reggeon_+} &= 0.9 \gev^{-2} \,.
\end{split}
\end{equation}
\newline

\noindent 
$\bullet$ reggeons $\reggeon_-=\omega_R,\rho_R$ 
(see~(3.14), (3.15)
and section~6.3 of~\cite{Ewerz:2013kda}) 
\vspace*{.2cm}
\newline
\hspace*{0.5cm}\includegraphics[width=125pt]{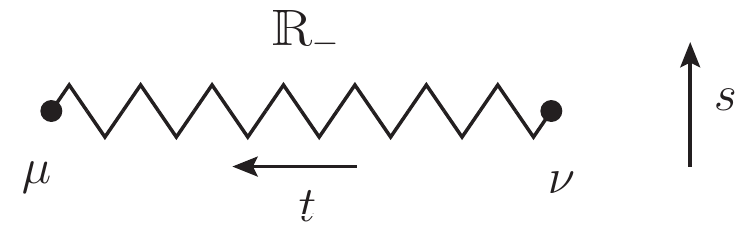} 
\begin{equation}
\label{eq: B.34}
i \Delta_{\mu\nu}^{(\reggeon_-)} (s,t) = i g_{\mu\nu} \,\frac{1}{M_-^2} \,
(-i s \alpha'_{\reggeon_-})^{\alpha_{\reggeon_-} (t)-1} \,,
\end{equation}
\be
\label{eq: B.35}
\begin{split}
\alpha_{\reggeon_-}(t) &= \alpha_{\reggeon_-} (0)+ \alpha'_{\reggeon_-} t \,,
\\
\alpha_{\reggeon_-} (0) &= 0.5475 \, \text{,}
\\
\alpha'_{\reggeon_-} &= 0.9 \gev^{-2} \, \text{,}
\\
M_- &= 1.41 \gev  \, \text{.}
\end{split}
\end{equation}
\newline

\noindent 
$\bullet$ odderon $\odderon$
(see~(3.16), (3.17)
and section~6.2 of~\cite{Ewerz:2013kda}) 
\vspace*{.2cm}
\newline
\hspace*{0.5cm}\includegraphics[width=125pt]{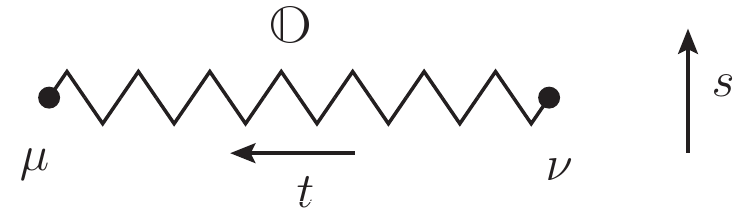} 
\begin{equation}
\label{eq: B.36}
i \Delta_{\mu\nu}^{(\odderon)} (s,t) = -i g_{\mu\nu} \,\frac{\eta_\odderon}{M_0^2} \,
(-i s \alpha'_{\odderon})^{\alpha_{\odderon} (t)-1} \,,
\end{equation}
\be\label{eq: B.37}
\begin{split}
\alpha_\odderon(t) &= 1 + \epsilon_\odderon + \alpha'_\odderon t \,,
\\
\eta_\odderon &= \pm 1 \,,
\\
\alpha'_\odderon &= 0.25 \gev^{-2} \, \text{.}
\end{split}
\end{equation}
We have set as default $\alpha'_\odderon =\alpha'_\pomeron$
in~\eqref{eq: B.37}.
\newline

\subsection*{Vertices}
In this section we list the vertices which are needed for discussing
the reaction~\eqref{eq: 1.1}. We use as in~\cite{Ewerz:2013kda}
the following rank-four tensor functions
\begin{align}
\label{eq: B.38}
\Gamma_{\mu\nu\kappa\lambda}^{(0)} (k_1,k_2) =\,& 
[(k_1\cdot k_2) g_{\mu\nu} - k_{2\mu} k_{1\nu}] 
\left[k_{1\kappa}k_{2\lambda} + k_{2\kappa}k_{1\lambda} - 
\frac{1}{2} (k_1 \cdot k_2) g_{\kappa\lambda}\right] \,,
\\
\label{eq: B.39}
\Gamma_{\mu\nu\kappa\lambda}^{(2)} (k_1,k_2) = \,
& (k_1\cdot k_2) (g_{\mu\kappa} g_{\nu\lambda} + g_{\mu\lambda} g_{\nu\kappa} )
+ g_{\mu\nu} (k_{1\kappa} k_{2\lambda} + k_{2\kappa} k_{1\lambda} ) 
\nn \\
& - k_{1\nu} k_{2 \lambda} g_{\mu\kappa} - k_{1\nu} k_{2 \kappa} g_{\mu\lambda} 
- k_{2\mu} k_{1 \lambda} g_{\nu\kappa} - k_{2\mu} k_{1 \kappa} g_{\nu\lambda} 
\\
& - [(k_1 \cdot k_2) g_{\mu\nu} - k_{2\mu} k_{1\nu} ] \,g_{\kappa\lambda} \,.
\nn
\end{align}
For $i=0,2$ we have
\be\label{eq: B.40}
\Gamma_{\mu\nu\kappa\lambda}^{(i)} (k_1,k_2) 
= \Gamma_{\mu\nu\lambda\kappa}^{(i)} (k_1,k_2) 
= \Gamma_{\nu\mu\kappa\lambda}^{(i)} (k_2,k_1) 
= \Gamma_{\mu\nu\kappa\lambda}^{(i)} (-k_1,-k_2) \,,
\ee
\be\label{eq: B.41}
\begin{split}
 k_1^\mu \Gamma_{\mu\nu\kappa\lambda}^{(i)} (k_1,k_2) &=0 \,,
\\
 k_2^\nu \Gamma_{\mu\nu\kappa\lambda}^{(i)} (k_1,k_2) &=0 \,,
\end{split}
\ee
\be\label{eq: B.42}
\Gamma_{\mu\nu\kappa\lambda}^{(i)} (k_1,k_2)\, g^{\kappa\lambda} =0 \,. 
\ee
The vertices read as follows.
\newline

\noindent 
$\bullet$ $\gamma V$, where $V=\rho,\omega,\phi, \rho'$ 
(see (3.23)-(3.25) and
section 4 of~\cite{Ewerz:2013kda})
\vspace*{.2cm}
\newline
\hspace*{0.5cm}\includegraphics[width=100pt]{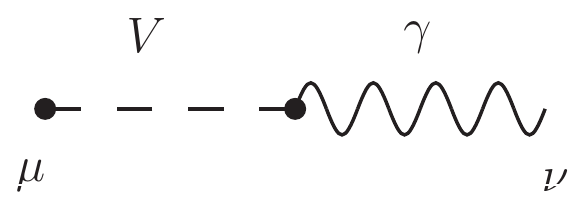} 
\begin{equation}
\label{eq: B.43}
 -i e \,\frac{m_V^2}{\gamma_V} \,g_{\mu\nu} \,,
\end{equation}
\be\label{eq: B.44}
e>0 \,,\quad \gamma_\rho >0\,,\quad \gamma_\omega>0 \,,\quad \gamma_\phi <0 \,,
\ee
\be\label{eq: B.45 d}
\frac{4 \pi}{\gamma_\rho^2} = 0.496 \pm 0.023 \,,
\quad
\frac{4 \pi}{\gamma_\omega^2} = 0.042 \pm 0.0015 \,,
\quad
\frac{4 \pi}{\gamma_\phi^2} = 0.0716 \pm 0.0017 \,.
\ee
A reliable determination of~$\gamma_{\rho'}$ is not known to us. 
\newline

\noindent 
$\bullet$ $\gamma p p$ 
(see (3.26)-(3.32) of~\cite{Ewerz:2013kda})
\vspace*{.2cm}
\newline
\hspace*{0.5cm}\includegraphics[height=85pt]{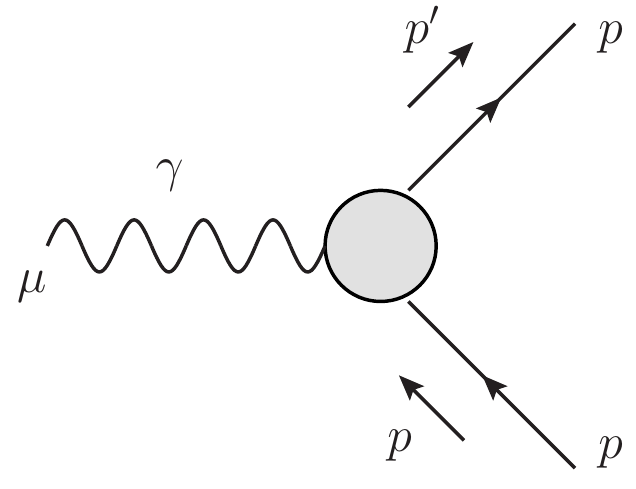} 
\begin{equation}\label{eq: B.46} 
i \Gamma_\mu^{(\gamma p p) } (p',p) = 
-ie \left[ \gamma_\mu F_1(t) + \frac{i}{2 m_p}\, \sigma_{\mu\nu} (p'-p)^\nu F_2(t)\right] \,,
\end{equation}
\be\label{eq: B.47}
e>0 \,, \qquad
t=(p'-p)^2 \,,
\ee
\be\label{eq: B.48}
\sigma_{\mu\nu} = 
\frac{i}{2} \left(\gamma_\mu \gamma_\nu - \gamma_\nu \gamma_\mu\right)\,,
\ee
\begin{align}\label{eq: B.49}
F_1(t) &= \left( 1 - \frac{t}{4m_p^2} \frac{\mu_p}{\mu_N}\right) 
\left( 1 - \frac{t}{4 m_p^2} \right)^{-1} G_D(t) \,,
\\
\label{eq: B.50}
F_2(t) &= \left(\frac{\mu_p}{\mu_N} - 1 \right) \left( 1 - \frac{t}{4 m_p^2}\right)^{-1} G_D(t) \,,
\end{align}
\be\label{eq: B.51}
\mu_N = \frac{e}{2 m_p} \,,
\qquad
\frac{\mu_p}{\mu_N} = 2.7928 \,,
\ee
\be\label{eq: B.52}
G_D(t) = \left(1 - \frac{t}{m_D^2} \right)^{-2} \,,
\qquad
m_D^2 = 0.71 \gev^2 \, \text{.}
\ee
The proton's Dirac and Pauli form factors and the dipole
form factor are denoted by~$F_1$, $F_2$, and $G_D$, respectively;
see for instance chapter~2 in~\cite{Close:2007zzd}. 

In many hadronic vertices we have,
realistically, to introduce form factors.
For simplicity we use for the proton
for momentum transfer squared $t<0$ in 
general the Dirac form factor~$F_1(t)$~\eqref{eq: B.49}.
For mesons we use for~$t<0$ a simple parametrisation of 
the pion's electromagnetic form factor
\begin{equation}
\label{eq: B.53}
F_M(t) = F_\pi(t) = \frac{m^2_0}{m^2_0-t}\,,
\qquad
m^2_0 = 0.50\gev^2\,;
\end{equation}
see~(3.34) of~\cite{Ewerz:2013kda}.
Further form factors are introduced and discussed below.
\newline

\noindent 
$\bullet$ $V\pi^+\pi^-$, $V=\rho, \omega, \rho'$
(see~\cite{Melikhov:2003hs} and section~4~of~\cite{Ewerz:2013kda})
\vspace*{.2cm}
\newline
\hspace*{0.5cm}\includegraphics[height=85pt]{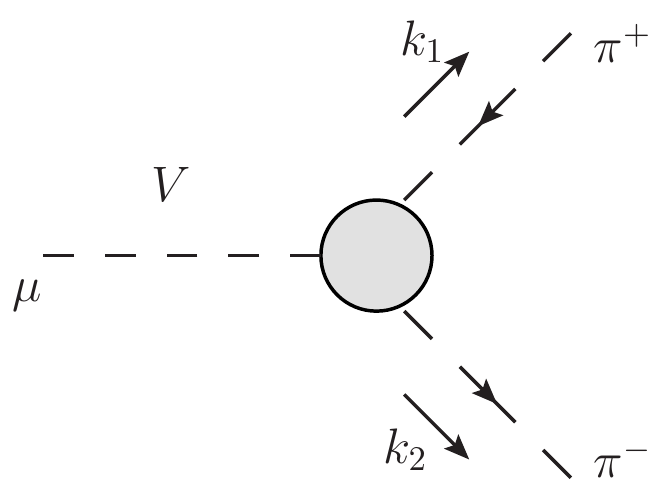} 
\begin{equation}\label{eq: B.54}
i \Gamma_\mu^{(V \pi\pi)} (k_1,k_2) = 
- \frac{1}{2} i g_{V \pi\pi} (k_1-k_2)_\mu \,,
\end{equation}
\begin{align}
\label{eq: B.55}
g_{\rho \pi\pi} &= 11.51 \pm 0.07 \,,\\
\label{eq: B.56}
g_{\omega \pi\pi} &= -0.35 \pm 0.10 \,.
\end{align}
As default values for the $\rho\pi\pi$ and
$\omega\pi\pi$ coupling constants we take those
from table~4~of~\cite{Melikhov:2003hs}. For
$g_{\rho'\pi\pi}$ we have no reliable estimate.
\newline

\noindent 
$\bullet$ $f_2\pi\pi$ 
(see~(3.37), (3.38) and section~5.1~of~\cite{Ewerz:2013kda})
\vspace*{.2cm}
\newline
\hspace*{0.5cm}\includegraphics[height=85pt]{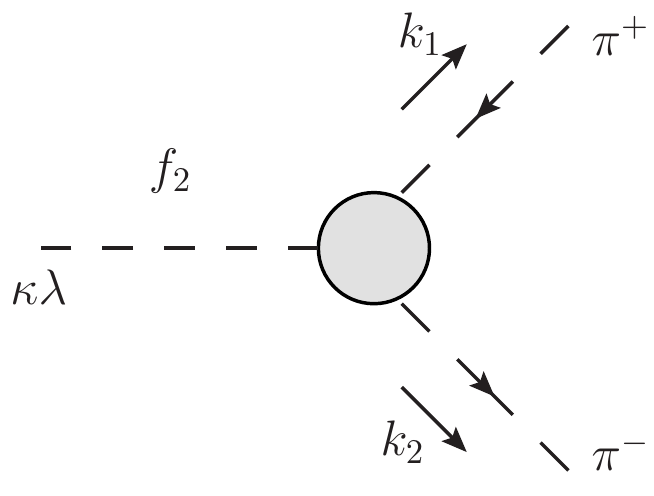} 
\hspace*{0.5cm}\includegraphics[height=85pt]{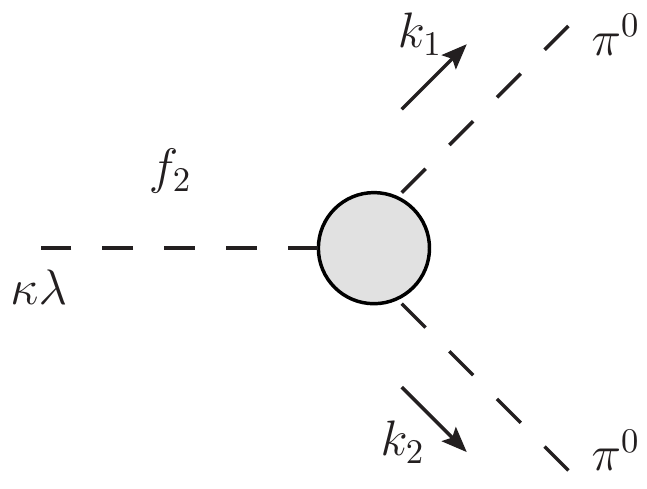} 
\be\label{eq: B.57}
i \Gamma_{\kappa \lambda}^{(f_2\pi\pi)} (k_1,k_2) 
= -i\, \frac{g_{f_2\pi\pi}}{2 M_0} 
\left[ (k_1-k_2)_\kappa (k_1-k_2)_\lambda - \frac{1}{4} g_{\kappa\lambda} (k_1-k_2)^2 \right] 
F^{(f_2\pi\pi)} (k^2) \,,
\ee
where $k=k_1+k_2$ and 
\be\label{eq: B.58}
g_{f_2\pi\pi} = 9.26 \pm 0.15 \,.
\ee
A convenient form for the $F^{(f_2\pi\pi)}$ form factor is
\begin{equation}
\label{eq: B.59}
F^{(f_2\pi\pi)}(k^2) = \exp \left(-\frac{k^2-m^2_{f_2}}{\Lambda^2_{f_2}}\right)
\end{equation}
which satisfies
$F^{(f_2\pi\pi)}(m_{f_2}^2)=1$. The parameter
$\Lambda_{f_2}$ is estimated to be
in the range 1 to $4 \gev$. 
\newline

\noindent 
$\bullet$ $f_2\gamma\gamma$ 
(see~(3.39), (3.40)
and sections~5.3, 7.2~of~\cite{Ewerz:2013kda})
\vspace*{.2cm}
\newline
\hspace*{0.5cm}
\includegraphics[height=85pt]{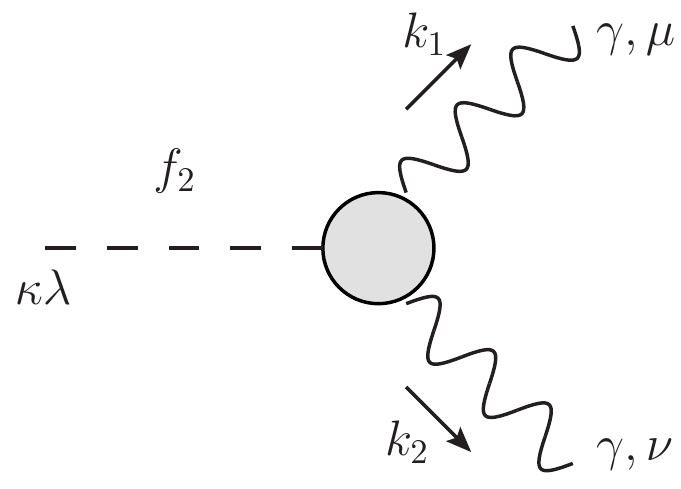} 
\begin{equation}\label{eq: B.60}
\begin{split}
i \Gamma_{\mu\nu\kappa\lambda}^{(f_2 \gamma \gamma)} (k_1,k_2) 
=\,& i F_M(k_1^2) F_M(k_2^2) F^{(f_2 \gamma \gamma)} (k^2) 
\\
& \times\left[
2 a_{f_2 \gamma\gamma} \Gamma_{\mu\nu\kappa\lambda}^{(0)}(k_1,k_2) 
- b_{f_2 \gamma\gamma} \Gamma_{\mu\nu\kappa\lambda}^{(2)}(k_1,k_2) 
\right] \,.
\end{split}
\end{equation}
Here $k=k_1+k_2$ and 
\begin{equation}\label{eq: B.61}
a_{f_2 \gamma\gamma} = \frac{e^2}{4 \pi} \, 1.45 \gev^{-3} \, \text{,}
\qquad
b_{f_2 \gamma\gamma} = \frac{e^2}{4 \pi} \, 2.49 \gev^{-1} \, \text{.}
\end{equation}
The form factor $F^{(f_2 \gamma \gamma)} (k^2)$ is
taken to be the same as for the $f_2 \pi\pi$ vertex
in~\eqref{eq: B.59},
\begin{equation}
\label{eq: B.62}
F^{(f_2 \gamma \gamma)} (k^2)=F^{(f_2 \pi \pi)} (k^2)\,.
\end{equation}
\newline

\noindent 
$\bullet$ $\pomeron pp$ 
(see~(3.43), (3.44)
and section~6.1~of~\cite{Ewerz:2013kda})
\vspace*{.2cm}
\newline
\hspace*{0.5cm}\includegraphics[height=85pt]{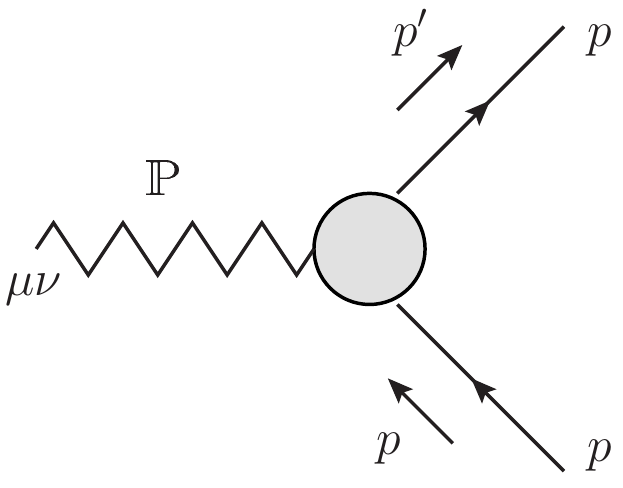} 
\be\label{eq: B.63}
\begin{split}
i \Gamma_{\mu\nu}^{(\pomeron pp)}  (p',p) = &
-i \,3 \beta_{\pomeron NN} F_1[(p'-p)^2]
\\ & \times 
\left\{ \frac{1}{2} \left[ \gamma_\mu (p'+p)_\nu + \gamma_\nu (p'+p)_\mu \right] 
-\frac{1}{4} \, g_{\mu\nu} (\slash{p}' + \slash{p}) \right\} \,,
\end{split}
\ee
\be \label{eq: B.64}
\beta_{\pomeron NN} = 1.87 \gev^{-1} \, \text{.}
\ee
\newline

\noindent 
$\bullet$ $\gamma\pi\pi$, $\gamma \gamma \pi \pi$, $\pomeron \pi\pi$, $\pomeron\gamma\pi\pi$
(see~(3.45), (3.46),
and section~7.1~of~\cite{Ewerz:2013kda})

\noindent
Our starting point here is the pion Lagrangian
including the $\pomeron\pi\pi$ coupling term as
written in (7.3) of~\cite{Ewerz:2013kda}.
Making a minimal substitution there in order
to include the coupling to the photon leads to
\begin{equation}
\label{eq: B.65}
\begin{split}
{\cal L}(x)=&
\left[(\partial_\mu-ieA_\mu(x)) \pi^-(x) \right]
(\partial^\mu+ieA^\mu(x)) \,\pi^+(x)
\\
&
+\frac{1}{2} \left(\partial_\mu \pi^0(x))(\partial^\mu \pi^0(x)\right)
- m^2_\pi \left( \pi^-(x) \pi^+(x) + \frac{1}{2} \pi^0(x) \pi^0(x) \right)
\\
&
+ 2 \beta_{\pomeron\pi\pi} \pomeron_{\kappa\lambda}(x)
\left( g^{\kappa\mu}g^{\lambda\nu}-
\frac{1}{4} g^{\kappa\lambda}g^{\mu\nu}\right)
\\
&\times
\bigg\{
\pi^-(x) (\partial_\mu+ieA_\mu(x))(\partial_\nu+ieA_\nu(x)) \,\pi^+(x)
\\
&\qquad 
+ \left[(\partial_\mu-ieA_\mu(x))(\partial_\nu-ieA_\nu(x)) \pi^-(x) \right] \pi^+(x)
\\
& \qquad
- \left[(\partial_\mu-ieA_\mu(x)) \pi^-(x)\right]
\left[(\partial_\nu+ieA_\nu(x))\pi^+(x)\right]
\\
& \qquad
-\left[(\partial_\nu-ieA_\nu(x))\pi^-(x)\right]
\left[(\partial_\mu+ieA_\mu(x))\pi^+(x)\right]
\\
&\qquad
+ \frac{1}{2}\pi^0(x) \partial_\mu \partial_\nu \pi^0(x)
+\frac{1}{2} (\partial_\mu \partial_\nu \pi^0(x)) \pi^0(x)
-(\partial_\mu \pi_0(x))(\partial_\nu \pi_0(x))
\bigg\}\,.
\end{split}
\end{equation}
The $\gamma\pi\pi$, $\gamma \gamma \pi \pi$,  $\pomeron\pi\pi$, and $\pomeron\gamma\pi\pi$ vertices following
from~\eqref{eq: B.65}, including a form factor for the pomeron coupling,
read as follows:
\vspace*{.2cm}
\newline
\hspace*{0.5cm}\includegraphics[height=85pt]{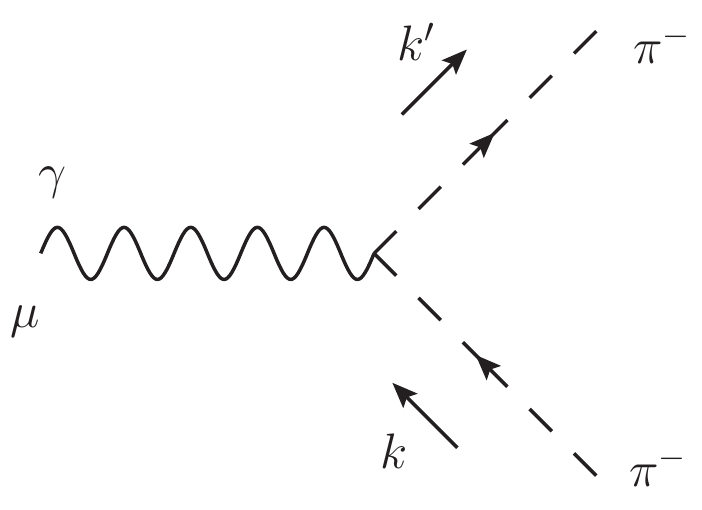} 
\be\label{eq: B.66}
i \Gamma_{\mu}^{(\gamma\pi\pi)}  (k',k) =
i\,e\, (k'+k)_\mu\,,
\ee
\vspace*{.2cm}
\hspace*{0.5cm}\includegraphics[height=85pt]{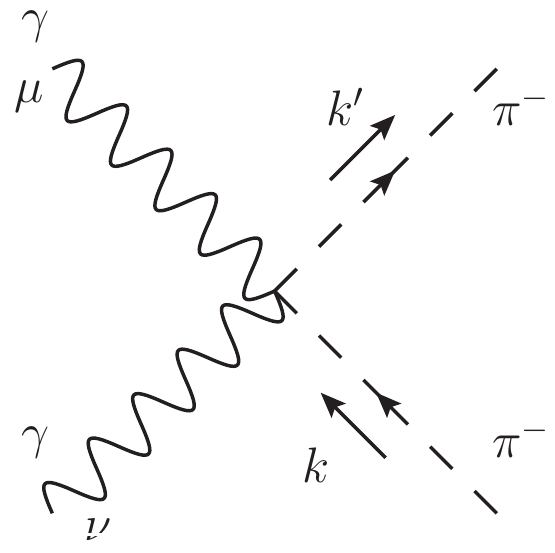} 
\be\label{eq: B.67a}
i \Gamma_{\mu\nu}^{(\gamma\gamma\pi\pi)}  (k',k) =
2 \,i\,e^2 g_{\mu\nu}  \,,
\ee
\vspace*{.2cm}
\hspace*{0.5cm}\includegraphics[height=85pt]{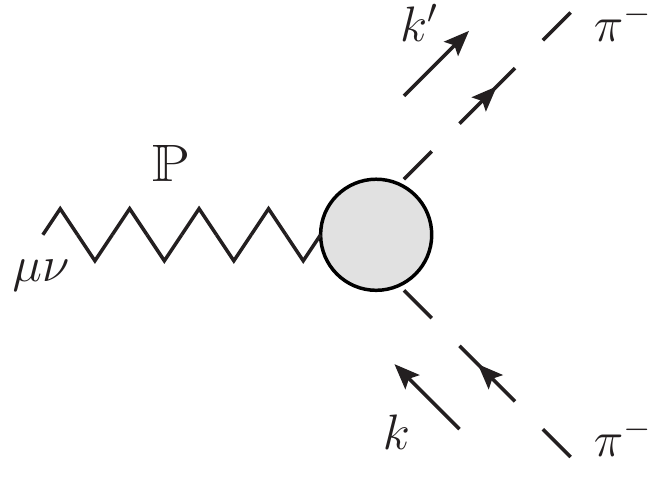} 
\hspace*{0.5cm}\includegraphics[height=85pt]{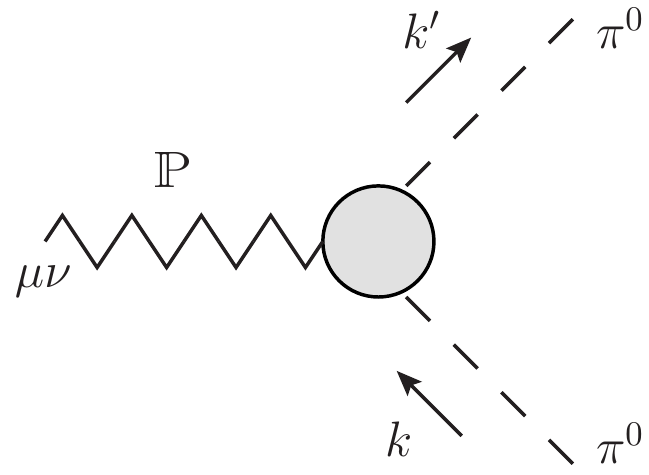} 
\be\label{eq: B.67}
i \Gamma_{\mu\nu}^{(\pomeron\pi\pi)} (k',k) 
= -i \, 2 \beta_{\pomeron\pi\pi} F_M [(k'-k)^2] 
\left[ (k'+k)_\mu (k'+k)_\nu - \frac{1}{4} \, g_{\mu\nu} (k'+k)^2 \right] \,,
\ee
\be\label{eq: B.68}
\beta_{\pomeron\pi\pi} = 1.76 \gev^{-1} \,,
\ee
\vspace*{.2cm}
\hspace*{0.5cm}\includegraphics[height=115pt]{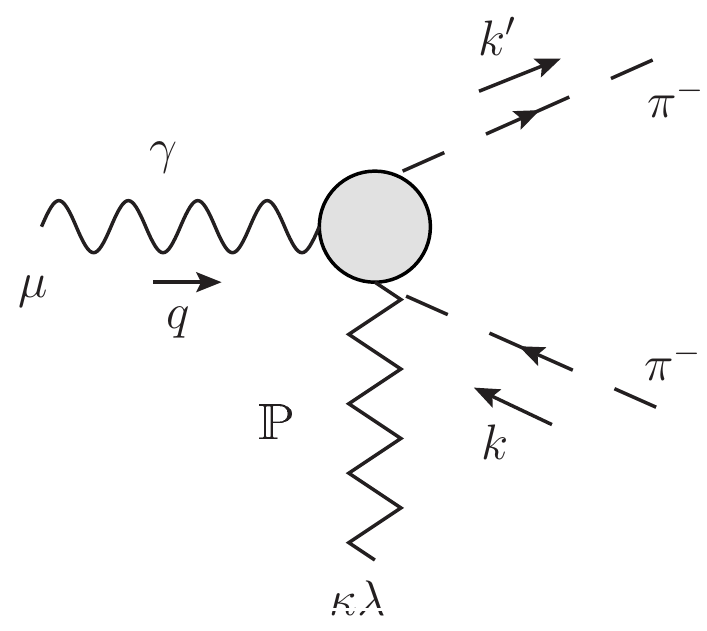} 
\be
\label{eq: B.69}
\begin{split}
i \Gamma_{\mu\kappa\lambda}^{(\pomeron\gamma\pi\pi)} (q,k',k) 
= & - 2\, i\, e \, \beta_{\pomeron\pi\pi} 
\left[ 2 g_{\kappa\mu} (k'+k)_\lambda  + 
	2 g_{\lambda\mu} (k'+k)_\kappa
-	 g_{\kappa\lambda} (k'+k)_\mu 
\right]  \\
& \times F_M [(k'-q-k)^2]\,.
\end{split}
\ee
\newline

\noindent 
$\bullet$ $f_{2R}pp$ 
(see~(3.49), (3.50), and
section~6.3 of~\cite{Ewerz:2013kda})
\newline
\vspace*{.2cm}
\hspace*{0.5cm}\includegraphics[height=85pt]{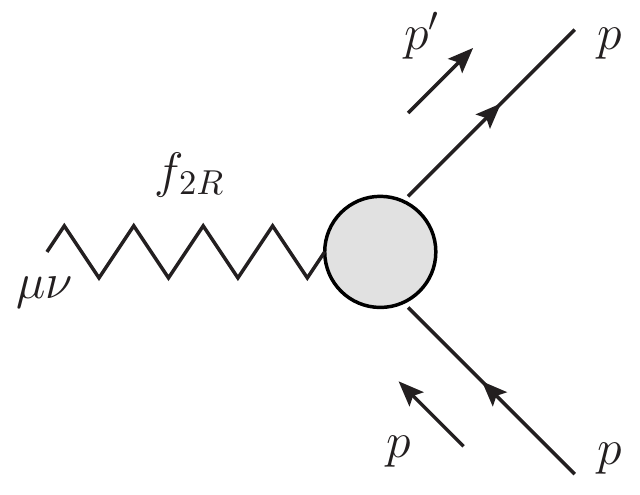} 
\be\label{eq: B.70}
\begin{split}
i \Gamma_{\mu\nu}^{(f_{2R} pp)} (p',p) =&\,
- i g_{f_{2R}pp} \frac{1}{M_0}
F_1[ (p'-p)^2] \\
& \times \left\{ \frac{1}{2} [ \gamma_\mu (p'+p)_\nu + \gamma_\nu (p'+p)_\mu ]
- \frac{1}{4} g_{\mu\nu} (\slash{p}' + \slash{p}) \right\} \,,
\end{split}
\ee
\be\label{eq: B.71}
g_{f_{2R}pp} = 11.04 \,.
\ee
\newline

\noindent 
$\bullet$ $a_{2R}pp$ 
(see~(3.51), (3.52), and
section~6.3 of~\cite{Ewerz:2013kda})
\vspace*{.2cm}
\newline
\hspace*{0.5cm}\includegraphics[height=85pt]{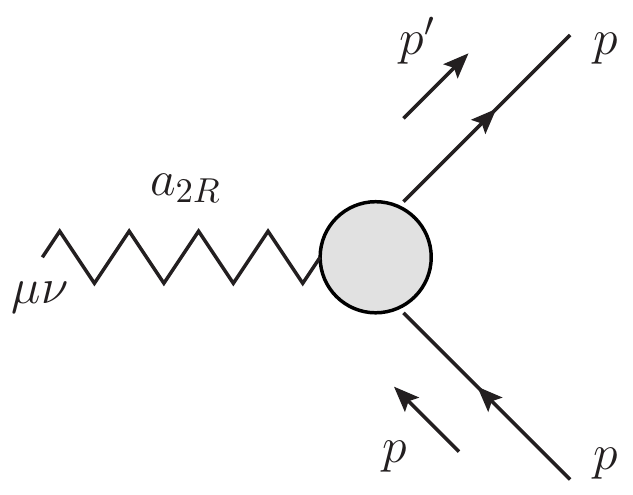} 
\be\label{eq: B.72}
\begin{split}
i \Gamma_{\mu\nu}^{(a_{2R} pp)} (p',p) =&\,
- i g_{a_{2R}pp} \frac{1}{M_0}
F_1[ (p'-p)^2]\\
&
\times \bigg\{ \frac{1}{2} [ \gamma_\mu (p'+p)_\nu + \gamma_\nu (p'+p)_\mu ]
- \frac{1}{4} g_{\mu\nu} (\slash{p}' + \slash{p}) \bigg\}\,,\\
\end{split}
\ee
\be\label{eq: B.73}
g_{a_{2R}pp} = 1.68 \,.
\ee
\newline

\noindent 
$\bullet$ $f_{2R}\pi\pi$, $f_{2R}\gamma\pi\pi$
\newline
The vertex for $f_{2R}\pi\pi$ is given in~(3.53),
(3.54), and section~7.1 of~\cite{Ewerz:2013kda}. 
The inclusion of the coupling to the photon is done
by minimal substitution in complete analogy to~\eqref{eq: B.65}.
We get, therefore,
\vspace*{.2cm}
\newline
\hspace*{0.5cm}\includegraphics[height=85pt]{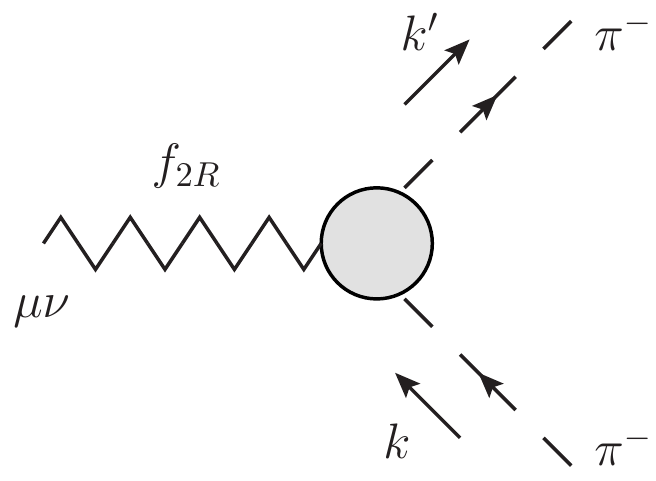} 
\hspace*{0.5cm}\includegraphics[height=85pt]{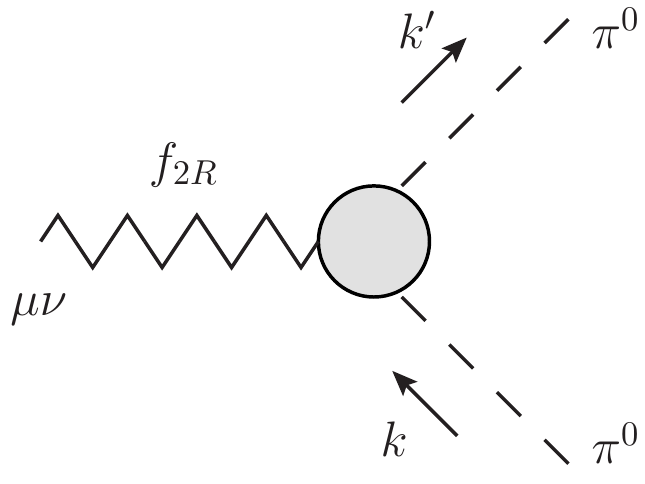} 
\be\label{eq: B.74}
i \Gamma_{\mu\nu}^{(f_{2R} \pi\pi)} (k',k) =
- i \,\frac{g_{f_{2R}\pi\pi}}{2 M_0} F_M[(k'-k)^2]
\left[ (k'+k)_\mu (k'+k)_\nu - \frac{1}{4} g_{\mu\nu} (k'+k)^2 \right]\,,
\ee
\be\label{eq: B.75}
g_{f_{2R}\pi\pi} = 9.30\,,
\ee
\vspace*{.2cm}
\hspace*{0.5cm}\includegraphics[height=115pt]{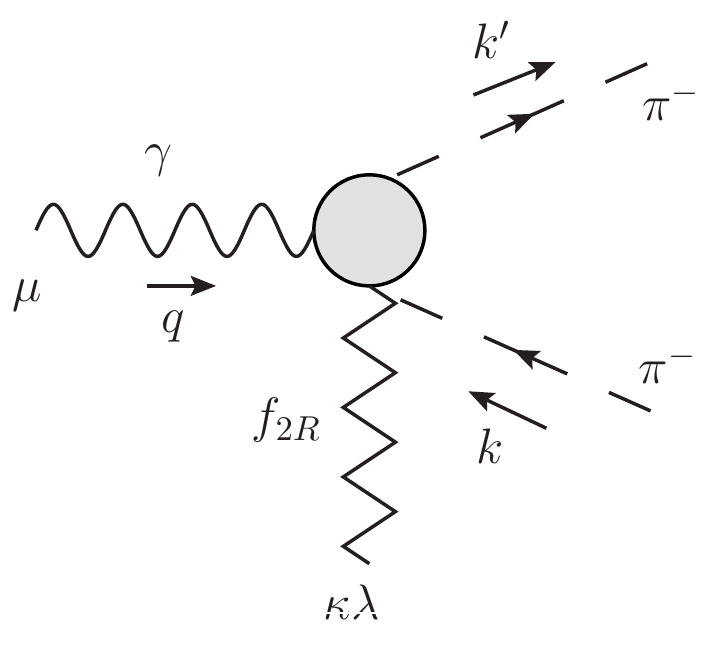} 
\be\label{eq: B.76}
\begin{split}
i \Gamma_{\mu\kappa\lambda}^{(f_{2R}\gamma \pi\pi)} (q,k',k) = & 
- i \,e \frac{g_{f_{2R}\pi\pi}}{2 M_0} F_M[(k'-q-k)^2] \\
& \times \left[ 2g_{\kappa\mu}(k'+k)_\lambda 
+2g_{\lambda\mu}(k'+k)_\kappa
-g_{\kappa\lambda}(k'+k)_\mu
\right]\,.
\end{split}
\ee
\newline

\noindent 
$\bullet$ $\rho_R \pi\pi$, $\rho_R \gamma \pi\pi$
\newline
The vertex for~$\rho_R \pi\pi$ is given in (3.63), (3.64) 
and section~7.1 of~\cite{Ewerz:2013kda}. The vertex for
$\rho_R \gamma \pi\pi$ is obtained from $\rho_R \pi\pi$
by the minimal substitution rule. This gives
\vspace*{.2cm}
\newline
\hspace*{0.5cm}\includegraphics[height=85pt]{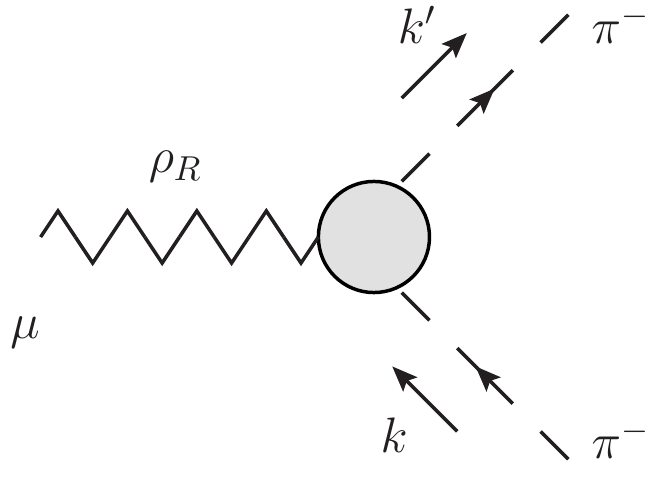} 
\be\label{eq: B.77}
i \Gamma_\mu^{(\rho_R \pi\pi)} (k',k) =
  \frac{i}{2} g_{\rho_R\pi\pi} F_M[ (k'-k)^2] (k'+k)_\mu\,,
\ee
\be\label{eq: B.78}
g_{\rho_R\pi\pi} = 15.63\,,
\ee
\hspace*{0.5cm}\includegraphics[height=115pt]{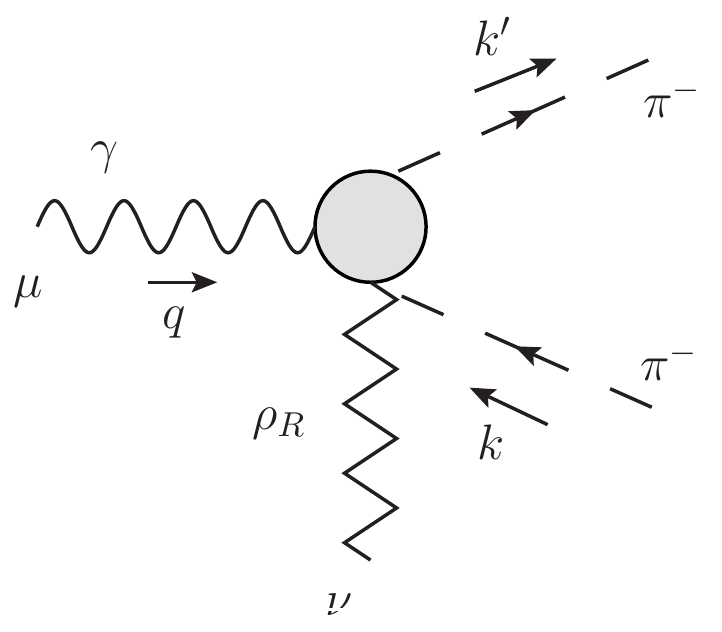} 
\be\label{eq: B.79}
i \Gamma_{\mu\nu}^{(\rho_R \gamma \pi\pi)} (q, k',k) 
= i\,e\, g_{\rho_R\pi\pi} F_M[ (k'-q-k)^2] g_{\mu\nu}\,.
\ee
\newline

\noindent 
$\bullet$ $\pomeron \rho\rho$ 
(see~(3.47), (3.48), and
section~7.2 of~\cite{Ewerz:2013kda})
\vspace*{.2cm}
\newline
\hspace*{0.5cm}\includegraphics[height=85pt]{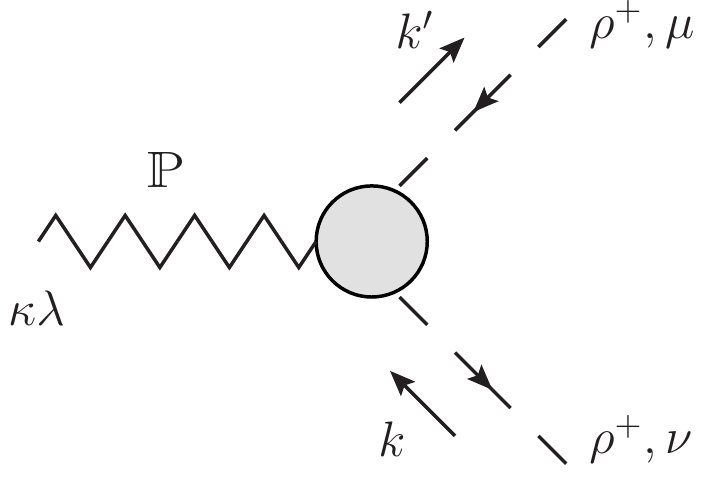} 
\hspace*{0.5cm}\includegraphics[height=85pt]{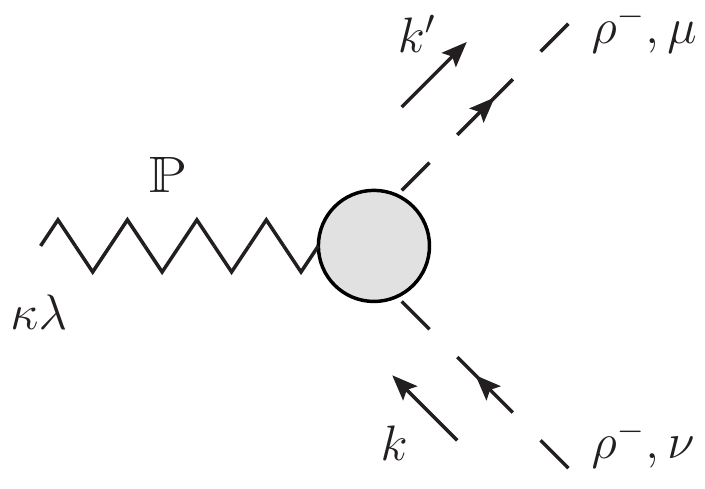} 
\hspace*{0.5cm}\includegraphics[height=85pt]{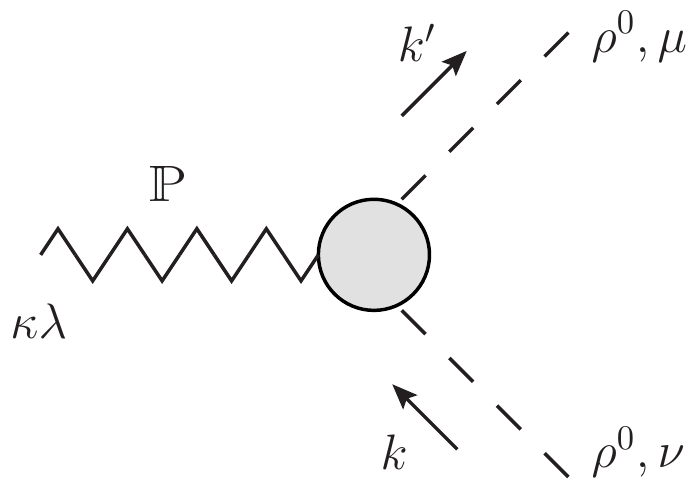} 
\be\label{eq: B.80}
\begin{split}
i \Gamma_{\mu\nu\kappa\lambda}^{(\pomeron\rho\rho)} (k',k) 
= &\,  iF_M[(k'-k)^2] \tilde{F}^{(\rho)}(k'^2) \tilde{F}^{(\rho)}(k^2) \\
& \times \left[2 a_{\pomeron\rho\rho} \Gamma_{\mu\nu\kappa\lambda}^{(0)}(k',-k) 
- b_{\pomeron\rho\rho}\Gamma_{\mu\nu\kappa\lambda}^{(2)}(k',-k) \right] \,,
\end{split}
\ee
\begin{equation}\label{eq: B.81}
\begin{split}
 a_{\pomeron\rho\rho} &\ge 0\,,\\
 2 m^2_\rho a_{\pomeron\rho\rho} + b_{\pomeron\rho\rho} &\ge 0\,.
\end{split}
\end{equation}
In section~7.2 of~\cite{Ewerz:2013kda} arguments for the 
following relation were presented: 
\be\label{eq: B.82}
2 m_\rho^2 a_{\pomeron\rho\rho} + b_{\pomeron\rho\rho} = 
4 \beta_{\pomeron\pi\pi} = 7.04 \gev^{-1} \, \text{.}
\ee
This should be considered as a default relation to
be checked and, if necessary, corrected by experiment.
In~\eqref{eq: B.80} we have introduced form factors~$\tilde{F}^{(\rho)}$
which can take into account that off-shell~$\rho$ mesons need not have
the same coupling to the pomeron as ``on-shell''~$\rho$ mesons.
A convenient form for $\tilde{F}^{(\rho)}$ is
\begin{equation} \label{eq: B.83}
\tilde{F}^{(\rho)}(k^2) = \left[ 1 + \frac{k^2(k^2-m^2_\rho)}{\Lambda^4_\rho} \right]^{-n_\rho}
\end{equation}
with $\Lambda_\rho$ a parameter in the range 2 to 5 GeV and $n_\rho > 0$.
Here we have
\begin{equation} \label{eq: B.84}
\tilde{F}^{(\rho)}(0) = \tilde{F}^{(\rho)}(m^2_\rho) = 1\,.
\end{equation}
\newline

\noindent 
$\bullet$ $\pomeron\omega\omega$ 
\vspace*{.2cm}
\newline
\hspace*{0.5cm}\includegraphics[height=85pt]{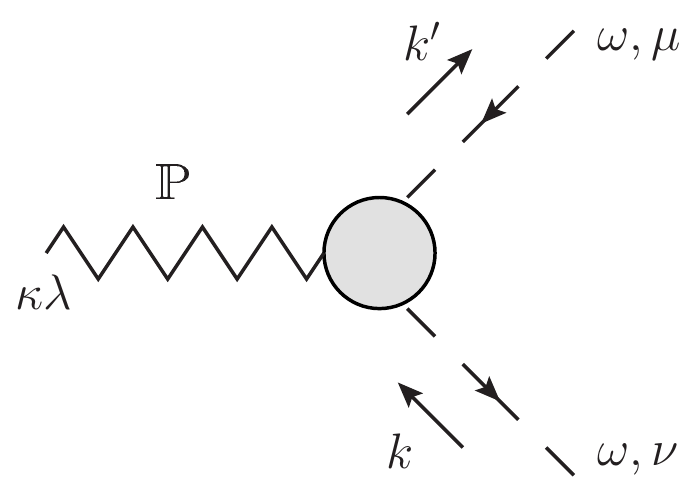}
\be\label{eq: B.85}
\begin{split}
i \Gamma_{\mu\nu\kappa\lambda}^{(\pomeron\omega\omega)} (k',k) 
= & \,i\,F_M[(k'-k)^2] \tilde{F}^{(\omega)}(k'^2) \tilde{F}^{(\omega)}(k^2) \\
& \times \left[2 a_{\pomeron\omega\omega} \Gamma_{\mu\nu\kappa\lambda}^{(0)}(k',-k) 
- b_{\pomeron\omega\omega}\Gamma_{\mu\nu\kappa\lambda}^{(2)}(k',-k) \right] \,.
\end{split}
\ee
As default values we may choose here
\begin{equation}\label{eq: B.86}
\begin{split}
 a_{\pomeron\omega\omega} &= a_{\pomeron\rho\rho}\,,\\
 b_{\pomeron\omega\omega} &= b_{\pomeron\rho\rho}\,,\\
 \tilde{F}^{(\omega)}(k^2) &= \tilde{F}^{(\rho)}(k^2)\,.
\end{split}
\end{equation}
\newline

\noindent 
$\bullet$ $\pomeron\rho'\rho'$ 
\newline
Here our ansatz reads as for $\pomeron\rho\rho$~\eqref{eq: B.80}
with~$\rho$ replaced by $\rho'$.
We have no good estimates for the values of the  
$\pomeron\rho'\rho'$ parameters. Thus, reasonable 
default values are again the $\pomeron\rho\rho$ parameters
\begin{equation}
\label{eq: B.86a}
\begin{split}
 a_{\pomeron\rho'\rho'} &= a_{\pomeron\rho\rho}\,,\\
 b_{\pomeron\rho'\rho'} &= b_{\pomeron\rho\rho}\,,\\
 \tilde{F}^{(\rho')}(k^2) &= \tilde{F}^{(\rho)}(k^2)\,.
\end{split}
\end{equation}
\newline

\noindent 
$\bullet$ $f_{2R}\rho\rho$, $f_{2R}\omega\omega$, $f_{2R}\rho'\rho'$ 
(see~(3.55), (3.56), and
section~7.2 of~\cite{Ewerz:2013kda})
\vspace*{.2cm}
\newline
\hspace*{0.5cm}\includegraphics[height=85pt]{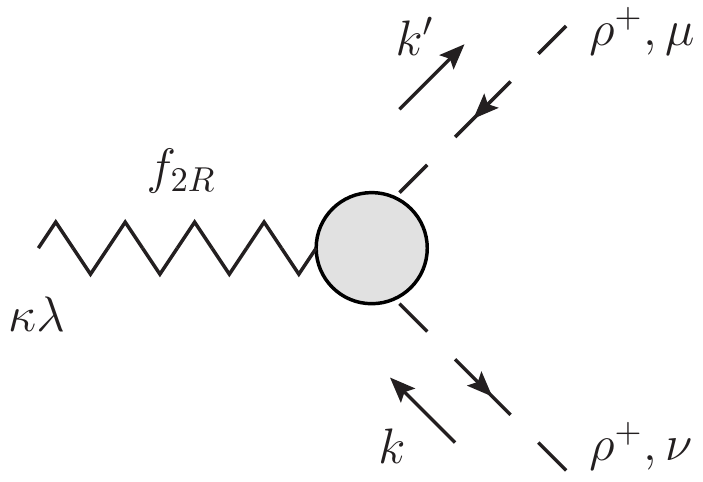} 
\hspace*{0.5cm}\includegraphics[height=85pt]{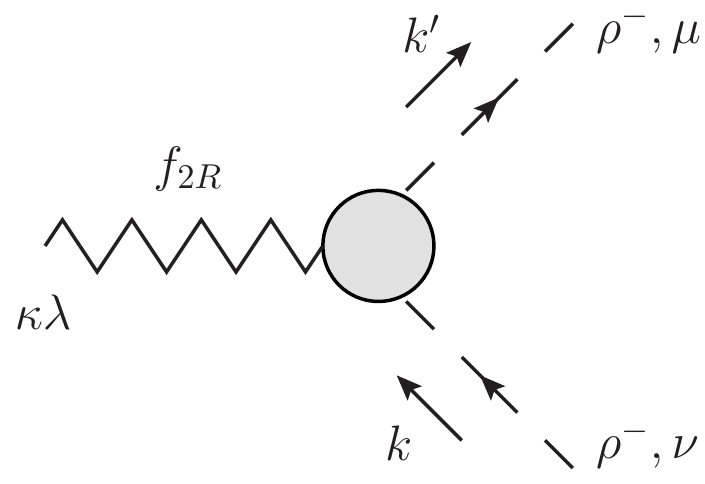} 
\hspace*{0.5cm}\includegraphics[height=85pt]{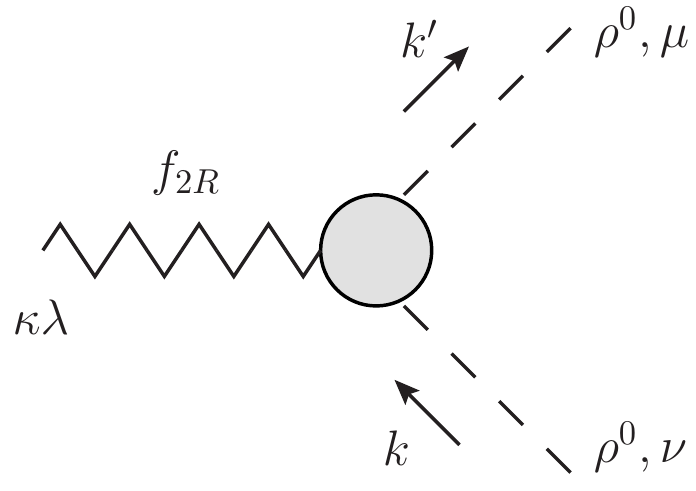} 
\begin{align}\label{eq: B.87}
i \Gamma_{\mu\nu\kappa \lambda}^{(f_{2R} \rho\rho)} (k',k) =& \,
i F_M[(k'-k)^2] \tilde{F}^{(\rho)}(k'^2) \tilde{F}^{(\rho)}(k^2)
\nn \\ 
&\times
\left[ 2 a_{f_{2R}\rho\rho} \Gamma_{\mu\nu\kappa\lambda}^{(0)}(k',-k) -
b_{f_{2R}\rho\rho} \Gamma_{\mu\nu\kappa\lambda}^{(2)}(k',-k) \right]\,,
\end{align}
\be\label{eq: B.88}
a_{f_{2R}\rho\rho} = 2.92\gev^{-3}\,,\qquad
b_{f_{2R}\rho\rho} = 5.02\gev^{-1}\,.
\ee
For the $f_{2R}\omega\omega$ and the $f_{2R}\rho'\rho'$ couplings
our ansatz is as for $f_{2R}\rho\rho$ but with parameters 
$a_{f_{2R}\omega\omega}$, 
$b_{f_{2R}\omega\omega}$ and
$a_{f_{2R}\rho'\rho'}$, 
$b_{f_{2R}\rho'\rho'}$, respectively. For the~$\omega$ we have, of course,
only the analogue of the rightmost of the above diagrams. For lack of
other information we set as default
\begin{equation}
\label{eq: B.89}
\begin{split}
 a_{f_{2R}\omega\omega} &= a_{f_{2R}\rho'\rho'} = a_{f_{2R}\rho\rho}\,,\\
 b_{f_{2R}\omega\omega} &= b_{f_{2R}\rho'\rho'} = b_{f_{2R}\rho\rho}\,.
\end{split}
\end{equation}
\newline

\noindent 
$\bullet$ $a_{2R} \omega \rho$ 
(see~(3.57), (3.58), and
section~7.2 of~\cite{Ewerz:2013kda})
\vspace*{.2cm}
\newline
\hspace*{0.5cm}\includegraphics[height=85pt]{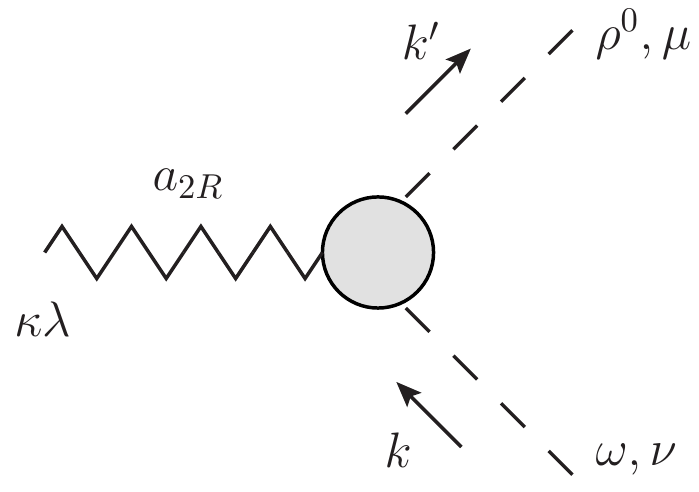} 
\begin{align}\label{eq: B.91}
i \Gamma_{\mu\nu\kappa\lambda}^{(a_{2R} \omega\rho)} (k',k) =&\,
i F_M[(k'-k)^2] \tilde{F}^{(\rho)}(k'^2) \tilde{F}^{(\omega)}(k^2) 
\nn\\ 
&\times
\left[ 2 a_{a_{2R}\omega\rho} \Gamma_{\mu\nu\kappa\lambda}^{(0)}(k',-k) -
b_{a_{2R}\omega\rho} \Gamma_{\mu\nu\kappa\lambda}^{(2)}(k',-k) \right]\,,
\end{align}
\begin{equation}\label{eq: B.92}
\left| a_{a_{2R}\omega\rho} \right|= 2.56\gev^{-3}\,,\qquad
\left| b_{a_{2R}\omega\rho} \right| = 4.68\gev^{-1}\,.
\end{equation}
\newline

\noindent 
$\bullet$ $\omega_R pp$ 
(see~(3.59), (3.60), and
section~6.3 of~\cite{Ewerz:2013kda})
\vspace*{.2cm}
\newline
\hspace*{0.5cm}\includegraphics[height=85pt]{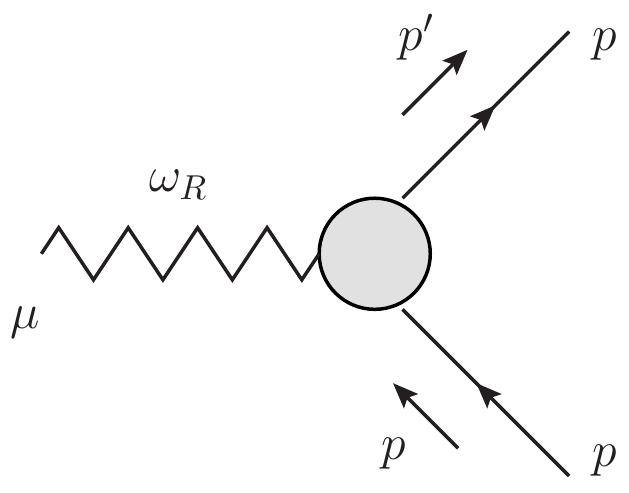} 
\be\label{eq: B.93}
i \Gamma_\mu^{(\omega_R pp)} (p',p) = - i g_{\omega_Rpp} F_1[ (p'-p)^2] \gamma_\mu\,,
\ee
\be\label{eq: B.94}
g_{\omega_Rpp} = 8.65\,.
\ee
\newline

\noindent 
$\bullet$ $\rho_R pp$ 
(see (3.61), (3.62), and
section~6.3 of~\cite{Ewerz:2013kda})
\vspace*{.2cm}
\newline
\hspace*{0.5cm}\includegraphics[height=85pt]{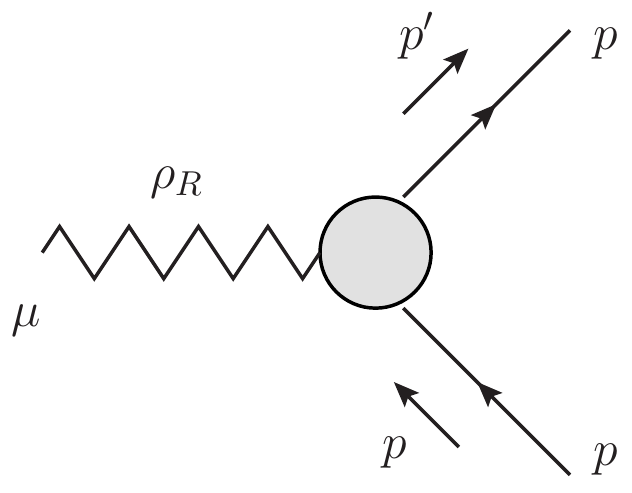} 
\be\label{eq: B.95}
i \Gamma_\mu^{(\rho_R pp)} (p',p) = - i g_{\rho_Rpp} F_1[ (p'-p)^2] \gamma_\mu\,,
\ee
\be\label{eq: B.96}
g_{\rho_Rpp} = 2.02\,.
\ee
\newline

\noindent 
$\bullet$ $\rho_R\rho f_2$ 
(see~(3.66), (3.67), and
section~7.2 of~\cite{Ewerz:2013kda})
\vspace*{.2cm}
\newline
\hspace*{0.5cm}\includegraphics[height=85pt]{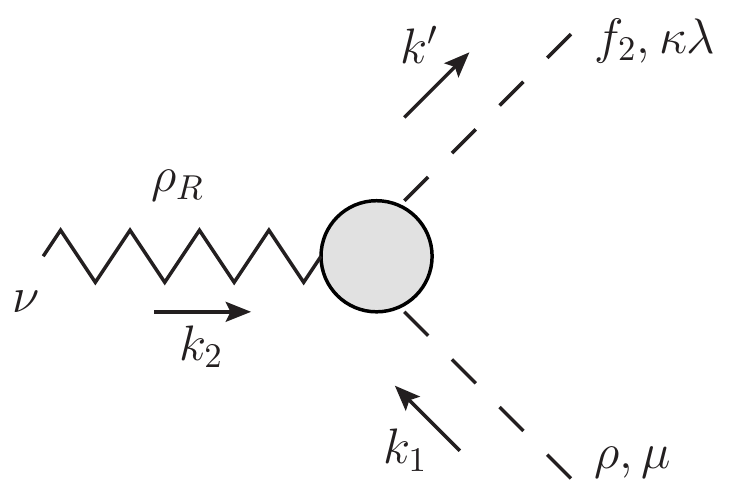} 
\begin{align}\label{eq: B.97}
i \Gamma_{\mu\nu\kappa\lambda}^{(\rho_R \rho f_2)} (k',k_1) =&\,
i F_M(k_2^2) F^{(f_2\pi\pi)}(k'^2) \tilde{F}^{(\rho)}(k_1^2)
\nn \\ 
& \times
\left[ 2 a_{\rho_R\rho f_2} \Gamma_{\mu\nu\kappa\lambda}^{(0)}(k_1,k_2) -
b_{\rho_R\rho f_2} \Gamma_{\mu\nu\kappa\lambda}^{(2)}(k_1,k_2) \right] \,,
\end{align}
where $k'=k_1 + k_2$ and 
\be\label{eq: B.98}
a_{\rho_R\rho f_2} = 2.92\gev^{-3}\,, \qquad
b_{\rho_R\rho f_2} = 5.02\gev^{-1}\,. 
\ee
\newline

\noindent 
$\bullet$ $\omega_R\omega f_2$ 
(see~(3.65) and
section~7.2 of~\cite{Ewerz:2013kda})
\vspace*{.2cm}
\newline
\hspace*{0.5cm}\includegraphics[height=85pt]{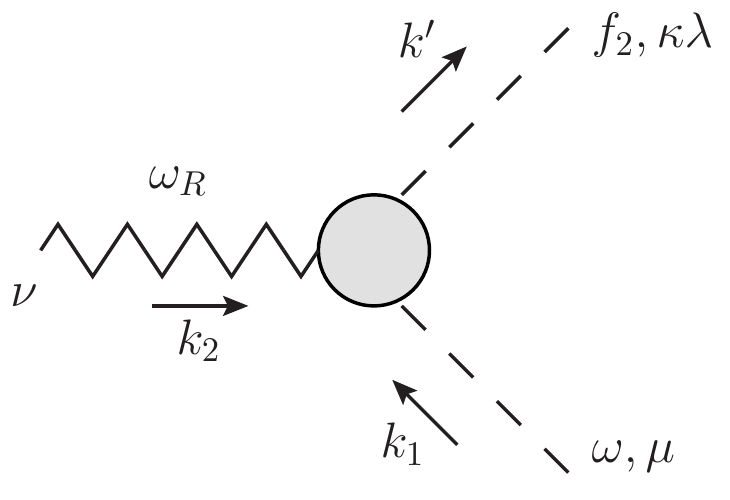} 
\begin{align}\label{eq: B.99}
i \Gamma_{\mu\nu\kappa\lambda}^{(\omega_R \omega f_2)} (k',k_1) = &\,
i F_M(k_2^2)  F^{(f_2\pi\pi)}(k'^2) \tilde{F}^{(\omega)}(k_1^2)
\nn\\ 
&\times
 \left[ 2 a_{\omega_R\omega f_2} \Gamma_{\mu\nu\kappa\lambda}^{(0)}(k_1,k_2) -
b_{\omega_R\omega f_2} \Gamma_{\mu\nu\kappa\lambda}^{(2)}(k_1,k_2) \right] 
\end{align}
with $k'=k_1 + k_2$\,. As default values we choose according to
(7.31) to (7.36) of~\cite{Ewerz:2013kda}
\begin{equation}\label{eq: B.100}
\begin{split}
a_{\omega_R\omega f_2} &= a_{\rho_R \rho f_2} = 2.92\gev^{-3}\,,\\
b_{\omega_R\omega f_2} &= b_{\rho_R \rho f_2} = 5.02\gev^{-1}\,.
\end{split}
\end{equation}
\newline

\noindent 
$\bullet$ $\odderon pp$ 
(see~(3.68), (3.69), and
section~6.2 of~\cite{Ewerz:2013kda})
\vspace*{.2cm}
\newline
\hspace*{0.5cm}\includegraphics[height=85pt]{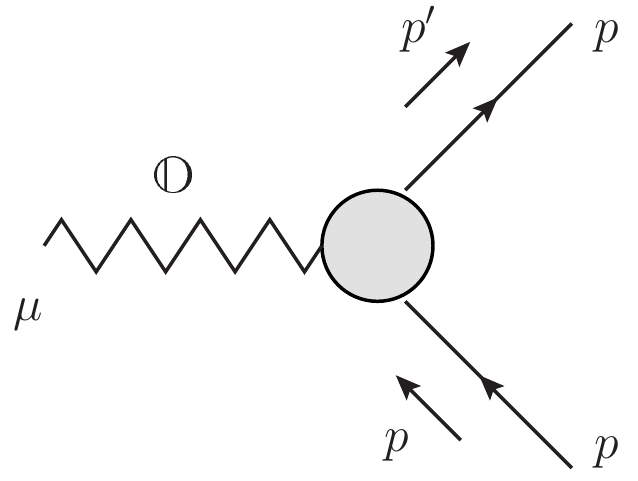} 
\be\label{eq: B.101}
i \Gamma_\mu^{(\odderon pp)} (p',p) 
=- i \,3\, \beta_{\odderon pp} M_0 F_1[ (p'-p)^2] \gamma_\mu\,.
\ee
The coupling parameter $\beta_{\odderon pp}$ has dimension $\gev^{-1}$. 
\newline

\noindent
$\bullet$ $\odderon\gamma f_2$ 
(see~(3.70), (3.71), and
section~6.2 of~\cite{Ewerz:2013kda})
\vspace*{.2cm}
\newline
\hspace*{0.5cm}\includegraphics[height=85pt]{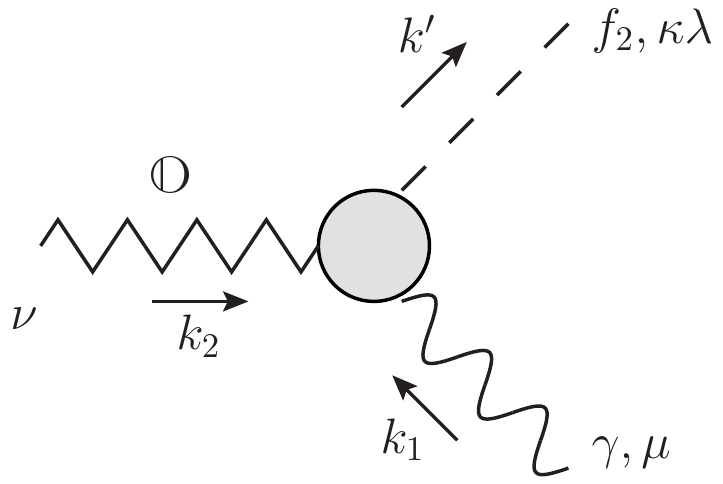} 
\begin{align}\label{eq: B.103}
i \Gamma_{\mu\nu\kappa\lambda}^{(\odderon \gamma f_2)} (k',k_1) =&\,
i F_M(k_1^2) F_M(k_2^2) F^{(f_2\pi\pi)}(k'^2)
\nn \\ 
&\times
\left[ 2 a_{\odderon \gamma f_2} \Gamma_{\mu\nu\kappa\lambda}^{(0)}(k_1,k_2) -
b_{\odderon \gamma f_2} \Gamma_{\mu\nu\kappa\lambda}^{(2)}(k_1,k_2) \right]
\end{align}
with $k'=k_1 + k_2$ and 
\be\label{eq: B.104}
a_{\odderon \gamma f_2} = e \hat{a}_{\odderon \gamma f_2}\,, \qquad
b_{\odderon \gamma f_2} = e \hat{b}_{\odderon \gamma f_2}\,.
\ee
Here $\hat{a}_{\odderon \gamma f_2}$ and $\hat{b}_{\odderon \gamma f_2}$ are
(unknown) odderon coupling parameters of hadronic scale 
with dimensions $\gev^{-3}$ and $\gev^{-1}$, respectively. 
\newline

In table~\ref{tab: B.1} we list all parameters of the model, the corresponding equations
and references. For many parameters the default values given are rather well constrained
from other sources as indicated. For parameters where there is an asterisk $*$ in the last column
we have no good estimates. The default values quoted there are just educated
guesses used to produce plots for the cross sections and asymmetries shown in
section~\ref{sec: Results}. 

\clearpage 
\begin{longtable}{cccccc}
%\centering
%\begin{tabular}
\toprule
 parameters & see eqs. & default value & constraint 	& ref. &\\
 \midrule
$m_\rho$ & \eqref{B.3}-\eqref{eq: B.18} & $775.26\mev$ & $775.26\pm0.25\mev$ & \cite{Beringer:1900zz}& \\
$m_\omega$ & \eqref{B.3}-\eqref{eq: B.18} & $782.65\mev$ & $782.65\pm0.12\mev$ & \cite{Beringer:1900zz}& \\
$\Gamma_\omega$ & \eqref{B.3}-\eqref{eq: B.18} & $8.49\mev$ & $8.49\pm0.08 \mev$ & \cite{Beringer:1900zz}& \\
$b_{\rho\omega}$ & \eqref{eq: B.17} & $3.5\times 10^{-3}$ & $(3.5\pm 0.5)\times 10^{-3}$ & \cite{Melikhov:2003hs}& \\
$g_{\rho\pi\pi}$ & \eqref{eq: B.17}, \eqref{eq: B.55} & $11.51$ & $11.51\pm 0.07$ & \cite{Melikhov:2003hs}& \\
$g_{\omega\pi\pi}$ & \eqref{eq: B.17}, \eqref{eq: B.56} & $-0.35$ & $-0.35\pm 0.10$ & \cite{Melikhov:2003hs}& \\
$m_{\rho'}$ & \eqref{eq: B.21} & $1465 \mev$ & $1465\pm 25 \mev$ & \cite{Beringer:1900zz}& \\
$\Gamma_{\rho'}$ & \eqref{eq: B.21} & $400 \mev$ & $400\pm 60 \mev$ & \cite{Beringer:1900zz}& \\
$g_{\rho'\pi\pi}$ & \eqref{eq: B.54} & $0.5$ & & &$*$\\
$m_{f_2}$ & \eqref{eq: B.28} & $1275.1 \mev$ & $1275.1\pm 1.2 \mev$ & \cite{Beringer:1900zz}& \\
$\Gamma_{f_2}$ & \eqref{eq: B.28} & $185.1 \mev$ & $185.1^{+2.9}_{-2.4} \mev$ & \cite{Beringer:1900zz}& \\
$\frac{\Gamma(f_2\rightarrow \pi\pi)}{\Gamma_{f_2}}$ & \eqref{eq: B.29}  & $84.8\%$ & $(84.8^{+2.4}_{-1.2})\%$& \cite{Beringer:1900zz}& \\
$\epsilon_\pomeron$ & \eqref{eq: B.31} & 0.0808 & 0.0808 & \cite{Ewerz:2013kda}& \\
$\alpha'_\pomeron$ & \eqref{eq: B.31} & $0.25\gev^{-2}$ & $0.25\gev^{-2}$ & \cite{Ewerz:2013kda}& \\
$\alpha_{\reggeon_+}(0)$ & \eqref{eq: B.33} & $0.5475$ & $0.5475$ & \cite{Ewerz:2013kda}& \\
$\alpha'_{\reggeon_+}$ & \eqref{eq: B.33} & $0.9\gev^{-2}$ & $0.9\gev^{-2}$ & \cite{Ewerz:2013kda}& \\
$\alpha_{\reggeon_-}(0)$ & \eqref{eq: B.35} & $0.5475$ & $0.5475$ & \cite{Ewerz:2013kda}& \\
$\alpha'_{\reggeon_-}$ & \eqref{eq: B.35} & $0.9\gev^{-2}$ & $0.9\gev^{-2}$ & \cite{Ewerz:2013kda}& \\
$M_-$ & \eqref{eq: B.35} & $1.41\gev$ & $1.41\gev$ & \cite{Ewerz:2013kda}& \\
$\eta_\odderon$ & \eqref{eq: B.37} & $-1$ & $\pm 1$ & \cite{Ewerz:2013kda}& $*$\\
$\epsilon_\odderon$ & \eqref{eq: B.37} & $0$ & $\epsilon_\odderon \le \epsilon_\pomeron$& \cite{Ewerz:2013kda}& $*$\\
$\alpha'_\odderon$ & \eqref{eq: B.37} & $0.25$ & & \cite{Ewerz:2013kda}& $*$\\
$\gamma_\rho$ & \eqref{eq: B.45 d} & $\big[\frac{1}{4\pi}0.496\big]^{-1/2}$& $\big[\frac{1}{4\pi}(0.496\pm 0.023)\big]^{-1/2}$& \cite{Ewerz:2013kda}& \\
$\gamma_\omega$ & \eqref{eq: B.45 d} & $\big[\frac{1}{4\pi}0.042\big]^{-1/2}$& $\big[\frac{1}{4\pi}(0.042\pm 0.0015)\big]^{-1/2}$& \cite{Ewerz:2013kda}& \\
$\gamma_\phi$ & \eqref{eq: B.45 d} & $-\big[\frac{1}{4\pi}0.0716\big]^{-1/2}$& $-\big[\frac{1}{4\pi}(0.0716\pm 0.0017)\big]^{-1/2}$& \cite{Ewerz:2013kda}& \\
$\gamma_{\rho'}$ & \eqref{eq: B.45 d} & $\gamma_\rho$ & & & $*$\\
$\frac{\mu_p}{\mu_N}$ &   \eqref{eq: B.51} & $2.7928$ & $2.7928$ & \cite{Ewerz:2013kda}& \\
$m_D^2$  &  \eqref{eq: B.52} & $0.71 \gev^2$ & $0.71 \gev^2$& \cite{Ewerz:2013kda}& \\
$m^2_0$ & \eqref{eq: B.53} & $0.50\gev^2$ & $0.50\gev^2$ & \cite{Ewerz:2013kda}& \\
$g_{f_2\pi\pi}$ & \eqref{eq: B.58} & $9.26$ & $9.26 \pm 0.15$ & \cite{Ewerz:2013kda}& \\
$\Lambda_{f_2}$ & \eqref{eq: B.59} & $1.8 \gev$ & $1-4 \gev$ & & $*$\\
$a_{f_2\gamma\gamma}$ & \eqref{eq: B.61} & $\frac{e^2}{4\pi} 1.45\gev^{-3}$ & $\frac{e^2}{4\pi} 1.45\gev^{-3}$ & \cite{Ewerz:2013kda}& \\
$b_{f_2\gamma\gamma}$ & \eqref{eq: B.61} & $\frac{e^2}{4\pi} 2.49\gev^{-1}$ & $\frac{e^2}{4\pi} 2.49\gev^{-1}$ & \cite{Ewerz:2013kda}& \\
$\beta_{\pomeron NN}$ & \eqref{eq: B.64} & $1.87\gev^{-1}$ & $1.87\gev^{-1}$ & \cite{Ewerz:2013kda}& \\
$\beta_{\pomeron\pi\pi}$ & \eqref{eq: B.68} & $1.76\gev^{-1}$ & $1.76\gev^{-1}$ & \cite{Ewerz:2013kda}& \\
$g_{f_{2R} p p}$ & \eqref{eq: B.71} & $11.04$ & $11.04$ & \cite{Ewerz:2013kda}& \\
$g_{a_{2R} p p}$ & \eqref{eq: B.73} & $1.68$ & $1.68$ & \cite{Ewerz:2013kda}& \\
$g_{f_{2R} \pi\pi}$ & \eqref{eq: B.75} & $9.30$ & $9.30$ & \cite{Ewerz:2013kda}& \\
$g_{\rho_R \pi\pi}$ & \eqref{eq: B.78} & $15.63$ & $15.63$ & \cite{Ewerz:2013kda}& \\
$a_{\pomeron \rho\rho}$ & \eqref{eq: B.80}, \eqref{eq: B.81} &  $0.45\gev^{-3}$ & & \cite{Ewerz:2013kda}& $*$\\
$b_{\pomeron \rho\rho}$ & \eqref{eq: B.80}, \eqref{eq: B.81} & $6.5 \gev^{-1}$ &  & \cite{Ewerz:2013kda}& $*$\\
$2 m^2_\rho a_{\pomeron \rho\rho}+b_{\pomeron \rho\rho}$ & \eqref{eq: B.82} & $7.04\gev^{-1}$ & $7.04\gev^{-1}$  & \cite{Ewerz:2013kda}& \\
$\Lambda_\rho$ & \eqref{eq: B.83} & $2 \gev$ & $2-5\gev$  & & $*$\\
$n_\rho$ & \eqref{eq: B.83} & $0.4$ & $n_\rho > 0$ & & $*$\\
$\Lambda_\omega$ & \eqref{eq: B.86} & $2 \gev$ & $2-5\gev$  & & $*$\\
$n_\omega$ & \eqref{eq: B.86} & $0.4$ & $n_\omega > 0$ & & $*$\\
$\Lambda_{\rho'}$ & \eqref{eq: B.86a} & $2 \gev$ & $2-5\gev$  & & $*$\\
$n_{\rho'}$ & \eqref{eq: B.86a} & $0.4$ & $n_{\rho'} > 0$ & & $*$\\
$a_{\pomeron \omega\omega}$ & \eqref{eq: B.86}& $0.45\gev^{-3}$ & & & $*$\\
$b_{\pomeron \omega\omega}$ & \eqref{eq: B.86}& $6.5\gev^{-1}$ & & & $*$\\
$a_{\pomeron \rho'\rho'}$ & \eqref{eq: B.86a}&  $0.45\gev^{-3}$ & & & $*$\\
$b_{\pomeron \rho'\rho'}$ & \eqref{eq: B.86a}& $6.5\gev^{-1}$ & & & $*$\\
$a_{f_{2R} \rho\rho}$ & \eqref{eq: B.88} & $2.92\gev^{-3}$& $2.92\gev^{-3}$ & \cite{Ewerz:2013kda}& \\
$b_{f_{2R} \rho\rho}$ & \eqref{eq: B.88} & $5.02\gev^{-1}$ & $5.02\gev^{-1}$ & \cite{Ewerz:2013kda}& \\
$a_{f_{2R} \omega\omega}$ & \eqref{eq: B.89}& $2.92\gev^{-3}$  & & & $*$\\
$b_{f_{2R} \omega\omega}$ & \eqref{eq: B.89}& $5.02\gev^{-1}$ & & & $*$\\
$a_{f_{2R} \rho'\rho'}$ & \eqref{eq: B.89}& $2.92\gev^{-3}$ & & & $*$\\
$b_{f_{2R} \rho'\rho'}$ & \eqref{eq: B.89}& $5.02\gev^{-1}$ & & & $*$\\
$a_{a_{2R} \omega\rho}$ & \eqref{eq: B.92} & $+2.56 \gev^{-3}$ & $\pm 2.56\gev^{-3}$ & \cite{Ewerz:2013kda}& \\
$b_{a_{2R} \omega\rho}$ & \eqref{eq: B.92} & $+4.68 \gev^{-1}$& $\pm 4.68\gev^{-1}$ & \cite{Ewerz:2013kda}& \\
$g_{\omega_R pp}$ & \eqref{eq: B.94} & $8.65$ & $8.65$ & \cite{Ewerz:2013kda}& \\
$g_{\rho_R pp}$ & \eqref{eq: B.96} & $2.02$ & $2.02$ & \cite{Ewerz:2013kda}& \\
$a_{\rho_{R} \rho f_2}$ & \eqref{eq: B.98} & $2.92\gev^{-3}$ & $2.92\gev^{-3}$ & \cite{Ewerz:2013kda}& \\
$b_{\rho_{R} \rho f_2}$ & \eqref{eq: B.98} & $5.02\gev^{-1}$ & $5.02\gev^{-1}$ & \cite{Ewerz:2013kda}& \\
$a_{\omega_R \omega f_2}$ & \eqref{eq: B.100} & $2.92\gev^{-3}$ & $2.92\gev^{-3}$ & \cite{Ewerz:2013kda}& \\
$b_{\omega_R \omega f_2}$ & \eqref{eq: B.100} & $5.02\gev^{-1}$ & $5.02\gev^{-1}$ & \cite{Ewerz:2013kda}& \\
$\beta_{\odderon pp}$ & \eqref{eq: B.101} & $0.18\gev^{-1}$ & & \cite{Ewerz:2013kda}& $*$\\
$\hat{a}_{\odderon\gamma f_2}$ & \eqref{eq: B.103}, \eqref{eq: B.104} & $0.5\gev^{-3}$  & & \cite{Ewerz:2013kda}& $*$\\
$\hat{b}_{\odderon\gamma f_2}$ & \eqref{eq: B.103}, \eqref{eq: B.104} & $0.5\gev^{-1}$ & & \cite{Ewerz:2013kda}& $*$\\
$M_0$ &            & $\equiv 1 \gev$ &      & & \\
\bottomrule 
%\vspace*{.2cm}
%\end{tabular}
%\captionsetup{width=\textwidth}
\caption{The parameters and their default values for the model.
An asterisk $*$ in the last column  means that we have no good estimate in this
case.
For most of the parameters constraints from experiment or theory are
available, for all other parameters example values are chosen. 
The listed default values are used 
to produce  the cross section and asymmetry plots
in section~\ref{sec: Results}.
The references given are only
meant to indicate where a complete discussion of the parameter in question,
with appropriate references, can be found.
The parameter $M_0\equiv 1 \gev$ is introduced for dimensional
reasons in various equations.
\label{tab: B.1}
}
\end{longtable}

\section{Behaviour of {\boldmath $C=-1$\unboldmath} exchanges for {\boldmath $t\rightarrow 0$ \unboldmath}}
\label{app C}

Consider the reaction~\eqref{eq: 1.1} in the $\pipi$ rest frame using the proton-Jackson
system; see appendix~\ref{app A}, figure~\ref{fig: AppendixA.1}. We have then 
from~\eqref{eq: A.8} 
\begin{align}
\label{eq: 3.5}
p^0 &=   \frac{1}{\mpipi} \left( p \cdot k \right) 
= \frac{1}{2 \mpipi} (s - m_p^2 +t) \, \text{,} \nonumber \\
p'^0 &=   \frac{1}{\mpipi} \left( p' \cdot k \right) 
= \frac{1}{2 \mpipi} (s - m_p^2 - \mpipi^2) \, \text{,} \nonumber \\
p^0  - p'^0&=   \frac{1}{2 \mpipi}  (\mpipi^2 +t) \, \text{.} 
\end{align}
For $s=\Wgp^2 \gg \mpipi^2, m_p^2, |t|$ we get,
neglecting terms of relative order $\mpipi^2 / s$, $m_p^2 / s$, $|t| / s$, 
the following: 
\begin{align}
\label{eq: 3.6}
p + p' &=  
\left(
\begin{array}{c}
p^0 + p'^0 \\
\mathbf{p'_{\bot}}  \\
p^0 + p'^0 
\end{array} \right) \, \text{,} \nonumber \\
p - p' &=  
\left(
\begin{array}{c}
p^0 - p'^0 \\
-\mathbf{p'_{\bot}}  \\
p^0 - p'^0 
\end{array} \right) \, \text{,}  \\
p + p' &\cong   \: \frac{p^0 + p'^0}{p^0 - p'^0} \left(p-p'\right) + \frac{2 p^0}{p^0 - p'^0} 
\left(
\begin{array}{c}
0 \\
\mathbf{p'_{\bot}}  \\
0 
\end{array} \right) \, \text{.}\label{eq: 3.7}
\end{align}
Now the $\gamma p p$ vertex in the diagram for the Primakoff contribution, see
figure~\ref{fig: 1}(c), gives at high energies a factor $(p + p')$, the emitted virtual
photon has momentum $p-p'$ and it is nearly real for $|t| \rightarrow 0$.
Gauge invariance at the $f_2 \gamma \gamma$ vertex implies then that the first
term on the r.\,h.\,s. of~\eqref{eq: 3.7} does not contribute. Thus, for $|t| \rightarrow 0$
the nearly real photon has effectively transverse linear polarisation due to the second term
on the r.\,h.\,s. of~\eqref{eq: 3.7}. From this argument we expect to get a factor
$\mathbf{p'_{\bot}}$ in the amplitude and a factor 
$|\mathbf{p'_{\bot}}|^2 \cong |t|$ in the cross section. For the Primakoff 
effect we get also a factor $1 / |t|^2$ from the photon propagator leaving us with a 
behaviour of $d \sigma / d t \propto 1 / |t|$ for small $|t|$. 
The reggeon and the odderon contributions from figures~\ref{fig: 1}(b) and~\ref{fig: 1}(d),
respectively, behave as $d \sigma / d t \propto |t|$ for $|t| \rightarrow 0$ since 
the effective propagators have no singularity there.  

\section{Determination of the Monte Carlo weights}
\label{app D}

In this appendix relevant formulae for the weight determination in the event generator are listed.
The fivefold differential cross section of the reaction~\eqref{eq: 1.1} is given by
the product of a normalisation term, the spin-sum of all matrix elements discussed
above and
the three-body differential phase space element $\mathrm{d} \phi_3$,
\begin{align}
  \label{eq: D1}
 d\sigma^{\gamma p} &=  
 \frac{1}{4}\frac{1}{2 (s - m_p^2)} (\hbar c)^2 
 \left((-1)\sum_{\sIndPrim, \sInd} \mathcal{M}_{\mu, \sIndPrim, \sInd}^* \, \mathcal{M}_{\sIndPrim, \sInd}^{\mu}\right)
 d \phi_3 \, \text{,} \\
 \label{eq: D2}
 d \phi_3 &=  \frac{1}{(2 \pi) ^ 5}
 \frac{d^3k_1}{2 k_1^0}
 \frac{d^3k_2}{2 k_2^0}
 \frac{d^3p'}{2 p'^0} \,
 \delta^{(4)} (k_1 + k_2 + p' - p - q) 
  \,,
\end{align}
where the conversion constant $(\hbar c)^2$ is written out explicitly.

In this appendix we use the notation 
\begin{align}
\label{eq:D2a}
m_{\pi\pi} \equiv m_{\pi^+\pi^-} \,, \qquad m_q=0 \,,\qquad t_2= (q-k_2)^2 \,.
\end{align}

For our model it is convenient to express the three-body phase space element as
\begin{align}
\label{eq: D3}
d\phi_3 =  \frac{1}{(2 \pi)^5}\, \frac{1}{8} \, \frac{1}{2 m_{\pi \pi}} 
\, \frac{
\lambda^{1/2}(m_{\pi \pi}^2, m_{\pi}^2, m_{\pi}^2)
} {
\lambda^{1/2}(s, m_{p}^2, m_{q}^2)
 }
 \, dm_{\pi\pi} \, dt \, d\varphi_p \, 
d\cos\vartheta_{\pi_2} \, d\varphi_{\pi_2}
 \end{align}
with 
\begin{equation}
 \label{eq: D4}
 \lambda(x, y, z) =  x^2 - 2(y+z)x + (y-z)^2 \,,
\end{equation}
and with $\varphi_p$ being the azimuthal angle of the outgoing proton in the $\gamma p$ rest-frame, and
$\vartheta_{\pi_2}$ and  $\varphi_{\pi_2}$ being the angles of one $\pi$ in the $\pi^+ \pi^-$ rest-frame.
Furthermore the conditions for the physically accessible region must be taken 
into account, compare with \cite{Byckling}, p.\ 89 and p.\ 131,
\begin{align}
 \label{eq: D5}
G(s, t, m_{\pi\pi}^2, m_p^2, m_q^2, m_{p'}^2) & \le  0  \qquad \text{ and} 
\nn \\
G(m_{\pi\pi}^2, t_2, m_\pi^2, t, m_p^2, m_\pi^2) & \le  0  
\end{align}
with 
\begin{align}
\label{eq: D6}
G(x,y,z,u,v,w) = & \, x^2 y + x y^2 + z^2u + z u^2 + v^2 w + v w^2 + x z w + x u v \nonumber \\
& + y z w + y u w - x y ( z + u + v + w ) \nonumber \\
& - z u ( x + y + v + w ) - v w ( x + y + z + u)
 \,. 
\end{align}
The $\gamma p$ cross section in the ranges 
$\Delta m_{\pi\pi}=[m_{\pi\pi, \text{ min}}, m_{\pi\pi, \text{ max}}]$ 
and $\Delta t = [t_{\text{min}}, t_{\text{max}}]$ is obtained by integrating 
over $m_{\pi\pi}$, $t$ and all angles:
\begin{align}
\label{eq: D7}
\sigma^{\gamma p} (\Delta m_{\pi\pi}, \Delta t) =
  \underset{\Delta m_{\pi\pi}}{\int} \, \underset{\Delta t}{\int} \;\, 
\underset{\text{angles}}{\int}  & d \sigma^{\gamma p} \\
   = \underset{\Delta m_{\pi\pi}}{\int} \, \underset{\Delta t}{\int} \;\, \underset{\text{angles}}{\int} 
& \frac{1}{4}\frac{1}{2 (s - m_p^2)} (\hbar c)^2 \nonumber \\
& \times\left| \mathcal{M}(m_{\pi \pi}, t, \cos\vartheta_{\pi_2}, \varphi_{\pi_2}) \right|^2 \nonumber \\
& \times\frac{1}{(2 \pi)^5}\, \frac{1}{8} \, \frac{1}{2 m_{\pi \pi}} 
\, \frac{
\lambda^{1/2}(m_{\pi \pi}^2, m_{\pi}^2, m_{\pi}^2)
} {
\lambda^{1/2}(s, m_{p}^2, m_{q}^2)
 } \nonumber \\
& \times\theta\big(- G(s, t, m_{\pi\pi}^2, m_p^2, m_q^2, t) \big) \, \nonumber \\
& \times\theta\big(-G(m_{\pi\pi}^2, t_2, m_\pi^2, t, m_p^2, m_\pi^2) \big) \nonumber \\
& \times d m_{\pi\pi} \, dt \, d\varphi_p \, d\cos\vartheta_{\pi_2} \,d\varphi_{\pi_2}
 \,,  \nonumber
\end{align}
with $\left| \mathcal{M}(m_{\pi \pi}, t, \cos\vartheta_{\pi_2}, \varphi_{\pi_2}) \right|^2  = (-1)\sum_{s, s'} \mathcal{M}_{\mu, s', s}^* \mathcal{M}_{s', s}^{\mu}$
and using the equations~\eqref{eq: D1}-\eqref{eq: D6}. 
With Monte Carlo importance sampling this integral can be approximated by 
\begin{align}
\label{eq: D9}
\sigma^{\gamma p} (\Delta m_{\pi\pi}, \Delta t) \approx 
\frac{1}{N} \sum_{i=0}^{N} & \,
\frac{1}{g(m_{\pi\pi, i}, t_i, \varphi_{p,i}, \vartheta_{\pi_2,i}, \varphi_{\pi_2,i})} \\
& \times \frac{1}{4}\frac{1}{2 (s - m_p^2)} (\hbar c)^2  \nonumber \\
 &\times \left| \mathcal{M}(m_{\pi \pi, i}, t_i, \cos\vartheta_{\pi_2, i}, \varphi_{\pi_2, i}) \right|^2 \nonumber \\
& \times \frac{1}{(2 \pi)^5}\, \frac{1}{8} \, \frac{1}{2\, m_{\pi \pi_i}} 
\, \frac{
\lambda^{1/2}(m_{\pi \pi, i}^2, m_{\pi}^2, m_{\pi}^2)
} {
\lambda^{1/2}(s, m_{p}^2, m_{q}^2)
 }  \nonumber \\
& \times \theta\left(- G(s, t_i, m_{\pi\pi, i}^2, m_p^2, m_q^2, t_i) \right) \, \nonumber \\
& \times \theta\left(-G(m_{\pi\pi,i}^2, t_{2,i}, m_\pi^2, t_i, m_p^2, m_\pi^2) \right) 
 \,, \nonumber 
\end{align}
with $N$ kinematic points 
$x_i = [m_{\pi\pi,i}, t_i, \varphi_{P,i}, \cos\vartheta_{\pi_2,i}, \varphi_{\pi_2,i}]$ 
randomly chosen according to 
a suitable pre-sampling density distribution $g(m_{\pi\pi}, t, \varphi_{p}, \vartheta_{\pi_2}, \varphi_{\pi_2})$. 
The event weight $w_i$ of a randomly chosen event $i$ is then given 
by the product of all terms with index $i$ in the sum on the r.\,h.\,s.\ 
of \eqref{eq: D9} divided  by $N$, such that 
$\sigma^{\gamma p} (\Delta m_{\pi\pi}, \Delta t) \approx \sum_{i=0}^{N} w_i $.

\end{document}